\documentclass[journal=jctcce,manuscript=article]{achemso}
\usepackage{indentfirst}
\usepackage{amsmath,amssymb}
\usepackage{booktabs}
\usepackage{array}
\usepackage{graphicx}
\usepackage{xcolor}
\usepackage{siunitx}
\usepackage{algorithm}
\usepackage{algpseudocode}
\algrenewcommand\algorithmicrequire{\textbf{Input:}}
\algrenewcommand\algorithmicensure{\textbf{Output:}}
\makeatletter\newcommand{\eqlabelback}[3][]{{\edef\@currentlabel{\the\numexpr\value{equation}-#3\relax#1}\label{#2}}}\makeatother
\newcommand{\jet}{JET-MLIPs}
\DeclareSIUnit\step{step}\DeclareSIUnit\day{day}\DeclareSIUnit\atom{atom}\DeclareSIUnit\flop{FLOP}\DeclareSIUnit\SM{SM}
\usepackage{xr-hyper}
\usepackage[colorlinks=true,linkcolor=blue,citecolor=blue,urlcolor=blue]{hyperref}
\makeatletter\AtBeginDocument{\def\NR@gettitle#1{\def\@currentlabelname{}}}\makeatother
\usepackage{caption}
\usepackage{rotating}
\makeatletter\renewcommand\section{\@startsection{section}{1}{\z@}{-3.5ex \@plus -1ex \@minus -.2ex}{2.3ex \@plus .2ex}{\normalfont\large\bfseries}}\renewcommand\subsection{\@startsection{subsection}{2}{\z@}{-3.25ex \@plus -1ex \@minus -.2ex}{1.5ex \@plus .2ex}{\normalfont\fontsize{13.2}{16}\selectfont\bfseries}}\let\acs@section\section\let\acs@subsection\subsection\makeatother

\makeatletter\renewcommand*\acs@author@star{}\makeatother

\title{Accelerating Equivariant Interatomic Potentials\texorpdfstring{\\}{ }for Molecular Dynamics through\texorpdfstring{\\}{ }Radial Tabulation and Kernel Fusion}
\author{Jaekyun Yoo}
\email{worbs3237@gmail.com}
\affiliation{Independent Researcher}
\author{Suyeon Ju}
\email{suyeonju224@gmail.com}
\affiliation{Independent Researcher}
\begin{document}
\begingroup\setlength{\leftmargini}{0pt}\begin{abstract}
\emergencystretch=3em
Equivariant machine-learned interatomic potentials (MLIPs) achieve near-quantum-chemical accuracy, but their computational cost and GPU memory footprint restrict molecular dynamics (MD) at extended length and time scales. Even when the tensor product (TP) is accelerated, both bottlenecks remain governed by evaluating and storing per-edge radial weights in the NequIP convolution. We introduce \jet{} (Joint-Kernel Equivariant Convolution with Tabulated Radial Weights for MLIP simulation), a GPU backend applicable to pretrained NequIP-family potentials without retraining. \jet{} replaces the radial multilayer perceptron (MLP) with a piecewise cubic Hermite table and fuses table lookup, spherical harmonics, TP and neighbor aggregation into a single kernel, eliminating per-edge radial-weight and MLP-intermediate storage during force evaluation. Across 15 released potentials from SevenNet, MACE, NequIP-OAM, Nequix and EquFlash on a 5632-atom liquid electrolyte, LAMMPS MD attains 1.5--5.9$\times$ step-time speed-ups and 2.4--24$\times$ peak memory reductions relative to each family's existing TP-accelerated backend on a single H200 GPU while preserving single-precision (fp32) accuracy and dynamical stability. These reductions enable multi-nanosecond transport-property simulations of roughly $10^4$-atom systems on a single GPU and broaden access to large-scale equivariant MLIP MD on widely available GPUs.
\end{abstract}\endgroup

\section{Introduction}
\label{sec:intro}
Machine-learned interatomic potentials (MLIPs) have extended atomistic simulations to length and time scales far beyond the
reach of quantum-chemical calculations \cite{behler_bp_2007,bartok_gap_2010,behler_review_2021,unke_mlff_review_2021}.
Building on graph neural networks (GNNs), which learn atomistic representations through message passing over the atomic
graph \cite{schutt_schnet_2017,xie_cgcnn_2018}, the field has increasingly shifted toward universal MLIPs capable of driving
production molecular dynamics (MD) directly without task-specific training data
\cite{deng_chgnet_2023,chen_m3gnet_2022,yang_mattersim_2024,neumann_orb_2024}. Central to this progress is equivariant
message passing, pioneered by NequIP \cite{batzner_nequip_2022}, in which internal features carry directional information
that rotates coherently with the atomic structure. Many modern architectures and universal potentials now adopt this design
to achieve near-quantum-chemical accuracy \cite{musaelian_allegro_2023,batatia_mace_2022}.

Despite these advances, MLIP-based MD remains constrained for many practically important materials properties that require extended length and time scales. Predicting transport properties such as diffusion coefficients, ionic conductivities and viscosities requires trajectories of several to tens of nanoseconds \cite{ju_electrolyte_2025}, corresponding to $10^5$--$10^8$ model evaluations. Both the runtime and GPU memory of each evaluation scale with atom count, so device memory limits the largest simulation cell that a single GPU can hold. This ceiling is especially restrictive for systems requiring large simulation cells, including polycrystalline solids with grain boundaries \cite{yoo_phd_2025}, solid--solid interfaces \cite{ding_interface_2025} and liquid electrolytes \cite{ju_electrolyte_2025}. Reaching such extended scales with MLIP-based MD has consequently required thousands to tens of thousands of GPUs \cite{jia_100million_sc20_2020,kozinsky_scaling_2023,lu_100million_2021}, resources unavailable to most researchers \cite{fan_gpumd_2022}. In practice, MLIP studies therefore often remain limited to static property calculations or short trajectories \cite{riebesell_matbench_2023,jacobs_mlip_guide_2025}, whereas large-scale, long-timescale classical MD still relies largely on empirical interatomic potentials.

Optimizing the GPU execution of equivariant MLIPs to reduce both runtime and memory consumption offers a direct route to extending accessible MD scales without retraining or compromising accuracy. In NequIP-family architectures such as SevenNet \cite{park_scalable_2024}, MACE \cite{batatia_mace_2022} and NequIP-OAM \cite{tan_nequip_2025}, achieving these reductions depends primarily on accelerating the shared convolution operation, which dominates both evaluation time and memory usage as each atom aggregates information from its neighbors. Within this convolution, a small neural network called the radial multilayer perceptron (MLP) maps each interatomic distance to radial weights, and a tensor product (TP) then combines these weights with neighbor features and bond directions. In e3nn \cite{geiger_e3nn_2022}, the PyTorch library underlying these MLIPs, the TP dominates convolution runtime, accounting for 89\% of the inference time of 7net-l3i5 \cite{lee_flashtp_2025}. In our measurements, replacing the e3nn TP and neighbor aggregation with cuEquivariance likewise removes 89--91\% of the convolution time of 7net-nano-5.5 \cite{oh_lightweight_2026} and 7net-0 (SI Fig.~\ref{fig:perf_summary}a). Recent acceleration efforts have therefore focused on the TP, with cuEquivariance \cite{cuequivariance}, FlashTP \cite{lee_flashtp_2025} and OpenEquivariance \cite{bharadwaj_oeq_2025} providing specialized GPU kernels that serve as the existing TP-accelerated backends of these MLIPs.

While the TP has thus been accelerated, the radial MLP, which is the other part of the convolution, has received little
acceleration. cuEquivariance, FlashTP and OpenEquivariance all read the radial weights as an input that is computed outside
their kernels. By our count, however, the radial MLP rather than the TP carries most of the arithmetic computation for each
pair of atoms (Sec.~\ref{sec:cost}, SI Table~\ref{tab:models}). Once the force calculation is taken into account, the radial MLP is also memory-demanding, because the radial weights
and intermediates of every pair must stay in memory until the forces are computed (SI Fig.~\ref{fig:perf_summary}b). Fusing the
radial MLP into the kernel removes most of this memory, but it does not reduce step time by itself,
because the arithmetic of the MLP remains. Reducing step time as well as memory thus requires a cheaper way
to obtain the radial weights.

Tabulation offers such a way. The radial weights of a trained model are a fixed function of the
interatomic distance and can therefore be precomputed once, stored in a table and read by interpolation during MD.
DP Compress \cite{lu_dpcompress_2022} applied this idea to the invariant Deep Potential model, typically accelerating MD
by an order of magnitude while raising the largest system one GPU can hold by a similar factor. However, its property
accuracy was confirmed only in double precision (fp64). In single precision (fp32), the errors did not decrease with a
finer table, which the authors attributed to the limited significant digits. Recently, DPA4C \cite{li_dpa4c_2026} extended
radial tabulation to an equivariant MLIP, but only for models trained in its own architecture, which was designed together
with its GPU kernels. Whether the combination of radial tabulation and kernel fusion extends to existing equivariant
MLIPs, including released NequIP-family models, has not been verified.

Here, we present \jet{} (Joint-Kernel Equivariant Convolution with Tabulated Radial Weights for MLIP simulation), a GPU backend for MLIP inference that accelerates MD with released checkpoints of NequIP-family MLIPs without retraining.
It supports SevenNet, MACE, NequIP-OAM, Nequix \cite{koker_nequix_2025} and EquFlash
\cite{samsungds_equflash_2026} in fp32 precision, both in LAMMPS \cite{thompson_lammps_2022} and through the Atomic
Simulation Environment (ASE) calculator interface \cite{larsen_ase_2017}. Instead of evaluating the radial MLP on the fly, \jet{} looks up precomputed radial weights from a table using cubic Hermite interpolation inside a single fused kernel that also computes the spherical harmonics, TP and neighbor aggregation. A table lookup produces no MLP intermediates, so the force backward pass needs to retain neither radial weights nor hidden activations in global memory.

We evaluated \jet{} across 15 released models spanning these five families. We first assessed numerical accuracy because acceleration is meaningful only when it preserves the predictions of the original potential. Specifically, we verified that
\jet{} reproduces the original models within the intrinsic limits of fp32 arithmetic, judged against a precision budget
set by the numerical variations of existing fp32 backends (Sec.~\ref{sec:budget}). Across single-point fp64 reference checks for the five families, isolated fp64 table evaluations and MD validation on representative models, radial tabulation added negligible force error relative to fp32 rounding while preserving energy conservation and MD observables within the run-to-run spread of existing backends (Sec.~\ref{sec:accuracy}). With accuracy preserved, LAMMPS MD of a 5632-atom electrolyte (without the D3 dispersion
correction \cite{grimme_d3_2010,grimme_bj_2011}) ran every model 1.5--5.9$\times$ faster on a single NVIDIA H200 GPU than
each family's existing TP-accelerated backend (Table~\ref{tab:mace} and Sec.~\ref{sec:nequip}) while reducing
whole-run peak GPU memory by 2.4--24$\times$ across all models (Sec.~\ref{sec:memory}). We further decomposed these speed-ups into the
contributions of tabulation, kernel fusion and LAMMPS integration (Sec.~\ref{sec:perf}) and characterized their scaling
with system size across three dense solids as well as the liquid electrolyte (Sec.~\ref{sec:scaling}). Finally, we demonstrate the practical impact of this acceleration by computing the Green--Kubo shear viscosity of liquid LiPF$_6$ carbonate electrolytes and the Li self-diffusion and ionic conductivity of Cl/S-disordered argyrodite Li$_6$PS$_5$Cl over multi-nanosecond trajectories of roughly $10^4$ atoms on a single GPU (Sec.~\ref{sec:applications}). Together, these reductions in step time and memory footprint lower trajectory cost and relax hardware constraints on both
system size and GPU selection. Such backend-level acceleration is expected to make large-scale, long-timescale MLIP simulations routine, opening access to transport and microstructural phenomena that were previously out of reach.

\section{Theory and Methods}
\label{sec:method}
\subsection{NequIP convolution and force evaluation}
\label{sec:background}
\begin{figure}[tp]
\centering
\includegraphics{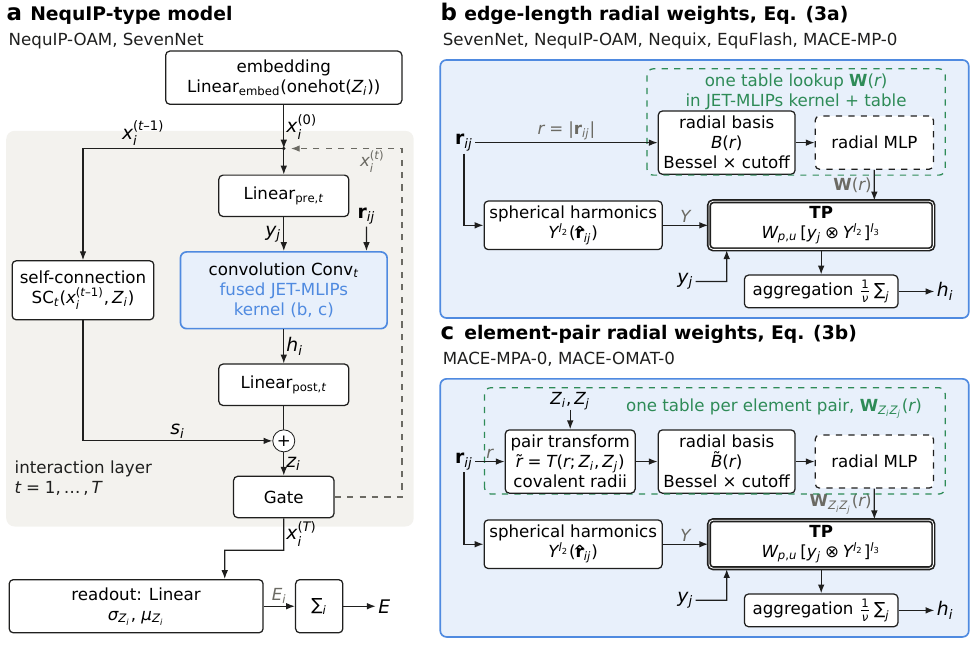}
\caption{\textbf{Interaction layer of a NequIP-type model and the convolution replaced by \jet{}.} (a) Embedding, interaction layer and
atomic-energy readout, illustrated for NequIP-OAM and SevenNet. Details outside the convolution differ between the two models. NequIP-OAM omits the self-connection in its first layer, uses a single linear readout layer and adds a Ziegler--Biersack--Littmark (ZBL) pair term to the atomic energies. (b) Convolution with edge-length radial weights. (c) Convolution with element-pair-dependent
radial weights, using one table per element pair. The radial multilayer perceptron (MLP, dashed outline) produces
$\mathbf W$, and the tensor product (TP, double outline) couples features with spherical harmonics. Blue shading denotes the
fused forward and force-backward kernel, and a single table lookup replaces the operations inside the green dashed frame.}
\label{fig:layer}
\end{figure}
The five MLIP families studied in this work, namely SevenNet \cite{park_scalable_2024}, MACE \cite{batatia_mace_2022}, NequIP-OAM \cite{tan_nequip_2025}, Nequix \cite{koker_nequix_2025} and EquFlash \cite{samsungds_equflash_2026}, all build on the equivariant message-passing architecture of NequIP \cite{batzner_nequip_2022} (Fig.~\ref{fig:layer}). In this architecture, the network embeds the atomic species $Z_i$ of each atom $i$, updates the atomic representations through $T$ interaction layers and sums the readout atomic energies $E_i$ to obtain the total potential energy $E$ (Fig.~\ref{fig:layer}a).
Within the network, each atom $i$ carries equivariant features $x_i$ organized into $m_l$ channels (indexed by $u$) for each rotation degree $l\le l_{\max}$ (SI Tables~\ref{tab:models} and~\ref{tab:family_models}). Each
channel represents a single degree-$l$ feature of $2l+1$ components with definite parity under spatial inversion, analogous to one hidden unit of a standard neural network
\cite{thomas_tfn_2018,geiger_e3nn_2022}. Under rotation, scalar features ($l=0$) remain invariant, whereas higher-degree features ($l>0$) rotate coherently with the atomic structure.

In each interaction layer $t$, a pre-convolution block (such as $\mathrm{Linear}_{\mathrm{pre},t}$ in Fig.~\ref{fig:layer}a) maps the layer-$t$ input features $x_j^{(t-1)}$ of each neighbor $j$ to $y_j$. The NequIP convolution (Fig.~\ref{fig:layer}b,c),
\begin{equation}
  h_i = \mathrm{Conv}_t\big(\{y_j,\mathbf r_{ij}\}_{j\in\mathcal N(i)}\big), \label{eq:conv}
\end{equation}
then combines the features $y_j$ of neighboring atoms $j\in\mathcal N(i)$ within the cutoff radius $r_c$ with the interatomic vectors
$\mathbf r_{ij}$ to produce the convolved features $h_i$. Although the operations before and after $\mathrm{Conv}_t$ differ across the five families, they all share the architecture of this convolution. Figure~\ref{fig:layer}a shows the NequIP-type layer of NequIP-OAM and SevenNet, and the layer equations are given in SI~\ref{app:layer_eqs}.

For each directed edge $j\to i$ with interatomic vector $\mathbf r_{ij}$ and distance $r=|\mathbf r_{ij}|$, the convolution $\mathrm{Conv}_t$ of Eq.~\eqref{eq:conv} evaluates
\begin{align}
  B_n(r) &= \frac{2}{r_c}\,\frac{\sin(c_n r)}{r}\, f_\text{cut}(r), \qquad n=1,\dots,N_B, \label{eq:bessel}\\
  \mathbf W(r) &= \text{MLP}\big(B(r)\big) \in \mathbb R^{N_W}, \qquad N_B\to H\to\dots\to H\to N_W,\ \text{SiLU},
  \refstepcounter{equation}\tag{\theequation a}\label{eq:mlpa}\\
  \mathbf W(r) &= \text{MLP}\big(\tilde B(r)\big), \qquad \tilde B_n(r) = \frac{2}{r_c}\,\frac{\sin(c_n \tilde r)}{\tilde r}\, f_\text{cut}(r), \qquad \tilde r = T(r;\, Z_i, Z_j), \tag{\theequation b}\label{eq:mlpb}\\
  m^{(p,u)}_{ij} &= W_{p,u}(r)\; \big[y^{(l_1,u)}_j \otimes Y^{l_2}(\hat{\mathbf r}_{ij})\big]^{l_3}, \label{eq:tp}\\
  h_i &= \frac{1}{\nu}\sum_{j\in\mathcal N(i)} m_{ij}. \label{eq:agg}
\end{align}
\eqlabelback[a,b]{eq:mlp}{2}%
Equation~\eqref{eq:bessel} expands the interatomic distance $r$ in a radial basis $B(r)=(B_1,\dots,B_{N_B})$ of $N_B$ Bessel functions with frequencies $c_n$, multiplied
by a smooth cutoff envelope $f_\text{cut}(r)$ that vanishes at $r_c$. A radial MLP (Eqs.~\eqref{eq:mlpa} and~\eqref{eq:mlpb}) then maps the basis vector through $n_h$ hidden layers of $H$ units with sigmoid linear unit (SiLU) activations
to the $N_W$ radial weights $\mathbf W(r)\in\mathbb R^{N_W}$ of the edge. In SevenNet,
NequIP-OAM, Nequix, EquFlash and MACE-MP-0, the MLP takes $B(r)$ directly so that $\mathbf W(r)$ depends on the edge length $r$ alone (Eq.~\eqref{eq:mlpa},
Fig.~\ref{fig:layer}b). In MACE-MPA-0 and MACE-OMAT-0, by contrast, a distance transform $T$ first rescales the Bessel argument to $\tilde r$ based on the covalent radii
of $(Z_i,Z_j)$ while leaving $f_\text{cut}(r)$ evaluated at $r$, yielding the modified basis $\tilde B(r)$ and element-pair-dependent radial weights (Eq.~\eqref{eq:mlpb}, Fig.~\ref{fig:layer}c).

Equation~\eqref{eq:tp} introduces angular dependence into the edge message $m^{(p,u)}_{ij}$ through a tensor product (TP) that couples the degree-$l_1$ input feature $y^{(l_1,u)}_j$ of neighbor $j$ in channel $u$ with the real spherical harmonic $Y^{l_2}$ of the unit edge vector
$\hat{\mathbf r}_{ij}=\mathbf r_{ij}/r$. The bracket $[\,\cdot\,]^{l_3}$ projects their tensor product $\otimes$ onto an output degree $l_3$ that satisfies $|l_1-l_2|\le l_3\le l_1+l_2$ and conserves parity under spatial inversion (SI Fig.~\ref{fig:sevennet_layer}a). Each valid degree triple $p=(l_1,l_2,l_3)$ defines one path of the TP
\cite{geiger_e3nn_2022}.
Whereas the linear layers before and after the convolution mix channels of the same degree, the TP itself acts on each channel $u$ independently. Each path $p$ thus uses one scalar radial weight $W_{p,u}(r)$ per channel $u$, giving $N_W=\sum_p m_{l_1(p)}$ radial weights per
edge.
Finally, Eq.~\eqref{eq:agg} collects $m^{(p,u)}_{ij}$ across all paths and channels into the edge message $m_{ij}$, sums over all neighbors $j\in\mathcal N(i)$ within $r_c$ and divides by the normalization constant $\nu$ stored in the checkpoint to yield the
convolution output $h_i$ of Eq.~\eqref{eq:conv}.

Each MD step requires one forward pass to evaluate $E$ followed by one backward pass to compute the atomic forces $\mathbf F=-\partial
E/\partial\mathbf r$, because the forces are the negative gradient of the total energy with respect to the atomic positions. The kernels presented in this work replace
both the forward and force-backward passes of Eqs.~\eqref{eq:bessel}--\eqref{eq:agg}, encompassing the edge embedding (defined as the evaluation of $B_n$, $f_\text{cut}$ and $Y^{l_2}$ from $\mathbf r_{ij}$) and the convolution of each interaction layer. By contrast, model training requires differentiating through the force backward pass with respect to the network parameters and lies outside the scope of this work.

\subsection{Computational bottlenecks in equivariant MLIPs}
\label{sec:cost}
\begin{figure}[tp]
\centering
\includegraphics{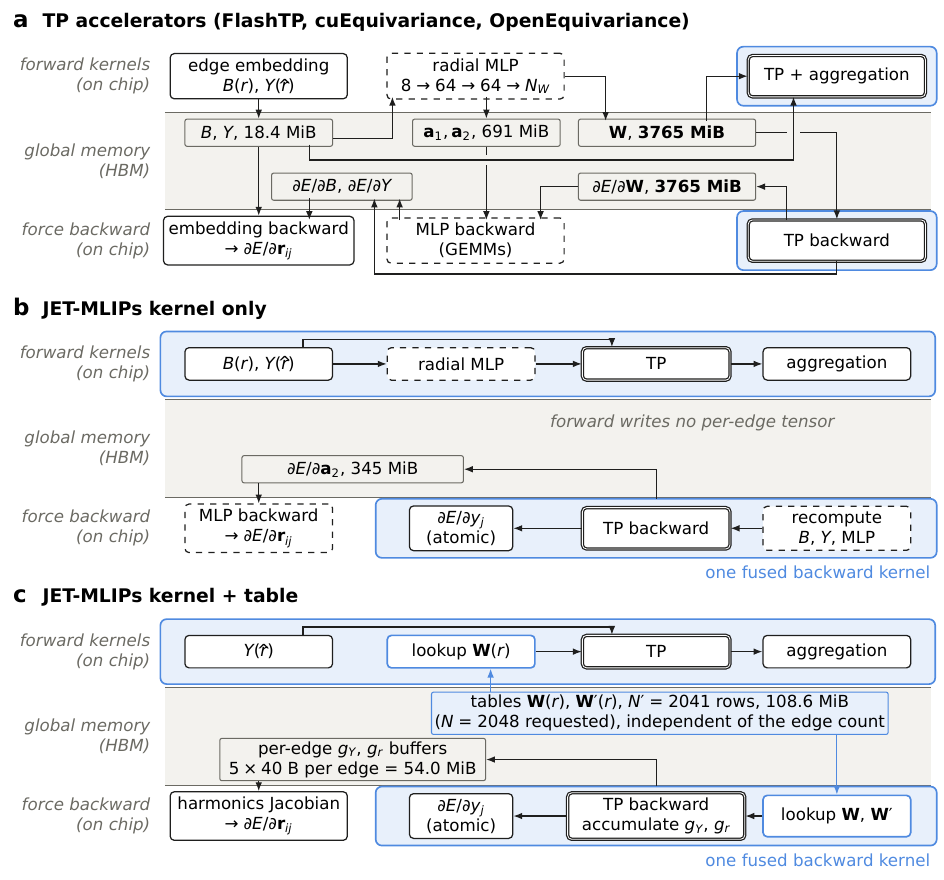}
\caption{\textbf{Data flow of one 7net-0 interaction layer with TP accelerators and with \jet{}.} (a) TP accelerators fuse the TP with gather and aggregation (blue boxes) but write the radial weights $\mathbf W$, their gradient $\partial E/\partial\mathbf W$, the radial basis $B$, spherical harmonics $Y$ and the MLP activations $\mathbf a_1$, $\mathbf a_2$ to global memory. (b) The fused kernel-only mode keeps $\mathbf W$ and $\partial E/\partial\mathbf W$ in registers. Only $\partial E/\partial\mathbf a_2$ crosses global memory to a second kernel that completes the MLP backward pass. (c) The default table replaces the MLP with a cubic Hermite lookup, and a short second kernel converts the buffered gradients with respect to the harmonics and edge length, $g_Y$ and $g_r$, into $\partial E/\partial\mathbf r_{ij}$. Tensor sizes are summed over the five interaction layers of 7net-0 on 5632 atoms, with $B$ and $Y$ computed once. Gray boxes denote per-edge tensors in global memory (HBM, high-bandwidth memory), dashed outlines denote the MLP, double outlines denote the TP and the force backward pass proceeds from right to left. GEMM denotes a dense matrix product.}
\label{fig:overview}
\end{figure}
Within the convolution, the radial MLP and the TP account for most of the per-edge work, with the radial MLP generating the radial weights $\mathbf W(r)$ and the
TP consuming them. On modern GPUs, the execution time of these stages depends not only on their arithmetic operation counts but also on the volume of per-edge intermediate data read from and written to global memory. In unfused implementations such as e3nn \cite{geiger_e3nn_2022}, evaluating the neighbor gather, TP and scatter as separate operations \cite{fey_pyg_2019} materializes per-edge feature and message tensors in global memory, and the TP stage consumes over 80\% of the model-call time \cite{lee_flashtp_2025} (SI~\ref{sec:time}). Existing TP accelerators, including FlashTP, cuEquivariance and OpenEquivariance, exploit the sparse Clebsch--Gordan coefficients and eliminate this TP memory traffic by fusing the gather, TP and scatter steps into a single kernel (Fig.~\ref{fig:overview}a).

Even after the TP is fused, however, the radial weights remain a major bottleneck because the radial MLP dominates the per-edge arithmetic while $\mathbf W(r)$ dominates the remaining per-edge memory. To quantify the arithmetic cost, we count the number of fused multiply--add (FMA) operations ($a\cdot b+c$, evaluated as a single instruction) per edge. Although the radial MLP of Eq.~\eqref{eq:mlp} takes only a scalar pair distance $r$ (expanded into $N_B$ basis values) as input, it must generate a separate weight $W_{p,u}(r)$ for every TP path and channel, producing a wide output vector of length $N_W$. Mapping the $N_B$ basis values through $n_h$ hidden layers of $H$ units to these $N_W$ radial weights requires one matrix--vector product per layer, with one FMA per matrix entry, corresponding to $N_BH$ FMA in the input layer, $H^2$ in each of the $n_h-1$ hidden-to-hidden layers and $HN_W$ in the output layer. The radial MLP thus performs $N_BH+(n_h-1)H^2+HN_W$ FMA per edge ($8\cdot64+1\cdot64^2+64\cdot960=\num{66048}$ for 7net-0), most of
them in the output layer due to the large $N_W$. Specifically, the output layer alone requires $H$ FMA ($64$ in 7net-0) to evaluate each scalar weight $W_{p,u}(r)$. By contrast, the TP of Eq.~\eqref{eq:tp} requires only a few FMA to evaluate the product $[y^{(l_1,u)}_j \otimes Y^{l_2}]^{l_3}$ weighted by each $W_{p,u}(r)$ (such as 6 FMA for the $(1,1,1)$ path), because most Clebsch--Gordan coefficients vanish by symmetry (SI Fig.~\ref{fig:sevennet_layer}b,c). Summed over all paths and channels of an interaction layer in 7net-0, the TP performs only \num{6848} FMA per edge, approximately one-tenth of the radial MLP, and the radial MLP similarly dominates across the other models (SI Table~\ref{tab:models}).

In memory, each radial weight vector $\mathbf W(r)$ contains $N_W$ fp32 values, occupying $4N_W$ bytes per edge in each layer (3840 bytes for 7net-0). When the radial MLP is evaluated as a separate step, the forward pass must write $\mathbf W(r)$ and the MLP hidden activations to global memory for every edge, and the force backward pass must read them back while producing the gradient $\partial E/\partial\mathbf W$ of the same size. The total footprint of these tensors scales with the number of edges times $N_W$, because they are stored for every directed edge rather than per atom and each atom in condensed-phase MD typically has dozens of neighbors within $r_c$. This per-edge scaling both slows execution through memory traffic and limits the maximum system size that fits on a single GPU. For 7net-0 on the 5632-atom electrolyte
(Fig.~\ref{fig:overview}a), the radial weights $\mathbf W$ alone occupy \SI{3765}{\mebi\byte} across the five interaction layers and up to
\SI{1036}{\mebi\byte} in a single layer.

Existing TP accelerators keep the radial MLP outside the fused kernel and accept $\mathbf W(r)$ as a precomputed input from global memory, which we attribute to two practical considerations. First, decoupling the TP from $\mathbf W(r)$ keeps the TP kernel independent of each model's radial basis and MLP architecture, delegating radial-weight evaluation to standard dense kernels and automatic differentiation. Second, fusing a multi-layer MLP directly into an already register-pressured TP kernel requires holding the MLP hidden activations and evaluating their force backward pass on-chip, which easily induces severe register spilling while still leaving the heavy per-edge MLP arithmetic intact. To eliminate both the memory traffic of $\mathbf W(r)$ and the arithmetic cost of the radial MLP, \jet{} combines fused convolution kernels with radial-weight tabulation (Sec.~\ref{sec:fused}).

\subsection{\jet{} convolution kernels}
\label{sec:fused}
\label{sec:table}

Similarly to existing TP accelerators \cite{cuequivariance,lee_flashtp_2025,bharadwaj_oeq_2025}, \jet{} fuses the neighbor gather, TP and scatter into a single kernel. Its kernel loops over the incident edges $j\in\mathcal N(i)$ of each destination atom $i$ while accumulating the output features $h_i$ in registers. Each kernel is compiled for the dimensions of the layer, and wide layers are partitioned into path groups when needed to fit the register budget (SI~\ref{sec:groups}). Unlike existing TP kernels, which take precomputed spherical harmonics $Y^{l_2}$ and radial weights $\mathbf W(r)$ from global memory (Fig.~\ref{fig:overview}a), \jet{} fuses the entire per-edge pipeline of Eqs.~\eqref{eq:bessel}--\eqref{eq:agg} into the kernel (Fig.~\ref{fig:overview}b, SI Algorithm~\ref{alg:fwd}). When the radial MLP is evaluated directly inside the kernel (the kernel-only mode, implemented for SevenNet in both fp32 and fp64 in SI~\ref{app:accuracy_full}), the forward pass computes the edge distance $r$, the spherical harmonics $Y^{l_2}(\hat{\mathbf r}_{ij})$, the radial basis $B(r)$ and the radial MLP for $\mathbf W(r)$ on the fly, accumulating the TP contributions directly in registers without writing $\mathbf W(r)$ or any other per-edge tensor to global memory.

In the force backward pass, whereas existing TP accelerators return $\partial E/\partial\mathbf W$ and $\partial E/\partial Y^{l_2}$ to global memory for external backward steps, the fused backward kernel of \jet{} reloads $y_j$, recomputes $Y^{l_2}(\hat{\mathbf r}_{ij})$ and keeps $\mathbf W(r)$ and $\partial E/\partial\mathbf W$ in registers (SI Algorithm~\ref{alg:bwd}). Across path groups, the kernel accumulates $\partial E/\partial y_j$ and, after applying the spherical-harmonics Jacobian in the kernel, the edge-vector gradient $\partial E/\partial\mathbf r_{ij}$ via GPU atomic additions. Propagating $\partial E/\partial\mathbf W$ through the radial MLP ($N_B=8$, two hidden layers of $H=64$ units in SevenNet) can be performed in two ways (SI~\ref{app:fused} and SI~\ref{app:memory}). In the split-kernel scheme used here (Fig.~\ref{fig:overview}b), the first kernel buffers the second hidden-layer gradient $\partial E/\partial\mathbf a_2$ per edge in global memory and a second kernel completes the MLP backward pass, whereas an alternative recomputation scheme avoids this buffer by repeating the MLP evaluation and backward pass in each path group. Although fusing the MLP eliminates $\mathbf W(r)$ and $\partial E/\partial\mathbf W$ from global memory, both schemes still incur the full arithmetic cost of the radial MLP and must either buffer or recompute its hidden activations during the backward pass.

To eliminate both the radial-MLP arithmetic and the backward-pass overhead of the kernel-only mode, \jet{} replaces the in-kernel radial basis and MLP of Eqs.~\eqref{eq:bessel} and~\eqref{eq:mlp} with a cubic Hermite spline lookup in its kernel + table mode (Fig.~\ref{fig:overview}c). This replacement is possible because the radial weights
$\mathbf W(r)$ of a trained model are a fixed function of the edge length $r$ alone in Eq.~\eqref{eq:mlpa} or of $r$ for each element pair $(Z_i,Z_j)$ in Eq.~\eqref{eq:mlpb},
building on radial tabulation in invariant potentials \cite{lu_dpcompress_2022,lysogorskiy_pace_2021} and in the co-designed architecture of DPA4C \cite{li_dpa4c_2026}. On each edge, the fused forward kernel looks up $\mathbf W(r)$ directly from the table, while the fused backward kernel evaluates both $\mathbf W(r)$ and its radial derivative $\mathbf W'(r)$ from the same piecewise cubic polynomial to obtain the radial gradient $g_r$. Across path groups, this kernel sums $g_r$ and the per-edge gradient with respect to the spherical harmonics, $g_Y$, into a buffer $(g_Y,g_r)$ of $(l_{\max}+1)^2+1$ values per edge, which a second kernel converts once per edge into $\partial E/\partial\mathbf r_{ij}$ via the spherical-harmonics Jacobian. This buffer is the sole per-edge global-memory intermediate of the tabulated convolution (Fig.~\ref{fig:overview}c). Evaluating $\mathbf W(r)$ and $\mathbf W'(r)$ from the same polynomial also ensures that the evaluated forces remain the exact negative gradient of the
tabulated energy, avoiding the systematic energy drift caused by non-conservative force errors (Sec.~\ref{sec:budget}).

At deployment, \jet{} constructs one table per layer by evaluating $\mathbf W(r)$ and
$\mathbf W'(r)$ in fp64 on $N'$ uniform grid nodes of spacing $h=r_c/(N'-1)$ over $[0,r_c]$ and storing the resulting table in fp32, occupying only $N'\times 4N_W$ fp32 values per layer in global memory regardless of system size. For models with element-pair-dependent radial weights (Eq.~\eqref{eq:mlpb}, such as MACE-MPA-0 and MACE-OMAT-0), one table is built per element pair in each layer and selected via a per-edge pair index, with table memory scaling with the number of element pairs present in the system (SI~\ref{app:limits}). Inside the kernels, loading four fp32 values for each edge and weight index $c$ supplies the cubic Hermite
interpolant of $W_c(r)$ in the forward pass and its radial derivative $W'_c(r)$ in the backward pass, with interpolation errors scaling as $\mathcal O(h^4)$ in $W_c(r)$ and $\mathcal O(h^3)$ in $W'_c(r)$ \cite{birkhoff_hermite_1967}. To reach the fp32 rounding floor within the precision budget of Sec.~\ref{sec:budget}, the table construction precomputes adjacent-node differences of $W_c$ in fp64 before casting to fp32 to avoid cancellation error, aligns a grid node with the cutoff onset radius $r_\text{on}$ in SevenNet and EquFlash ($N'\le N$) so that the second-derivative discontinuity of $f_\text{cut}(r)$ falls on a node boundary, and represents the inverse grid spacing $1/h=(N'-1)/r_c$ as the sum of two fp32 numbers to prevent systematic edge-length rescaling (SI~\ref{app:table_impl}).

\subsection{Precision budget}
\label{sec:budget}
Kernel fusion evaluates the exact expressions of the model and alters only floating-point rounding and summation order, whereas radial tabulation approximates the radial MLP with a piecewise cubic polynomial. We judge both modifications against the intrinsic numerical variations of the unmodified existing fp32 backends, defining these baseline variations as the precision budget. An accelerated backend meets this budget when its deviations remain comparable to those of the existing fp32 backend itself. At the single-call level, the primary scale is the fp32 rounding error relative to an fp64 evaluation with the same parameters, where the fp32 total energy is also bounded by the finite resolution of fp32 numbers \cite{goldberg_fp_1991}. A second static scale is the spread between repeated fp32 calls on the same configuration, caused by non-associative GPU floating-point addition across varying thread summation orders \cite{demmel_reproducible_2015,shanmugavelu_nonassoc_2024}. Evaluating the radial table alone in fp64 further tests whether its isolated interpolation error lies below both fp32 scales. Over an MD trajectory, chaotic dynamics amplifies single-step differences exponentially \cite{allen_tildesley_2017,stoddard_ford_1973,posch_hoover_1988}, causing trajectories started from the same initial state to diverge for every backend alike. Trajectory-level agreement is therefore assessed through energy conservation in constant-energy (NVE) runs \cite{fu_esen_2025} and through thermodynamic and structural observables, which must differ between backends by amounts comparable to repeated runs of the existing fp32 backend. This budget serves as an empirical acceptance scale on the tested configurations rather than a test of statistical equivalence (SI~\ref{app:accuracy_full} and~\ref{app:md}).

\subsection{Model adapters and MD integration}
\label{sec:md}
To run the kernels of Sec.~\ref{sec:fused} in pretrained models without retraining, \jet{} provides model adapters for each of the five families (SI~\ref{app:lammps}). For a given checkpoint, the adapter replaces the edge embedding, radial MLP, TP and neighbor aggregation of each interaction layer with the fused convolution while leaving the remaining modules unchanged, including the species embedding, linear projections and energy readout. The converted model runs directly through each family's ASE calculator \cite{larsen_ase_2017} for single-point and relaxation workflows, and through LAMMPS \cite{thompson_lammps_2022} interfaces for MD simulations. In LAMMPS, SevenNet runs through either its native \texttt{e3gnn} pair style or the machine-learning interatomic potential (ML-IAP) interface, whereas the other four families run through ML-IAP. To reduce per-step data-transfer and kernel-launch overhead during MD, the integration uploads the LAMMPS neighbor list to the GPU only on rebuild steps and selects the current in-cutoff edges on the GPU at each timestep. It then replays the model evaluation as a single CUDA graph, launching the recorded sequence of GPU operations as a unit. A GPU pair style \texttt{jet/d3} (JET-D3) is also provided to evaluate the D3 dispersion correction in LAMMPS alongside any of the model pair styles (SI~\ref{sec:d3}), although all main-text timings exclude D3 unless noted.

\subsection{Computational details}
\label{sec:protocol}
To compare kernels and MD integration on a common workload, we benchmark the models on a 5632-atom liquid electrolyte of LiPF$_6$ in
ethylene carbonate (EC) and ethyl methyl carbonate (EMC), with cell construction and hardware specifications given in
SI~\ref{app:protocol}.
The cubic simulation cell ($L=\SI{38.74}{\angstrom}$) contains 64 LiPF$_6$, 128 EC and 256 EMC formula units, corresponding to an EC:EMC mass ratio of 30:70 and a salt concentration of \SI{1.83}{\mole\per\liter} at its equilibrated density of \SI{1.361}{\gram\per\centi\metre\cubed}. The cell was equilibrated once by \SI{200}{\pico\second} of isothermal--isobaric MD with 7net-0 and the D3(BJ) dispersion correction, and the same configuration is used for every model and backend (SI~\ref{sec:system}).
Although this concentration exceeds the approximately \SI{1}{\mole\per\liter} of typical Li-ion electrolytes, the configuration (atom and edge counts in SI Table~\ref{tab:family_models}) serves here as a common test system for the accuracy evaluations of Sec.~\ref{sec:accuracy} and the 15-model performance comparison of Sec.~\ref{sec:nequip}, whereas the transport calculations of Sec.~\ref{sec:applications} use separate cells at the compositions described below. Across the 15-model comparisons, we group the models by the trainable-parameter count of the pretrained checkpoint into small ($<10^6$),
medium ($10^6$--$10^7$) and large ($>10^7$) models (SI Table~\ref{tab:family_models}).

Throughout Sec.~\ref{sec:results}, ``\jet{} kernel only'' denotes the fused kernels that evaluate the radial MLP on the fly, and ``\jet{} kernel + table'' denotes the fused kernels with the radial table ($N=2048$, or $N=4096$ for MACE). In the cross-family comparisons (Table~\ref{tab:mace}), we compare \jet{} kernel + table against each family's unaccelerated ``upstream model'' (e3nn for SevenNet, and the unconverted models for MACE and NequIP-OAM) and the ``existing TP-accelerated backend'' released by its developers. Specifically, this existing TP-accelerated backend is cuEquivariance through the Kokkos ML-IAP interface for SevenNet and MACE \cite{johansson_lammps_kokkos_2025}, OpenEquivariance through ML-IAP for NequIP-OAM, and the GGNN ML-IAP wrapper with cuEquivariance for EquFlash \cite{samsungds_equflash_2026}. For Nequix, the existing TP-accelerated backend denotes the nequix PyTorch model with the OpenEquivariance kernels of its package, which we ran in the \jet{} CUDA-graph wrapper without the \jet{} kernels. In these LAMMPS runs, \jet{} executes through \texttt{e3gnn} for SevenNet (LAMMPS stable 2Aug2023 update~3) and through its ML-IAP wrapper for the other families (LAMMPS 22Jul2025 update~6). CUDA-graph replay is used for all models except EquFlash V2. All benchmarks were executed on one NVIDIA H200 NVL GPU of the same node within the same measurement series (hardware, software versions and run protocol in SI~\ref{sec:hardware} and SI~\ref{sec:runprotocol}).

As our primary timing metric, we report the rebuild-free step time (Table~\ref{tab:mace} and SI Fig.~\ref{fig:perf_summary}c), defined as the median per-step time across the 45 ten-step intervals between steps 50 and 500. This window excludes
startup and CUDA-graph capture, while the median across ten-step intervals is insensitive to the 9--10 intervals that contain a neighbor-list rebuild (SI~\ref{sec:runprotocol}). The mean
step time including rebuilds, $(t_{500}-t_{50})/450$ with $t_n$ denoting the elapsed time at step $n$, is reported separately in SI~\ref{sec:rebuilds}. Unless a caption states otherwise, each table entry is the median of three independent runs, whereas each point in a system-size scan uses the first run of its configuration, following a selection rule fixed before the results were inspected.

We evaluate GPU memory consumption both over a full LAMMPS MD run (the whole-run peak, Table~\ref{tab:mace}) and for a standalone energy-and-force evaluation outside LAMMPS across system sizes (the model-call peak, Sec.~\ref{sec:memory}). For runs executed through Python and ML-IAP, we record the peak memory actively allocated by PyTorch (\texttt{torch.cuda.max\_memory\_allocated}). For SevenNet runs through the C++ \texttt{e3gnn} pair style, where internal PyTorch allocation counters are not exposed, we instead record the total GPU memory reported by \texttt{nvidia-smi}, which provides an upper bound on the model's memory use because it also includes the CUDA runtime context and cached memory reserved by the allocator (SI~\ref{sec:runprotocol}).

For the transport applications of Sec.~\ref{sec:applications}, we computed the shear viscosity of three liquids with 7net-nano-6.0 and D3(BJ) using \jet{} kernel + table and JET-D3 in LAMMPS. The three systems are pure EMC (670 molecules, \num{10050} atoms) and two LiPF$_6$ solutions in a solvent of 292 EC and 440 EMC molecules containing either 65 LiPF$_6$ (\num{10040} atoms, \SI{0.94}{\mole\per\liter}) or 148 LiPF$_6$ (\num{10704} atoms, \SI{2.02}{\mole\per\liter}). Each cell was equilibrated at \SI{298.15}{\kelvin} for \SI{1}{\nano\second} at constant pressure (\SI{1}{\bar}), rescaled to the mean volume of the second half of that run and equilibrated for a further \SI{100}{\pico\second} at constant volume. From this configuration, we ran three \SI{10}{\nano\second} constant-volume production replicas with independent initial velocities using a Nos\'e--Hoover thermostat at \SI{298.15}{\kelvin} and a \SI{1}{\femto\second} timestep. The shear viscosity was obtained from the plateau of the running Green--Kubo integral of the stress autocorrelation function, averaged over the five independent shear components sampled every \SI{2}{\femto\second}, with uncertainty given by the standard error of the mean across the three replicas (SI~\ref{app:applications}).

We also computed the Li self-diffusion coefficient $D$ of the argyrodite Li$_6$PS$_5$Cl with the same model, D3 correction and backend. Starting from $6\times6\times6$ supercells of the ordered structure (Materials Project mp-985592, \num{11232} atoms), we exchanged a fraction $x$ of the Cl atoms with the free S atoms not bonded to P. We considered $x=0.25$, 0.50 and 0.615, with $x=0.615$ matching the site disorder measured by neutron diffraction \cite{kraft_argyrodite_2017}. Each cell was annealed for \SI{50}{\pico\second} at \SI{600}{\kelvin} to redistribute Li, equilibrated for \SI{200}{\pico\second} at the target temperature and \SI{1}{\bar} with isotropic volume fluctuations and then fixed at the mean volume of the last \SI{100}{\pico\second}. Production runs of 4.5--\SI{20}{\nano\second} at 300--\SI{400}{\kelvin} were performed at constant volume with a Nos\'e--Hoover thermostat and a \SI{2}{\femto\second} timestep. We obtained $D$ from a linear fit to the mean square displacement (MSD) of the 5184 Li atoms between the lag time at which the MSD reaches \SI{14}{\angstrom\squared} and half the run length, and estimated its relative standard deviation following He et al. \cite{he_variance_2018}. The ionic conductivity was then estimated from the Nernst--Einstein relation $\sigma=n_\text{Li}e^2D/(k_BT)$ assuming a unit Haven ratio, where $n_\text{Li}$ is the Li number density, $e$ is the elementary charge, $k_B$ is the Boltzmann constant and $T$ is the temperature (SI~\ref{app:applications}).

\section{Results and Discussion}
\label{sec:results}
\subsection{Numerical accuracy and MD validation}
\label{sec:accuracy}
We first evaluated single-point energy and force errors relative to the fp64 reference for both \jet{} kernel + table and each family's existing fp32 TP-accelerated backend across ten MD frames of the 5632-atom electrolyte cell (Sec.~\ref{sec:protocol} and SI~\ref{sec:system}). Three models whose fp64 evaluations exceed GPU memory at 5632 atoms (7net-omni \cite{kim_optimizing_2026}, NequIP-OAM L and EquFlash V1) were tested on a 352-atom cell. EquFlash V2 exceeds GPU memory in fp64 even on the reduced cell and was therefore compared directly with its existing fp32 TP-accelerated backend (SI Table~\ref{tab:family_acc}). All \jet{} kernel + table evaluations use $N=4096$ grid nodes for MACE and $N=2048$ for the other four families based on the grid-convergence analysis that follows. Across SevenNet, NequIP-OAM and Nequix, the root-mean-square (rms) energy errors of both backends remain at or below \SI{6.1e-7}{\electronvolt\per\atom}, about \num{1e-7} of the per-atom energy and within twice the fp32 resolution of the total energy. The maximum is a constant shift of \jet{} kernel + table in NequIP-OAM L, which vanishes when the table is evaluated in fp64 and leaves the forces unchanged. The SI expresses these energy errors in units in the last place (ulp) of the fp32 total energy (SI~\ref{app:accuracy_full}). In MACE and EquFlash, fp32 summation of atomic energies elevates the baseline energy error and \jet{} kernel + table remains within 1.01 times the energy error of the existing TP-accelerated backend. Across the 14 models with an fp64 reference, \jet{} kernel + table achieves 0.54--1.02 times the rms force error of the existing TP-accelerated backend and 0.88--1.07 times its maximum force-component error (Table~\ref{tab:accratio}). The rms ratio lies within 0.97--1.02 for 11 of these models. The other three models (MACE-OMAT-0, NequIP-OAM S and MACE-MPA-0) exhibit lower rms force errors than their existing TP-accelerated backends (ratios of 0.54, 0.69 and 0.71). This improvement likely arises because precomputing the radial weights in fp64 avoids fp32 rounding in the radial MLP and distance transform (Sec.~\ref{sec:table}).

\begin{table}[tp]
\centering
\caption{\textbf{Force error of \jet{} kernel + table relative to the existing TP-accelerated backend.} For every model with an fp64
reference, the rms force error of \jet{} kernel + table against the fp64 reference is at most 1.02 times that of the existing fp32 TP-accelerated backend. The rms error of a
force component against the fp64 evaluation of the same model is given in \SI{e-6}{\electronvolt\per\angstrom}, followed by
the ratios of the rms and maximum errors, \jet{} kernel + table divided by the existing TP-accelerated backend (FlashTP for SevenNet, ten electrolyte MD frames, default $N$, one H200). SevenNet rms ratios use unrounded errors and carry
the 95\% confidence interval over frames, while the other ratios use the values of SI Table~\ref{tab:family_acc}. EquFlash V2 is omitted because its fp64 reference exceeds GPU memory.}
\label{tab:accratio}
\footnotesize\setlength{\tabcolsep}{4pt}
\begin{tabular}{lrcccc}
\toprule
model & atoms & existing & \jet{} & rms ratio & max ratio \\
\midrule
7net-nano-5.5 & 5632 & 1.42 & 1.42 & 1.01 [1.00, 1.01] & 0.89 \\
7net-nano-6.0 & 5632 & 1.51 & 1.48 & 0.98 [0.98, 0.99] & 0.89 \\
7net-0 & 5632 & 1.29 & 1.27 & 0.98 [0.98, 0.99] & 1.03 \\
7net-omni & 352 & 1.27 & 1.27 & 0.99 [0.96, 1.03] & 0.88 \\
\addlinespace
MACE-MP-0 medium & 5632 & 4.54 & 4.54 & 1.00 & 0.96 \\
MACE-MPA-0 medium & 5632 & 6.77 & 4.80 & 0.71 & 1.01 \\
MACE-OMAT-0 medium & 5632 & 8.96 & 4.86 & 0.54 & 0.95 \\
MACE-MP-0 large & 5632 & 4.82 & 4.67 & 0.97 & 0.99 \\
\addlinespace
NequIP-OAM S & 5632 & 7.70 & 5.33 & 0.69 & 0.99 \\
NequIP-OAM M & 5632 & 5.54 & 5.38 & 0.97 & 1.06 \\
NequIP-OAM L & 352 & 3.50 & 3.58 & 1.02 & 1.07 \\
\addlinespace
nequix-mp-1 & 5632 & 4.64 & 4.59 & 0.99 & 1.02 \\
nequix-oam-1 & 5632 & 4.63 & 4.57 & 0.99 & 0.99 \\
\addlinespace
EquFlash V1 & 352 & 1.77 & 1.81 & 1.02 & 0.88 \\
\bottomrule
\end{tabular}
\end{table}

\jet{} kernel + table exhibits smaller run-to-run variations than the existing TP-accelerated backends in repeated fp32 evaluations of the same configuration (Sec.~\ref{sec:budget}). Consecutive evaluations across the 15 models differ by at most \SIrange{1.1e-6}{5.5e-6}{\electronvolt\per\angstrom} in any force component with \jet{} kernel + table, compared with \SIrange{1.4e-6}{2.1e-5}{\electronvolt\per\angstrom} for the existing fp32 TP-accelerated backends (SI Table~\ref{tab:family_acc}). The \jet{} spread is smaller in every model. The forward pass of the \jet{} convolution requires no GPU atomic additions and is therefore deterministic, leaving the backward pass as the sole source of this variation within the convolution (SI~\ref{app:accuracy_full}).

\begin{figure}[tp]
\centering
\includegraphics{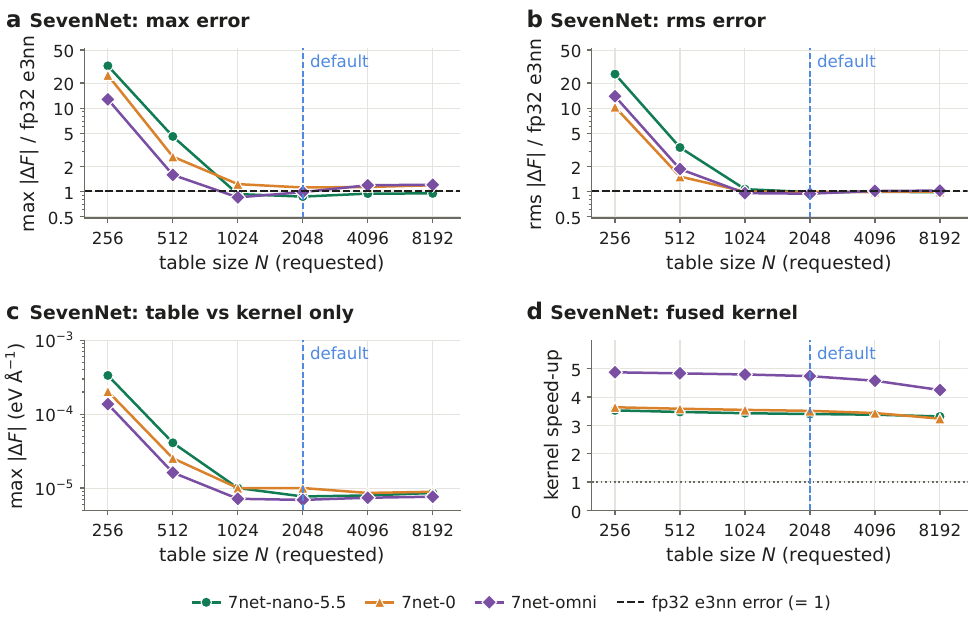}
\caption{\textbf{Radial-table accuracy and speed versus grid size $N$.} (a,b) Maximum and rms force error of fp32
\jet{} kernel + table against fp64 e3nn, divided by fp32 e3nn error. (c) Maximum force difference from \jet{} kernel only
(\si{\electronvolt\per\angstrom}). (d) Layer-summed forward plus force-backward kernel speed-up (kernel-only time divided by kernel + table time,
5632-atom electrolyte, one H200, except 352 atoms for 7net-omni in (a,b)). Dashed vertical lines mark default $N=2048$,
and horizontal lines mark parity.}
\label{fig:table}
\end{figure}

The radial table is the only analytical approximation in \jet{}, and its error is governed by the number of grid nodes $N$. The interpolation error of the table decreases as the fourth power of the grid spacing in the radial weights and as the third power in their derivatives (Sec.~\ref{sec:table} and SI~\ref{app:accuracy_full}), so that a twofold increase in $N$ reduces the error of the radial weights 16-fold and that of their derivatives 8-fold. The measured fp32 force error accordingly decreases with $N$ until it plateaus at the fp32 rounding floor of the existing backend (Fig.~\ref{fig:table}a,b and SI Fig.~\ref{fig:table_family}). All tested models converge by $N=1024$--2048 except MACE-MP-0 medium, which converges at $N=4096$ and sets the larger table size used for MACE. To isolate the table error, we chose SevenNet, whose ASE interface forms the edge vectors in fp64 and which therefore has the lowest fp32 rounding floor in single-point evaluations (SI~\ref{app:accuracy_full}). For the other families, the table was tested through the convergence of the total fp32 error with $N$ (SI Fig.~\ref{fig:table_family}). In SevenNet at $N=2048$, the maximum force difference between fp32 \jet{} kernel + table and kernel only is \SI{1.0e-5}{\electronvolt\per\angstrom} (Fig.~\ref{fig:table}c), matching the fp32 rounding floor. Isolating the interpolation error in fp64 yields rms force errors 9--55 times below the fp32 rounding floor in the electrolyte (SI Fig.~\ref{fig:g8}) and 43--188 times below it in LiCoO$_2$, diamond Si and amorphous SiO$_2$ (SI Table~\ref{tab:solids_acc}). The maximum force-component perturbation from tabulation in the electrolyte does not exceed \SI{1.8e-6}{\electronvolt\per\angstrom}, well below the \SIrange{3.1e-6}{6.4e-6}{\electronvolt\per\angstrom} run-to-run spread of fp32 FlashTP and e3nn.

In MD, chaotic dynamics amplifies single-step force differences exponentially over time. For 7net-nano-5.5 simulations started from identical initial positions and velocities, the rms atomic position difference between two trajectories grows by a factor of Euler's number every 110~fs for repeated FlashTP runs, 109~fs for repeated \jet{} kernel + table runs and 115~fs between the two backends (SI~\ref{app:chaos}). These nearly identical timescales show that switching from the existing TP-accelerated backend to \jet{} kernel + table causes trajectories to diverge no faster than repeating the simulation with the same backend. The tabulated forces remain the exact negative gradient of the tabulated energy (Sec.~\ref{sec:table}) and introduce no non-conservative force error, and NVE simulations confirm energy conservation. The magnitude of the energy drift stays below \SI{3e-5}{\milli\electronvolt\per\atom\per\pico\second} over \SI{100}{\pico\second} NVE simulations of 7net-nano-5.5 and 7net-0 across FlashTP, \jet{} kernel only and \jet{} kernel + table. Extending 7net-nano-5.5 to \SI{2}{\nano\second} in both \jet{} modes introduces no drift above this bound (SI~\ref{app:md}). Similarly, the drift ranges of \jet{} kernel + table and the existing TP-accelerated backend overlap over three-seed \SI{10}{\pico\second} NVE simulations of MACE-MP-0 medium and NequIP-OAM M (SI~\ref{app:family_nve}). Thermodynamic and structural observables across the \SI{100}{\pico\second} SevenNet trajectories also differ between backends by amounts comparable to repeated-run variations, confirming that \jet{} kernel + table supports stable MD simulations (SI Table~\ref{tab:md}). Overall, \jet{} kernel + table satisfies the criteria of Sec.~\ref{sec:budget} in single-point evaluations across all 15 models and in MD across the four models tested against a control backend, with the negligible energy shift of NequIP-OAM L as the sole exception. Full details of these static and dynamical tests are provided in SI~\ref{app:accuracy_full} and SI~\ref{app:md}.

\subsection{Step time across the 15 models}
\label{sec:nequip}
\begin{sidewaystable}[p]
\centering
\caption{\textbf{LAMMPS step time and peak memory of the 15 models on one H200.}
\jet{} kernel + table achieves the shortest step time and the lowest peak memory for every model. Speed-up divides the
step time of the existing TP-accelerated backend by that of \jet{} kernel + table, and the same-harness column reports this ratio when both backends execute inside the \jet{} CUDA-graph ML-IAP wrapper (Sec.~\ref{sec:perf}). Memory is the whole-run peak in GiB, where $\le$ marks
a whole-GPU upper bound (Sec.~\ref{sec:protocol}). Dashes denote unavailable or unmeasured entries, and OOM denotes out of memory.
Step times are rebuild-free medians over five runs, or three runs for the upstream model and the same-harness comparison (5632-atom electrolyte, NVE,
\SI{0.5}{\femto\second}, no D3, backends in Sec.~\ref{sec:protocol}). At this timestep,
\SI{43.2}{\milli\second\per\step} corresponds to \SI{1}{\nano\second\per\day}.}
\label{tab:mace}
\small\setlength{\tabcolsep}{5pt}
\begin{tabular}{l*{3}{rr}rr}
\toprule
 & \multicolumn{2}{c}{upstream model} & \multicolumn{2}{c}{existing TP-accelerated backend} & \multicolumn{2}{c}{\jet{} kernel + table} & & same \\
\cmidrule(lr){2-3}\cmidrule(lr){4-5}\cmidrule(lr){6-7}
model & \si{\milli\second\per\step} & GiB & \si{\milli\second\per\step} & GiB & \si{\milli\second\per\step} & GiB & speed-up & harness \\
\midrule
\multicolumn{9}{l}{SevenNet, TP accelerator cuEquivariance} \\
7net-nano-5.5 & 112.8 & -- & 19.9 & 2.67 & 3.5 & $\le$1.04 & 5.7 & 3.2 \\
7net-nano-6.0 & 143.8 & -- & 21.1 & 3.39 & 4.1 & $\le$1.06 & 5.2 & 3.4 \\
7net-0 & 328.5 & -- & 45.3 & 6.34 & 10.9 & $\le$1.34 & 4.1 & 2.6 \\
7net-omni & OOM & -- & 187.9 & 29.4 & 63.4 & $\le$3.88 & 3.0 & 2.4 \\
\midrule
\multicolumn{9}{l}{MACE, TP accelerator cuEquivariance} \\
MACE-MP-0 medium & 318.1 & 33.4 & 60.3 & 18.4 & 10.3 & 0.76 & 5.9 & 4.1 \\
MACE-MPA-0 medium & 325.0 & 34.5 & 64.9 & 19.5 & 13.0 & 3.07 & 5.0 & 3.5 \\
MACE-OMAT-0 medium & 325.1 & 34.5 & 64.9 & 19.5 & 12.9 & 3.07 & 5.0 & 3.5 \\
MACE-MP-0 large & 395.3 & 30.5 & 64.2 & 14.6 & 13.4 & 1.49 & 4.8 & 3.1 \\
\midrule
\multicolumn{9}{l}{NequIP-OAM, TP accelerator OpenEquivariance} \\
NequIP-OAM S & 9.1 & 1.01 & 6.9 & 0.71 & 2.0 & 0.29 & 3.5 & 1.9 \\
NequIP-OAM M & 145.4 & 16.2 & 44.8 & 5.92 & 11.7 & 0.73 & 3.8 & 3.2 \\
NequIP-OAM L & 561.2 & 57.9 & 139.6 & 15.3 & 37.9 & 1.47 & 3.7 & 3.3 \\
\midrule
\multicolumn{9}{l}{Nequix, PyTorch model with OpenEquivariance kernels in the CUDA-graph wrapper} \\
nequix-mp-1 & -- & -- & 74.0 & 11.8 & 25.0 & 1.21 & 3.0 & 3.0 \\
nequix-oam-1 & -- & -- & 74.0 & 11.8 & 24.9 & 1.21 & 3.0 & 3.0 \\
\midrule
\multicolumn{9}{l}{EquFlash, TP accelerator cuEquivariance in the GGNN ML-IAP wrapper} \\
EquFlash V1 & -- & -- & 868.7 & 114.7 & 365.2 & 10.7 & 2.4 & -- \\
EquFlash V2 & -- & -- & 1040.0 & 125.2 & 703.1 & 33.5 & 1.5 & -- \\
\bottomrule
\end{tabular}
\end{sidewaystable}

\begin{figure}[tp]
\centering
\includegraphics{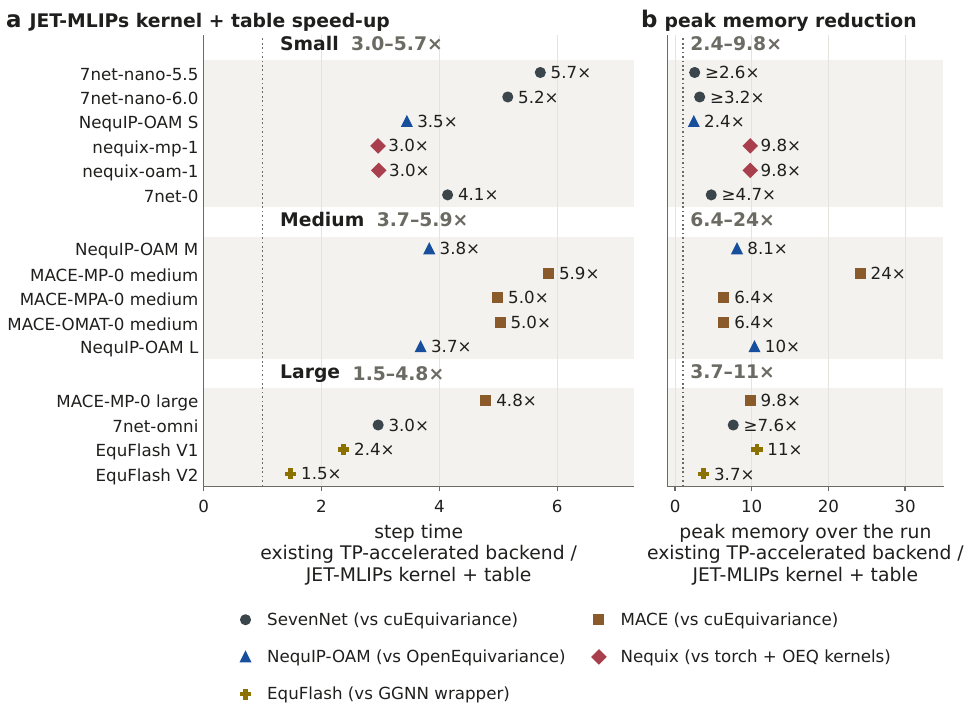}
\caption{\textbf{Speed-up and memory reduction across the 15 models.} (a) Rebuild-free LAMMPS step-time ratio,
existing TP-accelerated backend divided by \jet{} kernel + table. (b) Whole-run peak GPU memory ratio in the same sense, computed from unrounded peak PyTorch allocations (one run). Ratios for SevenNet represent lower bounds because its \jet{} memory uses the whole-GPU peak. In (b), MACE-MPA-0 and MACE-OMAT-0 show lower ratios than MACE-MP-0 medium because their radial tables are stored for each element pair (SI~\ref{app:limits}). Both panels use the 5632-atom electrolyte on one H200 without D3. Models are ordered by
parameter count into small ($<10^6$), medium ($10^6$--$10^7$) and large ($>10^7$) groups. Filled colors and markers identify families, and dotted lines mark parity. Labels use two significant figures. GGNN is the EquFlash package. OEQ denotes OpenEquivariance.}
\label{fig:family}
\end{figure}

We next compared LAMMPS MD step times across all 15 models on the 5632-atom electrolyte against each family's existing TP-accelerated backend (Table~\ref{tab:mace} and Fig.~\ref{fig:family}a). \jet{} kernel + table accelerates every model, achieving speed-ups of 3.0--5.7$\times$ in the small group, 3.7--5.9$\times$ in the medium group and 1.5--4.8$\times$ in the large group. In terms of rebuild-free throughput at a \SI{0.5}{\femto\second} timestep, this acceleration increases the number of models exceeding \SI{1}{\nano\second\per\day} on the 5632-atom cell from 3 to 12, reaching \SI{12.3}{\nano\second\per\day} for 7net-nano-5.5 and 4.0 and \SI{4.2}{\nano\second\per\day} for 7net-0 and MACE-MP-0 medium, respectively. For Nequix, which lacks a native LAMMPS interface, Table~\ref{tab:mace} compares \jet{} kernel + table against the released PyTorch model running its OpenEquivariance kernels inside the same \jet{} CUDA-graph wrapper. Within each parameter-size group, SevenNet and MACE achieve larger speed-ups than NequIP-OAM, Nequix and EquFlash. Parameter count measures model size rather than computational cost, which depends strongly on cutoff radius and neighbor count. For example, MACE-MP-0 large (15.8M parameters, \SI{4.5}{\angstrom} cutoff, 35.5 edges per atom) runs at \SI{13.4}{\milli\second\per\step} with \jet{} kernel + table, whereas NequIP-OAM L (9.6M parameters, \SI{6.0}{\angstrom} cutoff, 88.2 edges per atom) takes \SI{37.9}{\milli\second\per\step}.

In Table~\ref{tab:mace}, SevenNet compares \jet{} through the native \texttt{e3gnn} pair style against cuEquivariance through ML-IAP, so those ratios include differences between the two interfaces and their LAMMPS versions (Sec.~\ref{sec:protocol}). Using ML-IAP on both sides yields comparable speed-ups of 5.1, 4.7, 3.8 and 2.9$\times$ over cuEquivariance for 7net-nano-5.5, 7net-nano-6.0, 7net-0 and 7net-omni, respectively (SI Fig.~\ref{fig:perf_summary}c). In other respects, the SevenNet speed-ups in Table~\ref{tab:mace} are conservative. First, the table reports rebuild-free step times, whereas cuEquivariance incurs substantial additional overhead during neighbor-list rebuilds (SI~\ref{sec:rebuilds}). Second, the speed-ups over cuEquivariance increase further when the D3 dispersion correction is included (SI~\ref{sec:d3} and SI Fig.~\ref{fig:perf_summary}d).

Within SevenNet and MACE, the widest model exhibits the smallest speed-up, namely 3.0$\times$ for 7net-omni and 4.8$\times$ for MACE-MP-0 large. This trend reflects the smaller fraction of each step accelerated by radial tabulation and kernel fusion in wider architectures. In SevenNet, our operation count (Sec.~\ref{sec:cost}) shows that the ratio of radial-MLP to TP FMA operations decreases from 10.2 in the middle layer of 7net-nano-5.5 to 4.1 in the widest layer of 7net-omni, while reaching 18--39 in the first and last layers where fewer TP paths contribute to gated outputs (SI Table~\ref{tab:models} and SI~\ref{app:layer_eqs}). Tabulating the radial MLP therefore removes a smaller fraction of the per-edge arithmetic in 7net-omni. In MACE, GPU kernel profiling confirms a similar effect, showing that the fused convolution accounts for 44\% of the GPU kernel time per step in MACE-MP-0 large compared with 65\% in MACE-MP-0 medium (SI~\ref{sec:time}), leaving a larger share of the step in operations outside the convolution.

\subsection{Decomposition of the speed-up}
\label{sec:perf}
The end-to-end speed-ups of Sec.~\ref{sec:nequip} combine radial tabulation, kernel fusion and LAMMPS integration, and we next decomposed these contributions. SevenNet is the only family with a kernel-only mode and therefore isolates the effect of radial tabulation directly. Replacing the in-kernel radial basis and MLP with the cubic Hermite table removes approximately 70\% of the fused convolution block time, reducing it from 7.5 to \SI{2.3}{\milli\second} in 7net-nano-5.5 and from 27.3 to \SI{8.0}{\milli\second} in 7net-0 (SI Fig.~\ref{fig:perf_summary}a). In separate timings at 5632 atoms, the layer-summed fused convolution kernel runs 3.4--4.7$\times$ faster with the table than in kernel-only mode (Fig.~\ref{fig:table}d) and shortens an entire model call by 1.7--3.2$\times$ (SI~\ref{sec:callscaling}). Both gains are largest for 7net-omni due to its path-group count. Eliminating the in-kernel MLP hidden states leaves the tabulated kernel storing only interpolated weights, spherical harmonics and TP accumulators in registers, allowing each path group to hold more paths and channels within the register budget. In the widest layer of 7net-omni, the tabulated backward pass thus requires only 8 path groups instead of the 32 groups needed in kernel-only mode, each of which must re-traverse incident edges and re-evaluate the MLP (SI~\ref{sec:groups}). In contrast, existing TP accelerators leave the radial-MLP arithmetic intact, consistent with the operation count of Sec.~\ref{sec:cost} (SI~\ref{sec:time}).

Without radial tabulation, the speed-up is substantially smaller and relies strongly on LAMMPS integration. Through SevenNet's native \texttt{e3gnn} pair style, \jet{} kernel only runs 2.3, 2.0 and 1.5$\times$ faster than cuEquivariance for 7net-nano-5.5, 7net-nano-6.0 and 7net-0, respectively, but reaches only 0.9$\times$ its speed for 7net-omni (SI Fig.~\ref{fig:perf_summary}c). In standalone model calls outside LAMMPS, however, cuEquivariance is faster than kernel-only mode for 7net-0 and 7net-omni, showing that the end-to-end LAMMPS advantage of kernel-only mode stems from the integration rather than the convolution alone. On-GPU neighbor-list filtering and edge construction provide the largest single share of this integration gain (SI~\ref{app:integration} and SI Fig.~\ref{fig:perf_summary}g). For 7net-omni, where kernel-only mode trails cuEquivariance, introducing the radial table restores and extends the speed-up to 3.0$\times$.

To separate kernel gains from deployment overhead across families, we ran each pretrained model inside the same \jet{} CUDA-graph wrapper so that only the convolution backend differed (same-harness column of Table~\ref{tab:mace}). Under this control, \jet{} kernel + table remains 1.9--4.1$\times$ faster across SevenNet, MACE and NequIP-OAM, compared with 3.0--5.9$\times$ end-to-end against their existing TP-accelerated backends (while Nequix is evaluated inside the same wrapper at 3.0$\times$). For NequIP-OAM M, the 3.8$\times$ end-to-end speed-up thus factorizes into 1.2$\times$ from the wrapper and 3.2$\times$ from the kernels (SI~\ref{sec:intshare}). In MACE, the existing TP-accelerated ML-IAP deployment executes the model eagerly operation by operation from Python, whereas the \jet{} wrapper compiles the non-convolution blocks with \texttt{torch.compile} and replays the step as a CUDA graph (SI~\ref{app:companion}). Each end-to-end speed-up in Table~\ref{tab:mace} therefore reflects the product of a kernel factor (reported in the same-harness column) and a deployment factor. As convolution time falls, a substantial fraction of the remaining runtime shifts to operations outside the convolution (SI~\ref{sec:time} and SI Fig.~\ref{fig:timeline}), mirroring the fixed MD-step costs reported for compressed universal MLIPs \cite{li_dpa4c_2026} and indicating that the non-convolution modules set the next performance limit.

\subsection{System-size scaling across materials}
\label{sec:scaling}
\begin{figure}[tp]
\centering
\captionsetup{aboveskip=6pt}\includegraphics{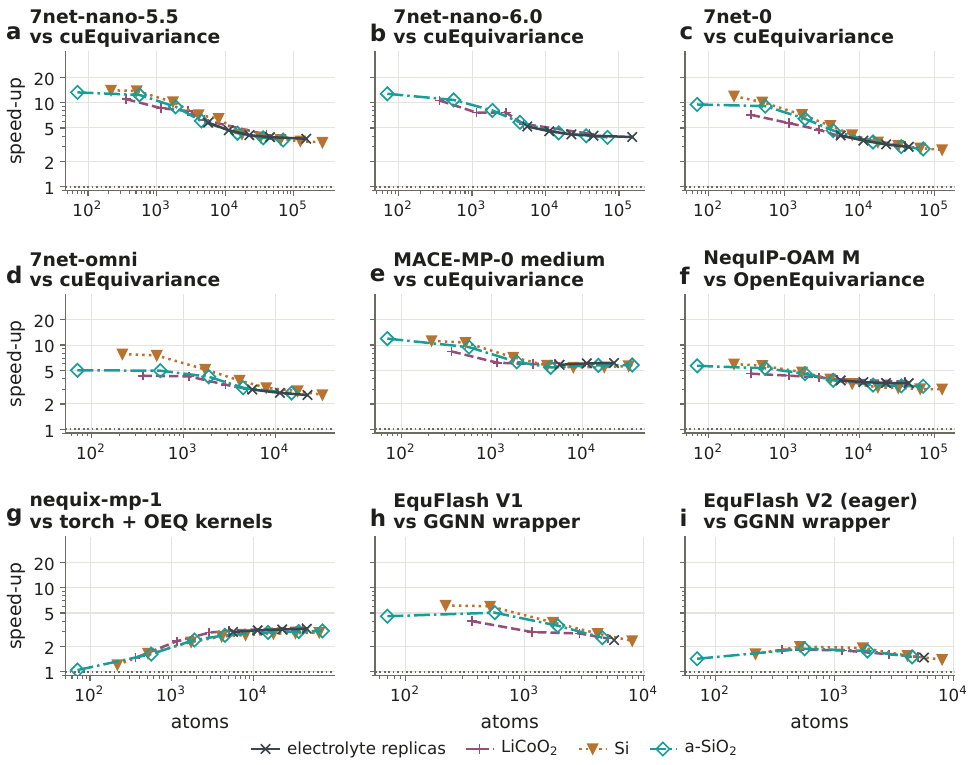}
\caption{\textbf{Speed-up versus system size across materials.} (a--d) Step time of the existing TP-accelerated backend divided by \jet{} kernel + table time
for SevenNet. (e--i) The same ratio for MACE, NequIP-OAM, Nequix and EquFlash, with models and comparators labeled
in each panel (default tables, one H200, rebuild-free LAMMPS steps, first run per point). Systems are indicated by crosses (electrolyte), plus signs (LiCoO$_2$), downward triangles (Si) and hollow teal diamonds (a-SiO$_2$). Runs use 500 steps,
except SevenNet Si and SiO$_2$, which use 460 rebuild-free steps. EquFlash V2 executes without graph replay. Static accuracy in these solids is verified for SevenNet only, and 7net-nano-6.0 omits the cuEquivariance Si benchmark. Curves terminate at
the largest measured system size accommodated by each existing TP-accelerated backend. OEQ denotes OpenEquivariance.}
\label{fig:scaling}
\end{figure}

To assess scaling beyond the 5632-atom cell, we scanned system size for nine models against their existing TP-accelerated backends across replicas of the electrolyte and three dense solids, namely LiCoO$_2$, diamond Si and amorphous SiO$_2$ (Fig.~\ref{fig:scaling}a--i and SI~\ref{sec:solids}). The speed-up persists up to the largest cell each existing TP-accelerated backend fits, remaining at or above 1.4$\times$ throughout, though its scaling with system size differs across models. Across all four SevenNet models, the speed-up over cuEquivariance decreases with size in every material (Fig.~\ref{fig:scaling}a--d). In electrolyte replicas, it declines from 5.8, 5.2, 4.1 and 3.0$\times$ at 5632 atoms (with 5.8$\times$ reflecting the first-run scan convention rather than the 5.7$\times$ five-run median in Table~\ref{tab:mace}, SI~\ref{sec:runprotocol}) to 3.7, 3.9, 3.0 and 2.6$\times$, respectively, at the largest sizes cuEquivariance accommodates (\num{152064} atoms for the two nano models, \num{45056} for 7net-0 and \num{22528} for 7net-omni). The decrease is steeper in the three dense solids, whose smallest cells contain 70--360 atoms, where the speed-up of 7net-nano-5.5 drops from 11.0--14.0$\times$ to 3.4--4.2$\times$ at the largest sizes (static accuracy in these solids is verified in Sec.~\ref{sec:accuracy}). Across the other families (Fig.~\ref{fig:scaling}e--i), NequIP-OAM M and EquFlash V1 and V2 (above the smallest cells) show a similar decrease with size. In contrast, MACE-MP-0 medium declines only up to a few thousand atoms before plateauing at 5.4--6.1$\times$, whereas nequix-mp-1 rises with size to 2.9--3.4$\times$ at the largest cells.

Step-time decomposition of 7net-nano-5.5 shows that the declining speed-up with size arises from fixed per-step overhead in the existing cuEquivariance backend rather than a loss of efficiency in \jet{} (SI~\ref{sec:intshare}). Specifically, host-side launches of the eager model and the ML-IAP interface impose an atom-count-independent floor, keeping the cuEquivariance step time nearly constant at approximately \SI{14}{\milli\second} for both nano models in LiCoO$_2$ cells below 2000 atoms. As the system grows, this fixed cost is amortized over more atoms and the speed-up approaches the ratio of the per-atom costs of the two backends. The same mechanism is expected for the other SevenNet checkpoints and NequIP-OAM M, whose existing TP-accelerated backends use the same interface, whereas the plateau observed for MACE-MP-0 medium above a few thousand atoms remains unresolved. Overall, small-cell speed-ups primarily reflect per-step launch overhead, whereas large-cell values reflect kernel throughput.

Power-law fits of step time against atom count ($t \propto N^{\alpha}$ for $N \ge 4000$ atoms) distinguish these three scaling regimes quantitatively (SI Fig.~\ref{fig:exponent}). Across all nine models, \jet{} kernel + table exhibits near-linear scaling with $\alpha =$ 0.87--0.98. Over the sizes each existing TP-accelerated backend accommodates, its exponent $\alpha$ is smaller for SevenNet and NequIP-OAM M (0.68--0.89), comparable for MACE-MP-0 medium and larger for nequix-mp-1, directly matching the decreasing, flat and rising speed-up trends. At the next scanned size beyond each limit, cuEquivariance runs out of memory in LAMMPS across all four SevenNet models while \jet{} kernel + table continues (Sec.~\ref{sec:memory} and SI~\ref{app:memory}), shifting the comparison at large scales from speed to feasibility.

\subsection{GPU memory scaling and accessible system size}
\label{sec:size}
\label{sec:memory}
GPU memory determines both the largest system a device can simulate and which hardware can run a given model, so we next evaluated memory consumption as the whole-run peak of the 15-model LAMMPS benchmarks (Sec.~\ref{sec:nequip}) and as the peak allocation of a standalone model call. Relative to the existing TP-accelerated backends, the whole-run peak falls by 2.4--9.8$\times$ in the small group, 6.4--24$\times$ in the medium group and 3.7--10.7$\times$ in the large group (Table~\ref{tab:mace} and Fig.~\ref{fig:family}b). For SevenNet, where \jet{} runs through \texttt{e3gnn} and reports total GPU memory in use (Sec.~\ref{sec:protocol}), the reductions over cuEquivariance represent lower bounds of at least 2.6$\times$ for 7net-nano-5.5, 3.2$\times$ for 7net-nano-6.0, 4.7$\times$ for 7net-0 and 7.6$\times$ for 7net-omni. Across all three groups, \jet{} kernel + table thus reduces both GPU memory and step time for every model.

\begin{figure}[tp]
\centering
\includegraphics{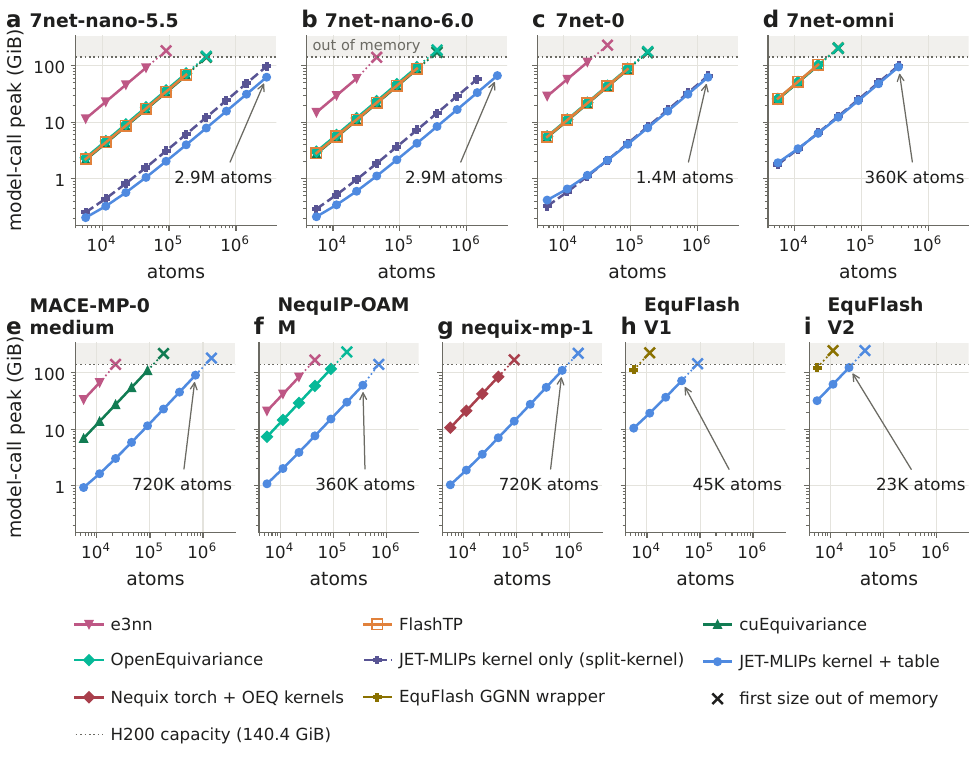}
\caption{\textbf{Model-call memory and accessible system size.} Peak PyTorch allocation in GiB against atom count
(5632-atom electrolyte replicas, one H200, forward plus forces). (a--d) SevenNet models. (e--i) MACE, NequIP-OAM,
Nequix and EquFlash models, labeled in each panel, using native interfaces without graph replay.
\jet{} kernel only uses the split-kernel scheme (dashed), and \jet{} kernel + table uses $N=4096$ for MACE, 2048 otherwise.
Backend markers are defined in the legend. Crosses mark the first out-of-memory size at its true atom count,
joined to the largest accommodated size by a dotted linear extrapolation, clipped at the memory-capacity line if below it.
The gray band marks out of memory, the horizontal dotted line marks the 140.4~GiB capacity of the 141~GB H200 and
panel annotations indicate the largest system size accommodated by \jet{} kernel + table. SevenNet TP-accelerator curves overlap within 10\%, as do
\jet{} curves in (c,d). In (d), e3nn exceeds GPU memory at 5632 atoms and therefore has no curve or marker, and in (h,i) GGNN accommodates only 5632 atoms. OEQ denotes OpenEquivariance.}
\label{fig:memory}
\end{figure}

To isolate model-call allocation independently of LAMMPS across system sizes, we measured peak PyTorch memory for a single energy-and-force evaluation across nine models (Fig.~\ref{fig:memory}). Tabulation lowers the allocation below every upstream and existing TP-accelerated backend, including an order-of-magnitude reduction for SevenNet. At 5632 atoms, tabulated calls of 7net-nano-5.5, 7net-0 and 7net-omni require 0.21, 0.42 and 1.92~GiB, respectively, which is 11--14$\times$ less than FlashTP (2.22, 5.48 and 26.1~GiB) and 11--13$\times$ less than cuEquivariance (2.16, 5.35 and 25.3~GiB, Fig.~\ref{fig:memory}a,c,d). Likewise, 7net-nano-6.0 requires 0.21~GiB versus 2.75~GiB for cuEquivariance (Fig.~\ref{fig:memory}b). Across the other families at 5632 atoms (Fig.~\ref{fig:memory}e--i), \jet{} kernel + table uses 0.93, 1.08 and 1.04~GiB for MACE-MP-0 medium, NequIP-OAM M and nequix-mp-1, respectively, corresponding to 7.4, 6.8 and 10.2$\times$ reductions relative to their existing TP-accelerated backends, while EquFlash V1 and V2 use 10.5 and 32.0~GiB (10.7 and 3.8$\times$ less than the GGNN wrapper). Within SevenNet, the recomputation scheme of \jet{} kernel only uses even less memory than kernel + table because it stores no per-edge buffer, at the cost of repeating the radial-MLP evaluation (SI~\ref{app:memory} and SI Table~\ref{tab:memory}).

These memory savings translate directly into much larger accessible system sizes. In standalone model-call scans on one H200, \jet{} kernel + table accommodates \num{2883584} atoms for 7net-nano-5.5 and 7net-nano-6.0, \num{1441792} for 7net-0 and \num{360448} for 7net-omni. cuEquivariance accommodates \num{180224} atoms for the two nano models, \num{90112} for 7net-0 and \num{22528} for 7net-omni. For MACE-MP-0 medium, NequIP-OAM M, nequix-mp-1, EquFlash V1 and EquFlash V2, it accommodates \num{720896}, \num{360448}, \num{720896}, \num{45056} and \num{22528} atoms, respectively, compared with \num{90112} atoms for the existing TP-accelerated backends of MACE-MP-0 medium and NequIP-OAM M, \num{45056} for the existing Nequix backend and \num{5632} for the GGNN wrapper. These values represent demonstrated sizes rather than exact capacity limits because the scan doubles atom count between consecutive points (model-call timings and additional SevenNet backends are given in SI~\ref{sec:callscaling} and SI Fig.~\ref{fig:call_scaling}). In full LAMMPS size scans across MACE, NequIP-OAM, Nequix and EquFlash V1 (SI Table~\ref{tab:family_oom}), the existing TP-accelerated backend runs out of memory or occupies 80--99\% of total GPU memory at system sizes where \jet{} kernel + table uses only 4--55\%.

The reduced memory footprint also makes these workloads feasible on a consumer GPU. On a 12~GB RTX 4070 SUPER under Windows Subsystem for Linux 2 (WSL2) (SI~\ref{app:consumer}), \jet{} kernel + table runs the 5632-atom electrolyte at 6.27, 1.54 and \SI{0.185}{\nano\second\per\day} for 7net-nano-5.5, 7net-0 and 7net-omni, respectively. In contrast, neither FlashTP nor cuEquivariance fits 7net-omni, and only FlashTP fits 7net-0 (SI Fig.~\ref{fig:perf_summary}e). At the \SI{0.5}{\femto\second} timestep of these runs, the consumer-GPU rates (from mean step times including rebuilds) exceed the highest H200 rates achieved by any existing TP-accelerated backend on the same metric, namely 2.24, 0.86 and \SI{0.132}{\nano\second\per\day} for the three models with OpenEquivariance. On the same 12~GB GPU and 5632-atom cell (SI Fig.~\ref{fig:perf_summary}f), \jet{} kernel + table runs MACE-MP-0 medium, MACE-MPA-0 medium and MACE-MP-0 large at 28.4, 41.2 and \SI{41.9}{\milli\second\per\step}, respectively, whereas cuEquivariance fits none of the three.

\subsection{Application to long-timescale transport properties}
\label{sec:applications}
To illustrate the long-timescale simulations made routine by this acceleration, we applied \jet{} kernel + table with JET-D3 to two transport properties requiring nanosecond trajectories of roughly $10^4$ atoms (computational details in Sec.~\ref{sec:protocol}, results in Tables~\ref{tab:viscosity} and~\ref{tab:argyrodite}). The first is the Green--Kubo shear viscosity of three carbonate liquids, each sampled in three independent \SI{10}{\nano\second} NVT replicas for a total of \SI{90}{\nano\second} of production MD. Based on the LAMMPS step times measured with D3 on the \SI{0.94}{\mole\per\liter} electrolyte cell (SI Fig.~\ref{fig:perf_summary}d), running three \SI{10}{\nano\second} viscosity replicas for a single liquid requires 129 days on one H200 with upstream e3nn and 76 days with FlashTP, compared with 3.9 days with \jet{} kernel + table and JET-D3. For the three liquids together, the same step times correspond to more than a year of dedicated H200 time with upstream e3nn and more than seven months with FlashTP, compared with 12 days with \jet{} kernel + table and JET-D3. In our production runs of the \SI{2.02}{\mole\per\liter} solution, the three replicas shared one H200 and each advanced at \SI{2.87}{\nano\second\per\day}, which corresponds to 3.5 days for the three \SI{10}{\nano\second} replicas. The corresponding speed-ups differ from those discussed above because these step times also include the D3 dispersion correction accelerated by JET-D3 (SI~\ref{sec:d3}).

The Green--Kubo shear viscosity of the liquid electrolytes at \SI{298.15}{\kelvin} is \SI{0.956\pm0.014}{\milli\pascal\second} for pure EMC, \SI{3.17\pm0.15}{\milli\pascal\second} for the \SI{0.94}{\mole\per\liter} LiPF$_6$ solution and \SI{9.27\pm0.33}{\milli\pascal\second} for the \SI{2.02}{\mole\per\liter} solution. The running Green--Kubo integral reaches its plateau at 30--\SI{75}{\pico\second} in pure EMC and 50--\SI{100}{\pico\second} in the \SI{0.94}{\mole\per\liter} solution, showing that the stress correlation time increases with salt concentration alongside the viscosity (SI~\ref{app:applications}). For comparison, a shear viscosity of \SI{3.2}{\milli\pascal\second} has been reported for a \SI{1}{\mole\per\liter} LiPF$_6$ solution in EC/EMC (3:7 by weight) at \SI{25}{\celsius} \cite{moganty_patent_2020}.

In the argyrodite Li$_6$PS$_5$Cl, the Li self-diffusion coefficient at \SI{400}{\kelvin} is \num{9.17e-7}, \num{7.54e-7} and \SI{7.09e-7}{\centi\metre\squared\per\second} for $x=0.25$, 0.50 and 0.615, respectively, with relative standard deviations of 5--7\% (Table~\ref{tab:argyrodite}). Arrhenius fits between 350 and \SI{400}{\kelvin} yield activation energies of $0.24\pm0.02$, $0.28\pm0.02$ and \SI{0.29\pm0.02}{\electronvolt} for the three compositions. The decrease of $D$ from $x=0.25$ to 0.615 agrees with a previous MLIP-based MD study that found the conductivity maximum near $x=0.25$ \cite{lee_disorder_2024}, whereas other simulations place the maximum near $x=0.5$ \cite{jeon_disorder_2024,chen_lpscl_2025}. At \SI{300}{\kelvin} and $x=0.615$, a \SI{20}{\nano\second} trajectory gives $D=\SI{4.33e-8}{\centi\metre\squared\per\second}$ (6.7\% relative standard deviation), corresponding to a Nernst--Einstein ionic conductivity of \SI{6.95}{\milli\siemens\per\centi\metre}. This value exceeds the 2--\SI{5}{\milli\siemens\per\centi\metre} measured for samples with $x$ between 0.52 and 0.62 \cite{kraft_argyrodite_2017,yu_lpscl_2018,schlenker_lpscl_2020,lee_disorder_2024}. The MSD is mildly sublinear across the fit window at all temperatures (log--log slope of 0.80--0.88), primarily from the cage contribution, and restricting the fit to the second half of the window lowers $D$ by approximately 1\%, well within the statistical uncertainty. As a transport-level consistency check, we repeated the \SI{400}{\kelvin} calculation at $x=0.615$ using FlashTP in place of \jet{} kernel + table with the same model, D3 implementation, initial structure and analysis. The two backends yield $D=\num{7.07e-7}$ and \SI{7.09e-7}{\centi\metre\squared\per\second}, differing by 0.2\% against a 5.5\% relative standard deviation, while their equilibrium volumes agree within 0.01\%.

\begin{table}[H]
\centering
\caption{\textbf{Shear viscosity of carbonate liquids computed with \jet{} kernel + table and JET-D3.} Shear viscosity $\eta$ at \SI{298.15}{\kelvin} with the standard error over three replicas (7net-nano-6.0 with D3(BJ), computational details in Sec.~\ref{sec:protocol}).}
\label{tab:viscosity}
\footnotesize\setlength{\tabcolsep}{5pt}
\begin{tabular}{lrc}
\toprule
liquid & atoms & $\eta$ (\si{\milli\pascal\second}) \\
\midrule
EMC & \num{10050} & $0.956\pm0.014$ \\
\SI{0.94}{\mole\per\liter} LiPF$_6$ in EC/EMC & \num{10040} & $3.17\pm0.15$ \\
\SI{2.02}{\mole\per\liter} LiPF$_6$ in EC/EMC & \num{10704} & $9.27\pm0.33$ \\
\bottomrule
\end{tabular}

\end{table}

\begin{table}[H]
\centering
\caption{\textbf{Li self-diffusion in Li$_6$PS$_5$Cl computed with \jet{} kernel + table and JET-D3.} Li self-diffusion coefficient $D$ at Cl/S site-exchange fraction $x$ with the activation energy $E_a$ from an Arrhenius fit between 350 and \SI{400}{\kelvin} (7net-nano-6.0 with D3(BJ), computational details in Sec.~\ref{sec:protocol}).}
\label{tab:argyrodite}
\footnotesize\setlength{\tabcolsep}{5pt}
\begin{tabular}{lcccc}
\toprule
 & \multicolumn{3}{c}{$D$ (\SI{e-7}{\centi\metre\squared\per\second})} & \\
\cmidrule(lr){2-4}
$x$ & \SI{350}{\kelvin} & \SI{375}{\kelvin} & \SI{400}{\kelvin} & $E_a$ (\si{\electronvolt}) \\
\midrule
0.25 & 3.46 & 5.88 & 9.17 & $0.24\pm0.02$ \\
0.50 & 2.39 & 4.54 & 7.54 & $0.28\pm0.02$ \\
0.615 & 2.13 & 4.17 & 7.09 & $0.29\pm0.02$ \\
\bottomrule
\end{tabular}
\end{table}

\subsection{Transferability and limits}
\label{sec:discussion}
Our fp32 accuracy results (Sec.~\ref{sec:accuracy}) appear to contradict DP Compress, which found that property errors of its compressed models in fp32 did not decrease with finer table spacing and therefore recommended fp64 \cite{lu_dpcompress_2022}. An error independent of table spacing indicates that tabulation is not its dominant source, and our radial table exhibits the same plateau once it reaches the fp32 rounding floor at $N=1024$--2048 (Fig.~\ref{fig:table}a,b). The difference lies in the evaluation criterion rather than the arithmetic. Our precision budget establishes the fp32 rounding floor of the unmodified existing backend against the fp64 reference and isolates the table error in fp64, where it lies 9--55 times below fp32 rounding in the electrolyte and 43--188 times below it in three dense solids. In comparison, DP Compress reported neither the uncompressed fp32 error nor the precision of its reference in the fp32 test. Its fp32 limit thus appears to reflect fp32 rounding amplified in derived properties rather than tabulation itself, whereas under the budget set by the existing fp32 backend the interpolation error lies well below inherent fp32 rounding.

Beyond the five families tested, radial tabulation applies wherever the radial weights depend solely on interatomic distance and element pair. Among the leading models of Matbench Discovery \cite{riebesell_matbench_2023}, this condition holds directly for the radial MLPs of GRACE \cite{lysogorskiy_grace_2025} and TACE \cite{xu_tace_2025}. In EquiformerV3 \cite{liao_eqv3_2026}, eSEN \cite{fu_esen_2025} and TECE \cite{xu_tece_2026}, only the radial functions can be tabulated because their attention or mixing depends on node features. Models whose radial inputs include node features directly, such as Prophet \cite{kairos_prophet_2026}, fall outside this approach. The same condition would extend beyond interatomic potentials. TrajCast, an autoregressive equivariant network that predicts atomic displacements and velocities directly instead of forces, builds its messages through a NequIP-type convolution whose radial weights depend only on edge length \cite{thiemann_trajcast_2026}. Although we have not tested TrajCast, radial tabulation and the fused forward kernel should carry over to its message-passing step, whereas its velocity-dependent update lies outside the present kernels and requires no force backward pass. Even where radial tabulation applies, shared equivariance alone does not guarantee compatibility with other architectures, such as the higher-degree representations of EquiformerV2 \cite{liao_eqv2_2024}. Each new architecture must therefore be verified against the supported layouts (SI~\ref{app:limits}) and architecture-dependent compilation limits (SI~\ref{sec:groups}) before reusing these kernels.

Within the tested families, our accuracy validation spans three levels of evidence across models. SevenNet was tested against the full precision budget of rounding, run-to-run spread and chaotic amplification, using FlashTP as the control for trajectory divergence and \jet{} kernel only as the control for the table. These dynamical tests cover the 5632-atom electrolyte cell on one H200 (Sec.~\ref{sec:accuracy}), while single-point force and energy errors extend validation to LiCoO$_2$, diamond Si and amorphous SiO$_2$ (SI Table~\ref{tab:solids_acc}). MACE-MP-0 medium and NequIP-OAM M were validated by single-point errors against the fp64 reference and by NVE drift against the existing TP-accelerated backend. The remaining MACE and NequIP-OAM models, Nequix and EquFlash V1 were tested by single-point errors against the fp64 reference, while EquFlash V2 (whose fp64 model exceeds GPU memory) was compared directly with its existing fp32 TP-accelerated backend. Nequix and EquFlash V1 were additionally checked for NVE drift with \jet{} alone without a control run of the existing TP-accelerated backend (SI~\ref{app:family_nve}), and MD trajectory divergence was measured only for SevenNet, whereas the other four families were tested by repeated single-point evaluations (SI Table~\ref{tab:family_acc}). The precision budget must be established per model because it is set by each model's own existing fp32 backend (Sec.~\ref{sec:accuracy}), making dynamical validation in dense solids and accuracy checks across additional electrolyte compositions the most direct extensions.

The performance benchmarks carry several scope limitations, as the timings cover one electrolyte composition and three solids and must be interpreted alongside their neighbor counts. Multi-GPU execution lies outside the present scope (SI~\ref{app:limits}), while per-edge-type cutoffs, NequIP-OAM XL and the 7net-l3i5 layer configuration remain unsupported and Allegro's edge-feature tensor products remain unevaluated. No single universal speed-up applies across workloads because measured gains depend jointly on model architecture, simulation cell and MD integration. For NequIP-OAM, the comparison with OpenEquivariance inside the same CUDA-graph wrapper illustrates this integration dependence directly (Sec.~\ref{sec:perf} and SI~\ref{sec:intshare}).

\section{Conclusions}
\label{sec:conclusions}
By combining fused convolution kernels with cubic Hermite tabulation of the deployed radial MLP, \jet{} accelerates MD and reduces GPU memory across 15 released checkpoints from five equivariant MLIP families without storing or recomputing per-edge MLP intermediates in the force backward pass. Within the precision budget set by each model's existing fp32 backend, \jet{} kernel + table retains single-point force accuracy across all models tested against the fp64 reference while keeping isolated fp64 interpolation errors well below fp32 rounding and run-to-run variation. In MD, it preserves NVE energy conservation and statistical observables within repeated-run spread (Sec.~\ref{sec:accuracy}). The sole exception is a negligible energy shift of approximately \SI{6e-7}{\electronvolt\per\atom} in NequIP-OAM L that vanishes in fp64 and leaves forces unchanged. On one H200, LAMMPS MD of the 5632-atom electrolyte without D3 runs 1.5--5.9$\times$ faster than the existing TP-accelerated backends while reducing whole-run peak GPU memory by 2.4--24$\times$, with the SevenNet memory ratios representing lower bounds (Secs.~\ref{sec:nequip} and \ref{sec:memory}). Together with JET-D3, these gains enable routine multi-nanosecond transport calculations of roughly $10^4$-atom liquid and solid electrolytes on a single GPU (Sec.~\ref{sec:applications}).

Within the scope outlined in Sec.~\ref{sec:discussion}, these results show that eliminating the radial MLP and its per-edge intermediates, rather than accelerating the TP alone, simultaneously relieves both the runtime and memory bottlenecks of models sharing the NequIP convolution. The kernel-only control isolates this contribution directly for SevenNet, whereas the end-to-end speed-ups in the other four families combine kernel fusion, radial tabulation and MD integration. Whereas DPA4C addressed this radial-weight bottleneck through a co-designed architecture \cite{li_dpa4c_2026}, \jet{} resolves it for existing pretrained checkpoints without retraining. By lowering both trajectory cost and memory requirements, \jet{} substantially relieves the GPU bottleneck of equivariant MLIP simulations and extends large-scale, nanosecond MD to widely accessible hardware, including consumer GPUs.

\section*{Data and Software Availability}
JET-MLIPs, which comprises the kernel library \texttt{jet\_mlip}, the adapters for SevenNet, MACE, NequIP, Nequix and
EquFlash, JET-D3 and the LAMMPS build scripts, is available at \url{https://github.com/yanselmo-jk/jet-mlips}
under the Creative Commons Attribution-NonCommercial-ShareAlike 4.0 International License (CC BY-NC-SA 4.0).
The files derived from LAMMPS (the JET-D3 pair style, the patches to LAMMPS ML-IAP and the LAMMPS pair styles changed by the SevenNet patch) remain under the GNU General Public License version 2 of LAMMPS.
Users who publish results obtained with \jet{} are asked to cite this article. The released packages are used unmodified or patched in place, GGNN
(EquFlash) is loaded and not copied, and no model weights are redistributed. JET-MLIPs is developed independently of
the developers of SevenNet, MACE, NequIP, Nequix and EquFlash.
The benchmark inputs, measurement scripts and raw data are available from the authors on request.

\begin{suppinfo}
Test systems and benchmark protocol, kernel implementation and LAMMPS integration, accuracy tests and MD validation, details of performance, memory and MD integration, application examples, supported layouts and limits, Tables S1--S7 and Figures S1--S9 (PDF).
\end{suppinfo}

\section*{Author Contributions}
J.Y. conceived the study and designed and implemented the fused kernels, the radial table, JET-D3 and the LAMMPS integration. J.Y. performed the benchmarks, the accuracy and MD validation and the transport simulations, analyzed the data and prepared the figures. J.Y. and S.J. wrote the manuscript. Both authors reviewed and approved the final version.

\section*{Notes}
The authors declare no competing financial interest.

\begin{acknowledgement}
This work received no external funding. Code development was assisted by Claude Fable 5.1 (Anthropic), and the writing of
this manuscript was assisted by Claude Fable 5.2 (Anthropic) and Gemini 4 Neon (Google).
The authors reviewed and verified all code, data and text and assume full responsibility for the content.
\end{acknowledgement}

\bibliography{refs}

\begin{tocentry}
\includegraphics{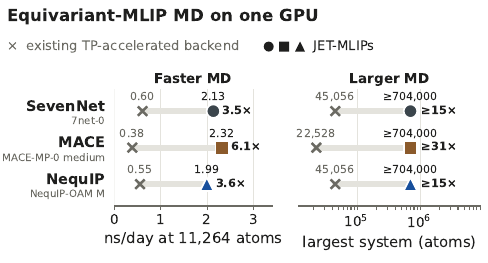}
\end{tocentry}
\end{document}

\maketitle
\renewcommand{\arraystretch}{1.15}
\section{Test systems and benchmark protocol}
\label{app:protocol}
This section describes the test systems and benchmark conditions of Sec.~\ref{sec:protocol}, in which one chemically intact 5632-atom electrolyte cell
serves both the timing benchmarks and the MD validation.

\subsection{Test systems}
\label{sec:system}
The common workload of the MD benchmarks is a LiPF$_6$ solution in ethylene carbonate (EC) and ethyl methyl carbonate (EMC), which has previously been modeled with machine-learning force fields \cite{si:magdau_ec_emc_2023} (main Sec.~\ref{sec:protocol} and Table~\ref{tab:mace}). Every backend receives identical atoms and edge counts. The primary cell contains 5632 atoms (64 LiPF$_6$, 128 EC and 256 EMC, or Li 64, P 64, F 384, C 1408, O 1152 and H 2560). A 352-atom cell of the same composition (4 LiPF$_6$, 8 EC and 16 EMC, cubic with $L=\SI{15.37}{\angstrom}$) serves the accuracy tests whose fp64 reference exceeds GPU memory at 5632 atoms. It also serves the small-cell timing of EquFlash V1 (Sec.~\ref{app:integration}).

The primary cell was packed from intact molecules and equilibrated by 200~ps of NPT MD with 7net-0 plus D3(BJ) using Perdew--Burke--Ernzerhof (PBE) parameters \cite{si:grimme_d3_2010,si:grimme_bj_2011} and cutoffs of \SI{9000}{bohr^2} for the energy and \SI{1600}{bohr^2} for the coordination number. Equilibration yields a density of \SI{1.361\pm0.007}{\gram\per\centi\metre\cubed} ($L=\SI{38.74}{\angstrom}$) with all 384 solvent molecules and 64 PF$_6^-$ anions intact. Timing runs start from a state reached after an additional \SI{10}{\pico\second} of NVT at \SI{300}{\kelvin} with 7net-0 (\jet{} kernel only) plus D3 and a \SI{0.5}{\femto\second} step.

To test scaling across system sizes and material classes, the size scans use periodic replicas of fixed cells. The electrolyte scans replicate the 5632-atom cell 2 to 512 times,
from \num{11264} to \num{2883584} atoms (for example $2\times2\times2$ copies for \num{45056} and $3\times3\times3$ for
\num{152064} atoms). The LiCoO$_2$ cells repeat an orthogonal 24-atom cell of the layered structure $5\times3\times1$,
$6\times4\times2$, $10\times6\times2$ and $13\times8\times2$ times (360, 1152, 2880 and 4992 atoms), with a $2\times2\times2$ replication of the 4992-atom
cell yielding \num{39936} atoms. The Si cells repeat the 8-atom cubic diamond cell $n$ times
along each axis for $n=3$--32 (216 to \num{262144} atoms). The amorphous SiO$_2$ cells repeat a 70-atom cell (23 Si and
47 O in a \SI{1}{\nano\metre} cube, melt-quenched with 7net-0) $n=1$--10 times along each axis (70 to \num{70000} atoms).

\subsection{Hardware and software}
\label{sec:hardware}
All main timings were measured on one node with three NVIDIA H200 NVL GPUs. Each GPU provides \SI{141}{\giga\byte} of memory (\SI{143771}{\mebi\byte} reported by nvidia-smi), sm\_90 compute capability and 132 streaming multiprocessors (SMs), operating with driver 595.99.02 under a \SI{600}{\watt} power limit. The host node is equipped with two AMD EPYC 9654 processors, \SI{251}{\gibi\byte} of RAM and Ubuntu 24.04. The transport calculations of main Sec.~\ref{sec:applications} were conducted on NVIDIA H200 or RTX 2080 Ti GPUs.

The SevenNet runs used Python 3.11, PyTorch 2.7.1+cu126 \cite{si:paszke_pytorch_2019}, CUDA 12.6, e3nn 0.6.0, SevenNet 0.13.0 (patched by JET-MLIPs for the \jet{} rows), cuEquivariance 0.9.1 \cite{si:cuequivariance}, OpenEquivariance 0.7.0 \cite{si:bharadwaj_oeq_2025} and FlashTP 0.1.0 \cite{si:lee_flashtp_2025} built for sm\_90. The other families used mace-torch 0.3.16 with PyTorch 2.7.1 and cuEquivariance 0.9.1 (MACE), NequIP 0.19.1 with PyTorch 2.11.0 and OpenEquivariance 0.7.0 (NequIP-OAM), nequix 0.4.5 with its torch backend and its OpenEquivariance kernels (Nequix) and GGNN (commit \texttt{a25dfa4}) with PyTorch 2.9.1 and cuEquivariance 0.6.0 (EquFlash).

All five families share the LAMMPS machine-learning interatomic-potential interface (ML-IAP) \cite{si:johansson_lammps_kokkos_2025}. The runs through ML-IAP used LAMMPS 22Jul2025 update~6 (SevenNet documents update~4) compiled with Kokkos, which provides portable parallel execution and device-data management for this interface \cite{si:trott_kokkos3_2022}. For the SevenNet pair style \texttt{e3gnn}, LAMMPS stable 2Aug2023 update~3 was compiled from the same source in two builds, one with the \jet{} pair styles and one with those of upstream SevenNet 0.13.0. Across all backends, TF32 (TensorFloat-32) matrix arithmetic was disabled, and each run used one MPI rank and one OpenMP thread.

\subsection{Run protocol and backends}
\label{sec:runprotocol}
This subsection provides benchmark protocol details not covered in main Sec.~\ref{sec:protocol}. Each family's existing TP-accelerated backend is defined there. For MACE, this backend is the cuEquivariance 0.9.1 model exported by MACE's script \texttt{mace\_create\_\allowbreak lammps\_model}, and for NequIP-OAM it is OpenEquivariance 0.7.0 enabled through NequIP's modifier \texttt{enable\_\allowbreak OpenEquivariance}. Additional SevenNet backends are described in Sec.~\ref{sec:lmp_backends}. Throughout this SI, upstream denotes the unmodified SevenNet 0.13.0 release, its backends (e3nn, FlashTP, cuEquivariance and OpenEquivariance) and its \texttt{pair\_style d3}. The upstream model of main Table~\ref{tab:mace}, in contrast, is the unaccelerated model of each family.

Each run started only when no GPU compute process was active and GPU utilization had remained at or below 3\% for \SI{3}{\second}. The other two GPUs of the node were idle during the small-cell, timeline, D3 and ablation series. During the other series, they ran additional benchmarks of this study but never upstream D3, whose host load slows neighboring runs. Runs repeated with and without concurrent jobs agree within 1\%.

LAMMPS timings used 500-step NVE runs at a \SI{0.5}{\femto\second} timestep from the \SI{300}{\kelvin} state with thermodynamic output every 10 steps. In the electrolyte, 9--10 of the 45 intervals between steps 50 and 500 contain a neighbor-list rebuild, whereas the SevenNet Si and amorphous SiO$_2$ scans used 460 steps without rebuilds. The H200 LAMMPS step times are medians of three runs. Five runs were used for the existing TP-accelerated and \jet{} kernel + table columns of main Table~\ref{tab:mace}, the cuEquivariance bars of Fig.~\ref{fig:perf_summary}c and the runs with upstream D3. Size scans report the first run per configuration. For 7net-nano-5.5 at 5632 atoms, the scan therefore gives a speed-up of 5.8$\times$ over cuEquivariance, compared with 5.7$\times$ from the five-run medians. Two additional repetitions of the electrolyte scans above 5632 atoms for 7net-nano-5.5 and 7net-0 confirm that rebuild-free step times vary by at most 1.1\% (\jet{}) and 0.2\% (cuEquivariance) of the median, altering the three-run speed-ups by at most 1.0\%.

Model-call timings used CUDA events after 10 warm-up calls in one process per backend and system size. Each timing is the mean GPU time over the calls of that run, and the model-call memory is the peak PyTorch allocation of one call. The whole-GPU peak is the maximum memory in use on the GPU reported by nvidia-smi, sampled every 200~ms during the 500-step run, minus the pre-run reading. It includes both the CUDA context and the allocator reserve. This measure is used in the size scans and for the \texttt{e3gnn} runs, for which no PyTorch allocation statistics were recorded. For the existing TP-accelerated backends, the whole-GPU reading increases progressively during execution owing to retained allocator cache that can be released without affecting step time. The comparison across families therefore uses the whole-run peak of Sec.~\ref{sec:protocol}.

\clearpage
\section{Kernel implementation and LAMMPS integration}
\label{app:kernels}
Building on the convolution equations and cost analysis of Secs.~\ref{sec:background} and~\ref{sec:cost}, this section details the layer shapes, the fused kernels of Sec.~\ref{sec:fused}, the radial table of
Sec.~\ref{sec:table} and the LAMMPS integration of Sec.~\ref{sec:md}.

\subsection{Tensor-product paths and layer shapes}
\label{app:shapes}
The shapes in this subsection are those of SevenNet, which is the family of the cost analysis in Sec.~\ref{sec:cost}. Table~\ref{tab:family_models} lists the sizes of the other families.

\paragraph{SevenNet interaction layer.}
\label{app:layer_eqs}
Within the SevenNet interaction layer (main Fig.~\ref{fig:layer}a), the convolution of main Eq.~\eqref{eq:conv} is embedded as follows.
A SevenNet model embeds the species $Z_i$ of each atom and applies $T$ interaction layers $t=1,\dots,T$ ($T=3$ or 5,
Table~\ref{tab:family_models}). Denoting the features of atom $i$ after layer $t$ by $x_i^{(t)}$, the model reads out the total energy
according to
\begin{align}
  x_i^{(0)} &= \mathrm{Linear}_{\mathrm{embed}}\big(\mathrm{onehot}(Z_i)\big), \label{eq:embed}\\
  s_i &= \mathrm{SC}_t\big(x_i^{(t-1)}, Z_i\big), \qquad y_i = \mathrm{Linear}_{\mathrm{pre},t}\big(x_i^{(t-1)}\big), \label{eq:si1}\\
  h_i &= \mathrm{Conv}_t\big(\{y_j,\mathbf r_{ij}\}_{j\in\mathcal N(i)}\big), \label{eq:conv_layer}\\
  x_i^{(t)} &= \mathrm{Gate}\big(z_i\big), \qquad z_i = \mathrm{Linear}_{\mathrm{post},t}(h_i) + s_i, \label{eq:gate}\\
  E &= \sum_i E_i, \qquad E_i = \sigma_{Z_i}\,\mathrm{Linear}_E\big(\mathrm{Linear}_h(x_i^{(T)})\big) + \mu_{Z_i}. \label{eq:readout}
\end{align}
Each $\mathrm{Linear}$ is an equivariant linear map mixing channels within each degree separately, and $\mathrm{SC}_t$ is a species-dependent linear self-connection. In Eq.~\eqref{eq:gate}, $z_i$ carries one extra gate scalar per $l>0$ channel. The $\mathrm{Gate}$ function applies parity-dependent activations (SiLU for even scalars, tanh for odd scalars) and scales each $l>0$ channel by its activated gate scalar. Since the final layer outputs scalars only, Eq.~\eqref{eq:readout} applies two linear maps followed by a species-wise scale $\sigma_{Z_i}$ and shift $\mu_{Z_i}$ to produce the atomic energies $E_i$. NequIP-OAM follows the same layout (main Fig.~\ref{fig:layer}a) with three differences. Its first layer omits $\mathrm{SC}_1$, and its readout uses one linear map plus a Ziegler--Biersack--Littmark (ZBL) pair term. It also applies the neighbor normalization $1/\sqrt{N_\text{avg}}$ before the TP, which is equivalent to $\nu=\sqrt{N_\text{avg}}$ in main Eq.~\eqref{eq:agg}.

For the checkpoints analyzed in main Sec.~\ref{sec:cost}, Table~\ref{tab:models} lists the input irreps, live paths, $N_W$ and per-edge FMA counts of each layer. Pruning unused outputs removes 4 of the 15 paths in the middle layer of 7net-nano-5.5. In fp32, the radial weights occupy $4N_W$ bytes per edge, which amounts to 1408, 3840 and \SI{12032}{\byte} in the middle layers of 7net-nano-5.5 ($N_W=352$), 7net-0 ($N_W=960$) and 7net-omni ($N_W=3008$), respectively. In the 5632-atom cell, the middle-layer weights of 7net-nano-5.5 and 7net-0 thus total \SI{515}{\mebi\byte} and \SI{1036}{\mebi\byte} (\num{383648} and \num{282982} edges, Table~\ref{tab:family_models}). The backward pass adds a $\partial E/\partial\mathbf W$ tensor of equal size.

\subsection{Fused forward and backward kernels}
\label{app:fused}
\begin{algorithm}[t]
\caption{Fused forward convolution of one layer.}
\label{alg:fwd}
\begin{algorithmic}[1]
\Require destination-sorted compressed sparse row (CSR) index of edges $j\to i$, edge vectors $\mathbf r_{ij}$, features $y_j$ and
radial table $\mathcal T$ (Sec.~\ref{sec:table})
\Ensure convolution output $h_i$ for each destination atom $i$
\For{each path group (SI~\ref{sec:groups}) and destination atom $i$}
  \State Initialize $\mathrm{acc}_{p,u}\gets0$ \Comment{output accumulators, registers}
  \For{each edge $j\to i$ in row $i$ of the CSR}
    \State Load $\mathbf r_{ij}$ \Comment{global memory to registers}
    \State Compute $r\gets|\mathbf r_{ij}|$ \Comment{shared memory}
    \State Compute $Y(\hat{\mathbf r}_{ij})$ \Comment{shared memory}
    \State Load $y_j$ \Comment{global memory to registers}
    \State Compute $\mathbf W(r)$ by Hermite lookup in $\mathcal T$ \Comment{global table to registers}
    \State Accumulate $W_{p,u}[y_j\otimes Y]_p$ into $\mathrm{acc}_{p,u}$ \Comment{Eq.~\eqref{eq:tp}, registers}
  \EndFor
  \State Write $h_i\gets\mathrm{acc}/\nu$ for the group's columns \Comment{global memory, one writer}
\EndFor
\end{algorithmic}
\end{algorithm}

\begin{algorithm}[t]
\caption{Fused force backward of one layer.}
\label{alg:bwd}
\begin{algorithmic}[1]
\Require inputs of Algorithm~\ref{alg:fwd} and $\partial E/\partial h_i$
\Ensure $\partial E/\partial y_j$ and $\partial E/\partial\mathbf r_{ij}$
\State Initialize $\partial E/\partial y_j\gets0$ \Comment{global memory}
\State Initialize $\mathcal G_{ij}\gets0$ \Comment{per-edge buffer for $(g_Y,g_r)$, global memory}
\Statex \textbf{Kernel 1}
\For{each path group and destination atom $i$}
  \State Load $\partial E/\partial h_i$ \Comment{global memory to registers}
  \State Divide the upstream gradient by $\nu$ \Comment{registers}
  \For{each edge $j\to i$ in row $i$ of the CSR}
    \State Load $\mathbf r_{ij}$ \Comment{global memory to registers}
    \State Compute $r\gets|\mathbf r_{ij}|$ \Comment{shared memory}
    \State Compute $Y(\hat{\mathbf r}_{ij})$ \Comment{shared memory}
    \State Load $y_j$ \Comment{global memory to registers}
    \State Compute $\mathbf W(r)$ by Hermite lookup in $\mathcal T$ \Comment{global table to registers}
    \State Compute $\mathbf W'(r)$ from the same interpolant \Comment{radial derivative, registers}
    \State Compute $g_y$ by TP backward \Comment{group's feature gradient, registers}
    \State Compute $g_Y$ by TP backward \Comment{group's harmonic gradient, registers}
    \State Compute $g_{W_{p,u}}$ by TP backward \Comment{group's weight gradient, registers}
    \State Accumulate $g_y$ into $\partial E/\partial y_j$ \Comment{shared staging, global atomic}
    \State Compute $g_r\gets\sum_{p,u}g_{W_{p,u}}W'_{p,u}$ \Comment{group's radial gradient, registers}
    \State Accumulate $(g_Y,g_r)$ into $\mathcal G_{ij}$ \Comment{global atomic, groups add}
  \EndFor
\EndFor
\Statex \textbf{Kernel 2}
\For{each edge $j\to i$ in the CSR}
  \State Load $\mathbf r_{ij}$ \Comment{global memory to registers}
  \State Compute $r\gets|\mathbf r_{ij}|$ \Comment{registers}
  \State Load $(g_Y,g_r)\gets\mathcal G_{ij}$ \Comment{global memory to registers}
  \State Write $\partial E/\partial\mathbf r_{ij}\gets(\partial Y/\partial\mathbf r_{ij})^{\top}g_Y+g_r\hat{\mathbf r}_{ij}$
    \Comment{global memory, one writer}
\EndFor
\end{algorithmic}
\end{algorithm}

\begin{figure}[!htbp]
\centering
\includegraphics{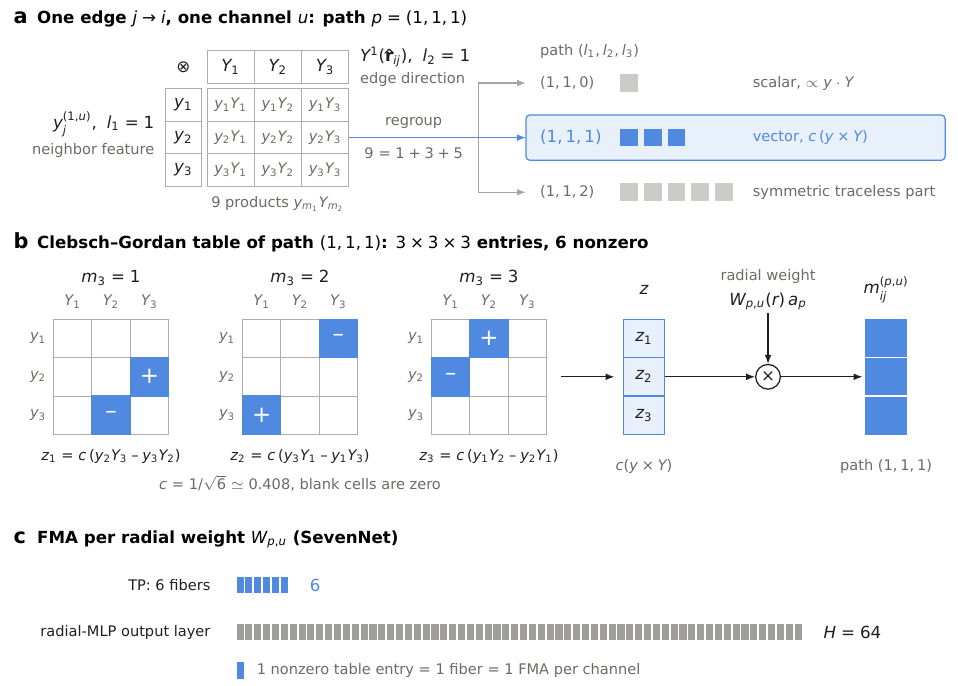}
\caption{\textbf{A tensor-product path as a sparse coefficient table.} \textbf{a}, For one edge $j\to i$ and one channel $u$, the degree-$1$ neighbor feature $y_j^{(1,u)}$ and edge harmonic $Y^1(\hat{\mathbf r}_{ij})$ yield nine products $y_{m_1}Y_{m_2}$. These regroup into degrees $0$, $1$ and $2$ with one, three and five components (one path for each output degree $l_3$), and the path $p=(l_1,l_2,l_3)=(1,1,1)$ retains the highlighted blue branch. \textbf{b}, The Clebsch--Gordan table of this path comprises one $3\times3$ slice for each output component $z_{m_3}$, with rows $y_{m_1}$ and columns $Y_{m_2}$ in the component order of e3nn. Each cell stores the coefficient of its product. Six of the 27 cells are nonzero and hold $+c$ or $-c$ with $c=1/\sqrt{6}\simeq0.408$ prior to the fixed path normalization $a_p$. The radial weight $W_{p,u}(r)$ scales the three components of $z=c(y\times Y)$ into the edge message $m_{ij}^{(p,u)}$. \textbf{c}, Each nonzero cell corresponds to one fiber and requires one fused multiply--add (FMA) per channel. The path thus requires six FMAs per radial weight, compared with $H=64$ FMAs to produce that weight in the output layer of SevenNet's radial MLP.}
\label{fig:sevennet_layer}
\end{figure}

Algorithms~\ref{alg:fwd} and~\ref{alg:bwd} outline the default table kernels. Both the table mode and the kernel-only mode, which exists for SevenNet only, assign one warp per destination atom over a destination-sorted CSR edge index and map lane $u$ to channel $u$ in 32-channel units. Each warp streams the incident edges in tiles, stages $r$ and $Y(\hat{\mathbf r}_{ij})$ in shared memory, gathers $y_j$ and applies the nonzero Clebsch--Gordan fibers (Fig.~\ref{fig:sevennet_layer}b,c). It multiplies by the radial weight once per path-and-channel unit rather than once per fiber. The per-channel accumulators remain in registers across the neighbor list and are written once without atomics, which renders the forward pass deterministic.

In kernel-only mode, the forward kernel replaces the table lookup with an in-register evaluation of Eqs.~\eqref{eq:bessel} and~\eqref{eq:mlp}. This evaluation uses accurate transcendental intrinsics (no fast-math approximations) and lane-parallel matrix--vector products against MLP weights held in shared memory. The backward kernel recomputes the edge embedding and forms $\partial E/\partial W$ in registers. Kernel-only mode propagates this gradient through the MLP to $\partial E/\partial\mathbf r_{ij}$ directly with per-edge atomics. Table mode instead reads $\mathbf W'(r)$ from the interpolant and applies the harmonics Jacobian once per edge in Kernel 2 (Algorithm~\ref{alg:bwd}, Sec.~\ref{sec:groups}). In both modes, $\partial E/\partial y_j$ is staged per warp in shared memory and accumulated into global memory with coalesced atomics, which introduces the run-to-run force variation measured in main Sec.~\ref{sec:accuracy}.

For multi-group layers in kernel-only mode, the deployment flag \texttt{--exact\_backward} selects between two variants. The \texttt{recompute} variant repeats the MLP backward in every group and needs no per-edge buffer. The default \texttt{split} variant atomically buffers $\partial E/\partial\mathbf a_2\in\mathbb R^{64}$ at \SI{256}{\byte} per edge, which amounts to \SI{93.7}{\mebi\byte} for the middle layer of 7net-nano-5.5 at 5632 atoms. A second kernel then completes the MLP backward pass once per edge before the buffer is freed (memory trade-off in Sec.~\ref{app:memory}).

\subsection{Path groups and build-time specialization}
\label{sec:groups}
A path group is the set of channel units that one warp processes in one pass over its atom's edges. Each channel unit is one TP path restricted to 32 consecutive channels with one channel per
lane. A path with fewer than 32 channels occupies one unit
and leaves the remaining lanes idle.
Each kernel build partitions the layer shapes into groups that satisfy the shared-memory and register limits of the target GPU.
Specialization is resolved at build time, when each layer shape becomes compile-time
tables of paths and fibers that hold only the nonzero Clebsch--Gordan terms of each path. The fiber loops are then fully
unrolled and the accumulators remain in registers.

The table kernels hold no MLP state, and their groups are therefore sized by the registers required for interpolation, harmonics and accumulators. Each additional group re-streams the incident edges of an atom and repeats the geometry and interpolation evaluations. The table backward pass therefore counts only the registers active in each group and permits a limit slightly above the 255-register ceiling when fewer groups outweigh the cost of register spilling. Consequently, every layer of 7net-nano-5.5 fits in one table group in each direction. On the H200, the widest layer of 7net-omni (64 paths, $N_W=3008$) uses 8 table backward groups against 32 in kernel-only mode (or 94 under kernel-only register accounting). The table backward pass offloads the harmonics Jacobian to Kernel 2 through a transient per-edge buffer of $[(l_\text{max}+1)^2+1]\times\SI{4}{\byte}$, corresponding to \SI{68}{\byte} for $l_\text{max}=3$ and \SI{40}{\byte} for $l_\text{max}=2$. This buffer is freed after the backward pass and increases the peak memory only in the table call of 7net-omni.

\subsection{Tabulated radial weights}
\label{app:table_impl}

The table uses uniform nodes $r_k=kh$ with $h=r_c/(N'-1)$, where $N'$ is the node count after the alignment described below. For each node and each column $c$, one entry stores four fp32 values, namely $W_c(r_k)$, $W_c'(r_k)$, the divided difference $(W_c(r_{k+1})-W_c(r_k))/h$ and $W_c'(r_{k+1})$. The kernel reads an entry in one 16-byte load, and one layer occupies $N'\times 4N_W\times\SI{4}{\byte}$.

Main Sec.~\ref{sec:table} introduces three numerical safeguards. First, the divided difference $(W(r_{k+1})-W(r_k))/h$ is precomputed in fp64, which avoids a cancellation that otherwise raises the maximum force error to \SI{1.4e-3}{\electronvolt\per\angstrom} at $N=8192$. Second, a node is aligned with the cutoff onset $r_\text{on}$, which is 5.0, 4.5 and \SI{5.5}{\angstrom} for 7net-nano-5.5, 7net-0 and 7net-omni, respectively. For this purpose, $N'$ is set to the largest integer in $(N/2, N]$ that places a node on $r_\text{on}$, which reduces the requested $N=2048$ to 2047, 2041 and 2041. Third, $(N'-1)/r_c$ is stored as the sum of two fp32 numbers. This removes a relative rounding of the edge length of up to \num{6e-8}, which adds a systematic energy bias but leaves the forces unchanged.

\subsection{Integration with LAMMPS}
\label{app:lammps}

\jet{} connects to LAMMPS \cite{si:thompson_lammps_2022} through one ML-IAP wrapper shared by all five families. For SevenNet, it also provides the pair style \texttt{e3gnn}, which remains the main SevenNet backend (main Sec.~\ref{sec:md}, with the two interfaces compared in Sec.~\ref{sec:intshare}). The paragraphs below describe graph replay, the GPU edge build and the family adapters, and Sec.~\ref{sec:d3} describes JET-D3.

\paragraph{CUDA-graph replay.} A CUDA graph replays a recorded sequence of GPU operations with one launch, which reduces the CPU launch cost \cite{si:ekelund_cudagraphs_2025}. The \texttt{pair\_style e3gnn} step comprises edge construction, the model forward, $\partial E/\partial\mathbf r$ and the force and virial reduction. It is captured once for each combination of local atom count and edge bucket and then replayed. Although the edge count is rounded up to a bucket, the destination CSR terminates at the actual edge count, preventing the kernels from processing padded entries. The ML-IAP wrapper described below replays its model call in the same way.

\paragraph{GPU edge build.} In \texttt{e3gnn}, LAMMPS's full neighbor list is uploaded only on rebuilds. On each step, three fixed-shape kernels inside the graph select the pairs within the cutoff from the current positions and build the destination CSR. They use explicitly rounded fp64 squared distances without FMA contraction, which produces an edge set bitwise identical to that of LAMMPS's host loop. If the edge count exceeds the captured capacity, the buffers are resized, the graph is recaptured and the step is recomputed.

\paragraph{ML-IAP wrapper and family adapters.}
\label{app:companion}
The table stores the whole $\mathbf W(r)$ including the cutoff and the basis normalization, which makes the kernel independent of how each code forms its radial basis. The family adapters of Sec.~\ref{sec:md} patch the installed package in place (SevenNet, MACE) or rewrite the loaded model (NequIP, Nequix, EquFlash). The MACE blocks outside the kernels are compiled with the PyTorch compiler \texttt{torch.compile}. In contrast, MACE's released ML-IAP model, which \texttt{mace\_create\_lammps\_model} creates with cuEquivariance, runs eagerly without compilation or CUDA-graph capture. The NequIP adapter supplies raw edge vectors directly to the kernels rather than precomputed spherical harmonics.

A single ML-IAP wrapper deploys every family, including SevenNet. It is a LAMMPS-side module separate from the adapters and requires each family to supply only its model forward, its cutoff and its index and vector types. On a single rank, the wrapper folds each ghost atom into its owner, eliminating the exchange of ghost features between layers. It masks capacity padding out of the convolution CSR and replays the entire model call, including input preparation, as a single CUDA graph. EquFlash V2 executes without graph replay, since a cuEquivariance launch invalidates stream capture under both cuEquivariance 0.6.0 and 0.9.1. Across multiple MPI ranks, every family falls back to per-layer ghost exchange without ghost folding or graph replay. This exchange uses the released wrappers for SevenNet, MACE, NequIP-OAM and EquFlash and a per-layer exchange hook of \jet{} for Nequix.

\clearpage
\section{Accuracy and MD validation}
\subsection{Reference tests and full accuracy tables}
\label{app:accuracy_full}
This subsection details the reference tests, fp64 comparisons and repeated-call checks supporting main Sec.~\ref{sec:accuracy}. Unless stated otherwise, the accuracy runs used the hardware and software of Sec.~\ref{sec:hardware} with TF32 disabled. They evaluated ten frames of the 5632-atom electrolyte, taken \SI{0.5}{\pico\second} apart from an NVE trajectory of 7net-nano-5.5 at \SI{300}{\kelvin}. For 7net-omni \cite{si:kim_optimizing_2026}, NequIP-OAM L and EquFlash, whose fp64 evaluations exceed GPU memory at 5632 atoms, ten frames of the 352-atom cell (Sec.~\ref{sec:system}) were used instead. SevenNet undergoes the most comprehensive validation, as it provides several fp32 backends (e3nn, FlashTP, cuEquivariance and OpenEquivariance) alongside the kernel-only mode, and its ASE interface constructs edge vectors in fp64, yielding the lowest single-point fp32 rounding floor. For the remaining families, \jet{} kernel + table is compared against the released fp32 TP-accelerated and fp64 models (Table~\ref{tab:family_acc}).

\paragraph{Reference tests.} A pure-PyTorch implementation of Eqs.~\eqref{eq:bessel}--\eqref{eq:agg} matches SevenNet's e3nn modules to $10^{-10}$ in fp64 on every layer of 7net-nano-5.5 and 7net-0. The kernel-only CUDA kernels, instantiated in fp64, pass the forward test against this reference at the same tolerance. In fp64, the kernels reproduce the e3nn forces to \SI{1.7e-14}{\electronvolt\per\angstrom} for 7net-nano-5.5 and 7net-omni and to \SIrange{1.3e-8}{2.8e-8}{\electronvolt\per\angstrom} for 7net-0 and its multi-fidelity counterpart 7net-mf-0 \cite{si:kim_sevennet_mf_2024} (evaluated in its PBE modality). The cause of the larger difference for the latter two models was not identified.

\paragraph{Differences between fp32 implementations.} In direct fp32 comparisons with e3nn (5632-atom electrolyte, 352 atoms for 7net-omni), the maximum force difference of \jet{} kernel only is \SIrange{1.00e-5}{1.18e-5}{\electronvolt\per\angstrom}, which is 1.7 to 4.3 times the \SIrange{2.3e-6}{6.4e-6}{\electronvolt\per\angstrom} of FlashTP, cuEquivariance and OpenEquivariance. Its energies agree with e3nn within \SI{1.7e-8}{\electronvolt\per\atom} and its stresses within \SI{1.0e-8}{\electronvolt\per\cubic\angstrom}. Against fp64, however, \jet{} is as accurate as e3nn (Table~\ref{tab:family_acc}). The fp32 rounding of e3nn itself is approximately $10^{-5}$~\si{\electronvolt\per\angstrom} in the maximum, comparable to the \jet{}--e3nn difference, so that direct fp32 differences reflect rounding in both implementations and cannot identify which is more accurate.

\begin{figure}[!htbp]
\centering
\includegraphics{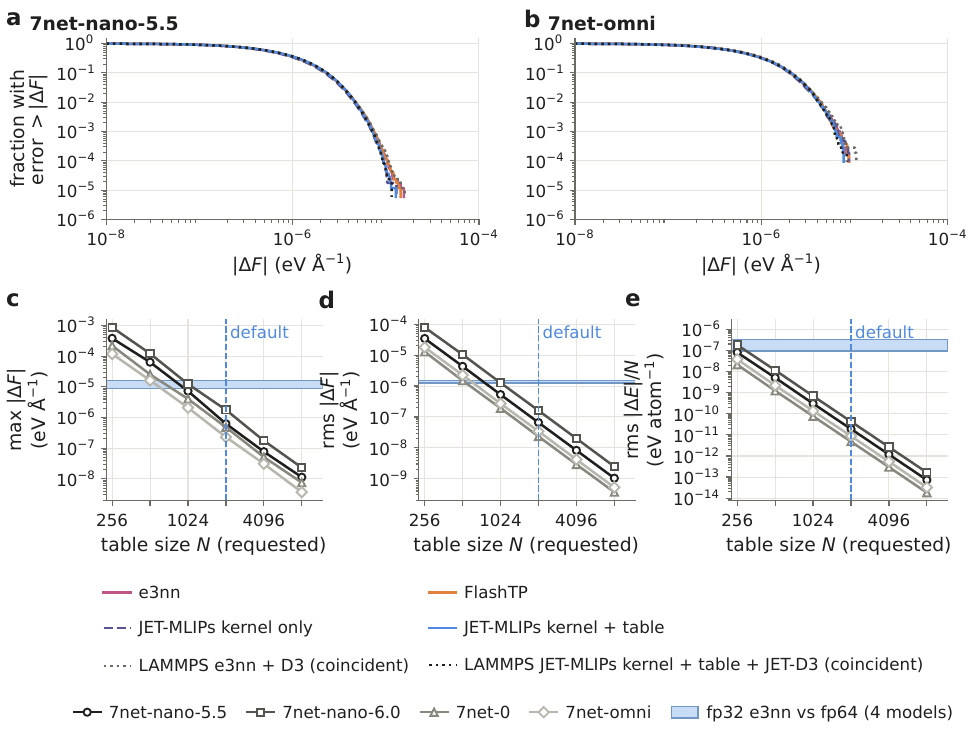}
\caption{\textbf{Force-error distributions against fp64.} (a,b) Fraction of force components exceeding $|\Delta F|$
for 7net-nano-5.5 and 7net-omni (ten electrolyte MD frames, 5632 and 352 atoms, respectively, one H200).
D3 deployments use fp64 e3nn + s-dftd3 as reference. (c--e) Maximum and rms force error
(\si{\electronvolt\per\angstrom}) and rms per-atom energy error (\si{\electronvolt\per\atom})
of the fp64 table against the fp64 radial MLP versus requested nodes $N$ on the aligned grid, using the same frames.
The light blue band spans four models' fp32 e3nn errors against fp64, and the dashed line marks default $N$.
Model and backend encodings are defined in the legends.}
\label{fig:g8}
\end{figure}

\paragraph{Single-point accuracy against fp64.}
\label{app:accuracy}
Table~\ref{tab:family_acc} lists the rms and maximum force-component errors and the per-atom energy errors underlying main Sec.~\ref{sec:accuracy} and main Table~\ref{tab:accratio}, and Fig.~\ref{fig:g8}a,b shows the force-error distributions for 7net-nano-5.5 and 7net-omni. At the total energy of the 5632-atom cell (\SI{-3.2e4}{\electronvolt}), one fp32 unit in the last place (ulp) \cite{si:goldberg_fp_1991} is \SI{1.95e-3}{\electronvolt} or \SI{3.5e-7}{\electronvolt\per\atom}. Across SevenNet, NequIP-OAM and Nequix, the rms energy errors remain near this resolution limit. In SevenNet, they are \SIrange{1.0e-7}{3.4e-7}{\electronvolt\per\atom} for \jet{} kernel + table and \SIrange{0.93e-7}{3.3e-7}{\electronvolt\per\atom} for fp32 e3nn. The sole deviation is a nearly constant offset of approximately \SI{-6e-7}{\electronvolt\per\atom} (1.7~ulp) for \jet{} kernel + table in NequIP-OAM L. This offset vanishes when the table is evaluated in fp64 and leaves the forces unchanged. In MACE and EquFlash, fp32 summation of atomic energies elevates the baseline energy errors to 80--800~ulp and 5~ulp, respectively, while \jet{} remains within 1.01 times the error of the existing backend.

For the forces of SevenNet, FlashTP matches fp32 e3nn in rms error and agrees within 8\% in maximum error. The LAMMPS runs with D3 use fp64 e3nn plus fp64 s-dftd3 \cite{si:ehlert_sdftd3_2024} as reference (Fig.~\ref{fig:g8}a,b). Against this reference, both \jet{} kernel + table with JET-D3 and upstream e3nn with upstream D3 remain within 0.99--1.05 times the rms error of fp32 e3nn, and both D3 implementations alone differ from s-dftd3 by \SI{4.4e-8}{\electronvolt\per\angstrom} rms. EquFlash V2 lacks an fp64 reference even at 352 atoms and was therefore compared directly with its existing fp32 TP-accelerated backend. The rms and maximum force differences are \SI{1.49e-6}{\electronvolt\per\angstrom} and \SI{1.13e-5}{\electronvolt\per\angstrom}, and the rms energy difference is \SI{1.49e-6}{\electronvolt\per\atom}, which is comparable to the fp32 rounding error of EquFlash V1.

In the remaining families, single-point maximum force errors are larger (\SI{1.1e-4}{\electronvolt\per\angstrom} in the 5632-atom cell against \SIrange{1.2e-5}{1.5e-5}{\electronvolt\per\angstrom} in SevenNet). They originate from each family's ASE interface, which constructs edge vectors from fp32 positions. Across a periodic boundary, subtracting coordinates of magnitude $L$ rounds an edge vector by half an ulp of $L$ (\SI{1.9e-6}{\angstrom} at $L=\SI{38.7}{\angstrom}$ and \SI{4.8e-7}{\angstrom} at $L=\SI{15.4}{\angstrom}$). In MACE, NequIP-OAM and Nequix, this rounding concentrates the maximum error on atoms possessing a bond shorter than \SI{1.75}{\angstrom} across a periodic boundary. In the 5632-atom cell, the maximum occurs on such an atom in all ten frames, whereas the largest error among the remaining atoms is 2 to 4 times smaller. \jet{} receives identical edge vectors and exhibits the same maxima. In contrast, the LAMMPS interfaces construct edge vectors from fp64 positions, so this artifact does not arise in MD.

\begin{figure}[!htbp]
\centering
\includegraphics{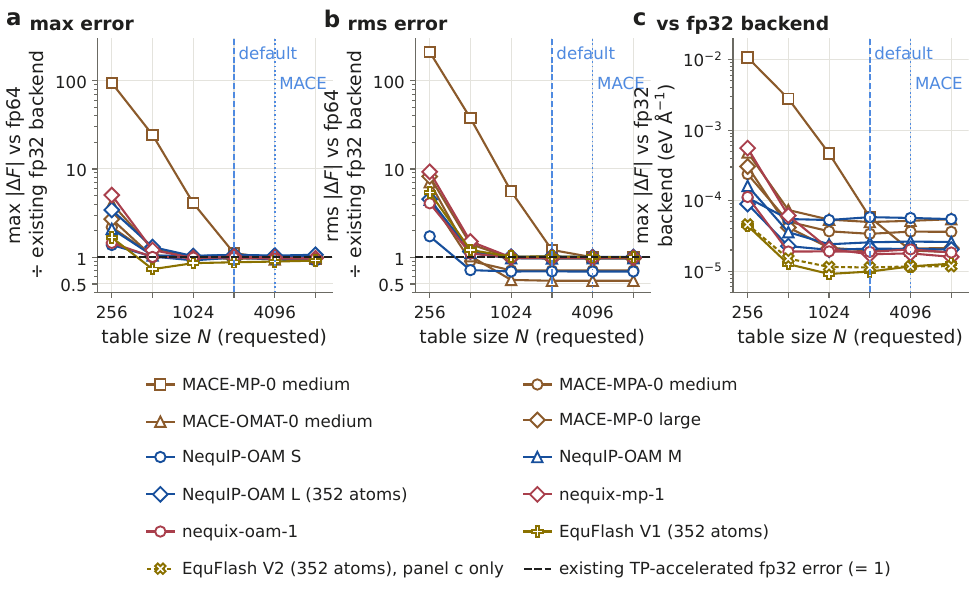}
\caption{\textbf{Radial table accuracy in the other families.} (a,b) Maximum and rms force errors of
fp32 \jet{} kernel + table against fp64 at table size $N$, divided by the fp32 errors of the existing TP-accelerated backend
(5632-atom electrolyte, 352 atoms for NequIP-OAM L and EquFlash, one H200).
(c) Maximum force difference between fp32 \jet{} kernel + table and the existing fp32 TP-accelerated backend of the same model. EquFlash V2 (dashed) appears in (c) only because its fp64 reference exceeds GPU memory.
Vertical dashed and dotted lines mark defaults 2048 and MACE 4096.}
\label{fig:table_family}
\end{figure}

\paragraph{Table error against size $N$.} For a cubic Hermite table, the interpolation error is bounded by $h^4\max|W_c^{(4)}|/384$ in $W_c(r)$ and by $\sqrt{3}\,h^3\max|W_c^{(4)}|/216$ in $W'_c(r)$. These bounds fall 16-fold and 8-fold per doubling of $N$, and the measured force error falls accordingly until it reaches the fp32 rounding floor (main Fig.~\ref{fig:table}a,b and Fig.~\ref{fig:table_family}a,b). SevenNet and the other families converge to their fp32 floors by $N=1024$--2048. In SevenNet, the errors over $N=2048$--8192 remain within 0.94--1.06 times the rms error and 0.85--1.22 times the maximum error of fp32 e3nn. The exception is MACE-MP-0 medium, whose rms error is 1.21 times the floor at $N=2048$ and converges at $N=4096$ (Table~\ref{tab:family_acc}). The tabulated convolution time increases by 6--15\% from $N=256$ to 8192.

Since the compiled table kernel supports fp32 only, isolated table accuracy was evaluated in fp64 by applying the same cubic Hermite formula in PyTorch to an fp64 table in place of the radial MLP. On the onset-aligned grid, the isolated table force error decreases eightfold per doubling of $N$ (Fig.~\ref{fig:g8}c--e). At $N=2048$, it reaches \SIrange{0.23e-7}{1.6e-7}{\electronvolt\per\angstrom} in rms and \SIrange{0.23e-6}{1.8e-6}{\electronvolt\per\angstrom} in maximum across the four SevenNet models, lying 9--55 and 9--38 times below the fp32 floor, respectively. In fp32, \jet{} kernel + table differs from \jet{} kernel only by at most \SIrange{6.9e-6}{1.0e-5}{\electronvolt\per\angstrom} in a force component and by \SIrange{0.7e-8}{2.1e-8}{\electronvolt\per\atom} in energy at $N=2048$ (main Fig.~\ref{fig:table}c). Across ten MD frames each of LiCoO$_2$ (360 atoms), diamond Si (216 atoms) and amorphous SiO$_2$ (560 atoms), both \jet{} modes remain within 0.93--1.07 times the rms error and 1.6 times the maximum error of fp32 e3nn, while the isolated fp64 table error lies 43--188 times below the fp32 floor (Table~\ref{tab:solids_acc}).

\paragraph{Determinism.} The forward pass of the \jet{} convolution accumulates without floating-point atomics and is deterministic across all five families, whereas atomic additions in the backward pass render the forces nondeterministic. On the ten 5632-atom electrolyte frames, repeated evaluations of \jet{} kernel + table differ by at most \SIrange{1.7e-6}{2.4e-6}{\electronvolt\per\angstrom} in any force component, or by up to \SI{2.6e-6}{\electronvolt\per\angstrom} when the kernel-only variants are included. The corresponding differences are \SIrange{3.1e-6}{5.9e-6}{\electronvolt\per\angstrom} for fp32 e3nn (7net-nano-5.5, 7net-nano-6.0 and 7net-0) and \SIrange{3.6e-6}{6.4e-6}{\electronvolt\per\angstrom} for FlashTP (all four models). Repeated e3nn and FlashTP evaluations can also vary by one ulp in total energy. Across the remaining eleven models, \jet{} kernel + table likewise exhibits smaller run-to-run force differences than the existing fp32 TP-accelerated backends (\SIrange{1.1e-6}{5.5e-6}{\electronvolt\per\angstrom} against \SIrange{1.4e-6}{2.1e-5}{\electronvolt\per\angstrom}, Table~\ref{tab:family_acc}). Residual energy variations of up to 13~ulp in MACE originate from nondeterministic GPU \texttt{scatter\_add} summation of atomic energies within MACE rather than from the convolution. Overall, \jet{} satisfies the single-call precision budget of main Sec.~\ref{sec:budget} across all 15 models, with the energy offset of NequIP-OAM L as the sole exception.

\subsection{MD validation}
\label{app:md}
\paragraph{Run-to-run divergence.}
\label{app:chaos}
For 7net-nano-5.5, we ran three 10-ps NVE trajectories with upstream FlashTP and three with \jet{} kernel + table from the identical equilibrated \SI{300}{\kelvin} state (positions and velocities) of the 5632-atom electrolyte (Table~\ref{tab:md}b). The runs used a \SI{0.5}{\femto\second} timestep without D3 and saved thermodynamic output every 10 steps and positions every 100 steps. Mixed backend pairs diverge at the same rate as same-backend pairs, with RMSD e-fold times of 109--115~fs. Over 2--10~ps (eight 1-ps blocks), the temperatures span 302.57--\SI{303.13}{\kelvin} for FlashTP and 302.67--\SI{303.25}{\kelvin} for \jet{}, with standard errors of 0.12--\SI{0.26}{\kelvin}. Over the same interval, the potential energies agree within \SI{8.9e-5}{\electronvolt\per\atom}, the pressures span \SIrange{12521}{12703}{\bar} and the energy drift remains within \SI{1.0e-4}{\milli\electronvolt\per\atom\per\pico\second}.

\begin{figure}[!htbp]
\centering
\includegraphics{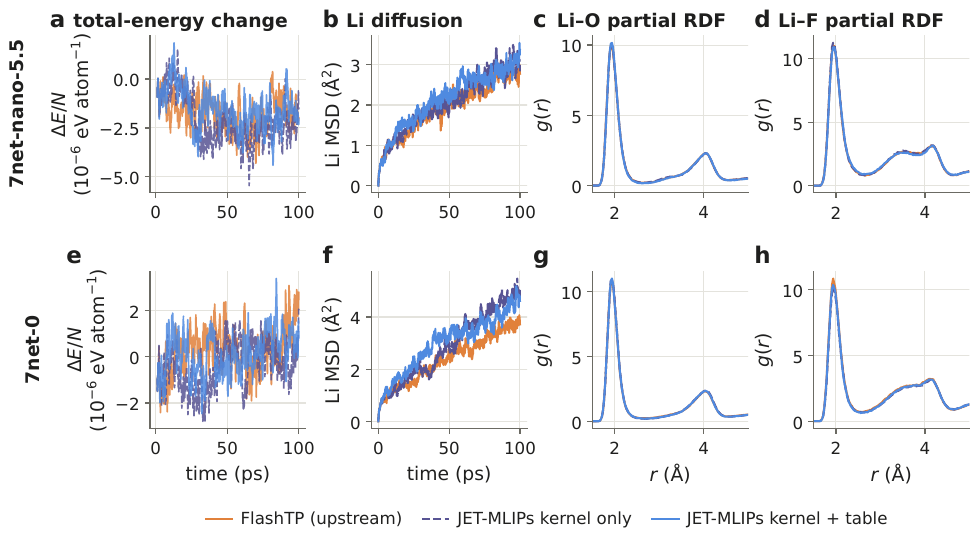}
\caption{\textbf{Energy conservation and Li coordination in NVE MD.} (a--d) 7net-nano-5.5. (e--h) 7net-0
(intact 5632-atom electrolyte, one H200, 100~ps NVE, \SI{0.5}{\femto\second} step, no D3).
Columns give total-energy change from its initial value per atom (\SI{e-6}{\electronvolt\per\atom}, 1-ps running mean),
Li mean square displacement (MSD, \AA$^2$), and dimensionless Li--O and Li--F partial radial distribution functions
(RDFs) versus distance (\AA). Backend encodings are defined in the legend.}
\label{fig:md}
\end{figure}

\begin{figure}[!htbp]
\centering
\includegraphics{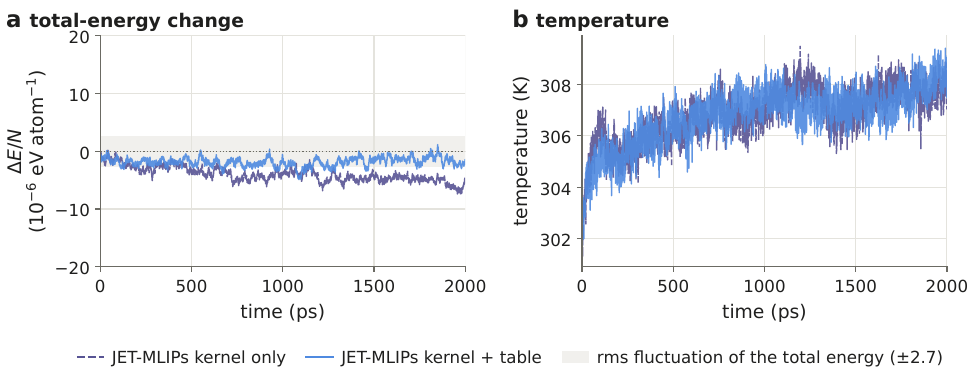}
\caption{\textbf{Two-nanosecond NVE simulation with 7net-nano-5.5.} (a) Total-energy change from its initial value per atom
(\SI{e-6}{\electronvolt\per\atom}). The gray band marks $\pm$\SI{2.7e-6}{\electronvolt\per\atom} about zero, which is the
root-mean-square fluctuation of the unsmoothed total energy about its linear fit. (b) Temperature. Both are 1-ps running means (intact 5632-atom electrolyte, one
H200, 2~ns NVE, \SI{0.5}{\femto\second} step, no D3, same initial positions and velocities for both backends).}
\label{fig:nve2ns}
\end{figure}

\paragraph{Energy conservation over 100~ps and 2~ns.}
For 7net-nano-5.5 and 7net-0, each model was initialized on the intact 5632-atom cell by minimization and \SI{10}{\pico\second} of NVT at \SI{300}{\kelvin}. We then ran \SI{100}{\pico\second} of NVE at \SI{0.5}{\femto\second} without D3 with upstream FlashTP, \jet{} kernel only and \jet{} kernel + table (Table~\ref{tab:md}a and Fig.~\ref{fig:md}). Every run drifts by at most \SI{3e-5}{\milli\electronvolt\per\atom\per\pico\second}, or less than \SI{3e-6}{\electronvolt\per\atom} over 100~ps, with no systematic sign. Extending the 7net-nano-5.5 runs to \SI{2}{\nano\second} with both \jet{} modes (Fig.~\ref{fig:nve2ns}) completes without atom loss and exhibits the same temperature increase of 3--\SI{4}{\kelvin} attributable to continued structural relaxation at constant total energy. Over 2~ns, the total-energy drift is \num{1.7e-7} (kernel + table) and \SI{-1.7e-6}{\milli\electronvolt\per\atom\per\pico\second} (kernel only). The accumulated change of the latter (\SI{3.4}{\micro\electronvolt\per\atom}) remains comparable to the rms fluctuation of \SI{2.7}{\micro\electronvolt\per\atom} about the linear fit.

\paragraph{Thermodynamic and structural properties.}
Over the second half of the 100-ps runs (Table~\ref{tab:md}a), the three backends agree within \SI{1.5e-4}{\electronvolt\per\atom} in potential energy, \SI{1.1}{\kelvin} in temperature and 0.4\% in pressure. The largest discrepancy reflects the exchange between potential and kinetic energy at constant total energy ($2\Delta E_\text{pot}/3k_B\approx\SI{1.2}{\kelvin}$). A repeated run of \jet{} kernel + table from the same state differs by up to \SI{5e-5}{\electronvolt\per\atom}, \SI{0.4}{\kelvin} and 0.9\% in pressure. Kernel only and kernel + table differ by at most \SI{6e-5}{\electronvolt\per\atom}, \SI{0.5}{\kelvin} and 0.4\%, indicating no tabulation bias above the run-to-run divergence. The first Li--O and Li--F RDF peaks agree within one bin of \SI{0.025}{\angstrom}, and the coordination numbers integrated to \SI{2.7}{\angstrom} agree within 0.06 (Fig.~\ref{fig:md}c,d,g,h). The large positive pressures (12500 and \SI{8400}{\bar}) result from omitting D3 at the volume equilibrated with 7net-0 plus D3. The sublinear Li MSD over 100~ps (Fig.~\ref{fig:md}b,f) is dominated by individual hops and is not used as an equivalence metric.

\paragraph{NVE drift of the other families.}
\label{app:family_nve}
 We also measured the NVE energy drift of the other families on the intact
5632-atom cell (Sec.~\ref{sec:system}). Each trajectory was equilibrated by \SI{10}{\pico\second} of NVT at \SI{300}{\kelvin} (Nos\'e--Hoover thermostat, \SI{0.05}{\pico\second} damping) using one of three initial velocity seeds. Velocities were then reinitialized at \SI{300}{\kelvin} with a fixed seed prior to NVE integration, using a \SI{0.5}{\femto\second} timestep throughout. The mean temperature of the NVE runs is \SIrange{295}{306}{\kelvin}.

In the 10~ps runs, MACE-MP-0 medium drifts by \num{-1.4e-4} to
\SI{+2.7e-4}{\milli\electronvolt\per\atom\per\pico\second} with \jet{} kernel + table and by \num{-1.6e-4} to
\SI{+2.4e-4}{\milli\electronvolt\per\atom\per\pico\second} with cuEquivariance. NequIP-OAM M drifts by \num{+3.4e-5} to
\SI{+1.5e-4}{\milli\electronvolt\per\atom\per\pico\second} with \jet{} kernel + table and by \num{-3.7e-5} to
\SI{+1.9e-4}{\milli\electronvolt\per\atom\per\pico\second} with OpenEquivariance. The seed-to-seed ranges of these 10~ps runs are \num{1.2e-4} to \SI{4.1e-4}{\milli\electronvolt\per\atom\per\pico\second} wide and set the resolution of the comparison. This resolution is approximately ten times coarser than the bound of the 100~ps SevenNet runs (Sec.~\ref{app:md}).

With \jet{} kernel + table, 10~ps runs of
nequix-mp-1 and EquFlash V1 with three seeds drift by \num{-1.2e-4} to \num{+1.2e-4} and by \num{-1.4e-4} to
\SI{+6.3e-6}{\milli\electronvolt\per\atom\per\pico\second}, respectively. The 100~ps runs (one seed) drift by
\num{-7.1e-5}, \num{-1.5e-5} and \SI{-1.2e-5}{\milli\electronvolt\per\atom\per\pico\second} for MACE-MP-0 medium,
NequIP-OAM M and nequix-mp-1, respectively. The standard deviation of the total energy is
\SIrange{2.3e-3}{3.7e-3}{\milli\electronvolt\per\atom} in every run.

\clearpage
\section{Performance, memory and MD integration}
\label{app:performance}
This section supports the performance results of Secs.~\ref{sec:nequip}--\ref{sec:memory}. It covers additional SevenNet backends and neighbor-list rebuilds, JET-D3, step timelines, model-call scaling, MD integration, larger cells and dense solids, memory behavior and consumer-GPU benchmarks. The run protocol and memory definitions are given in Secs.~\ref{sec:protocol} and~\ref{sec:runprotocol}.

\subsection{Additional SevenNet backends and neighbor-list rebuilds}
\label{sec:lmp_backends}
\begin{figure}[!htbp]
\centering
\includegraphics{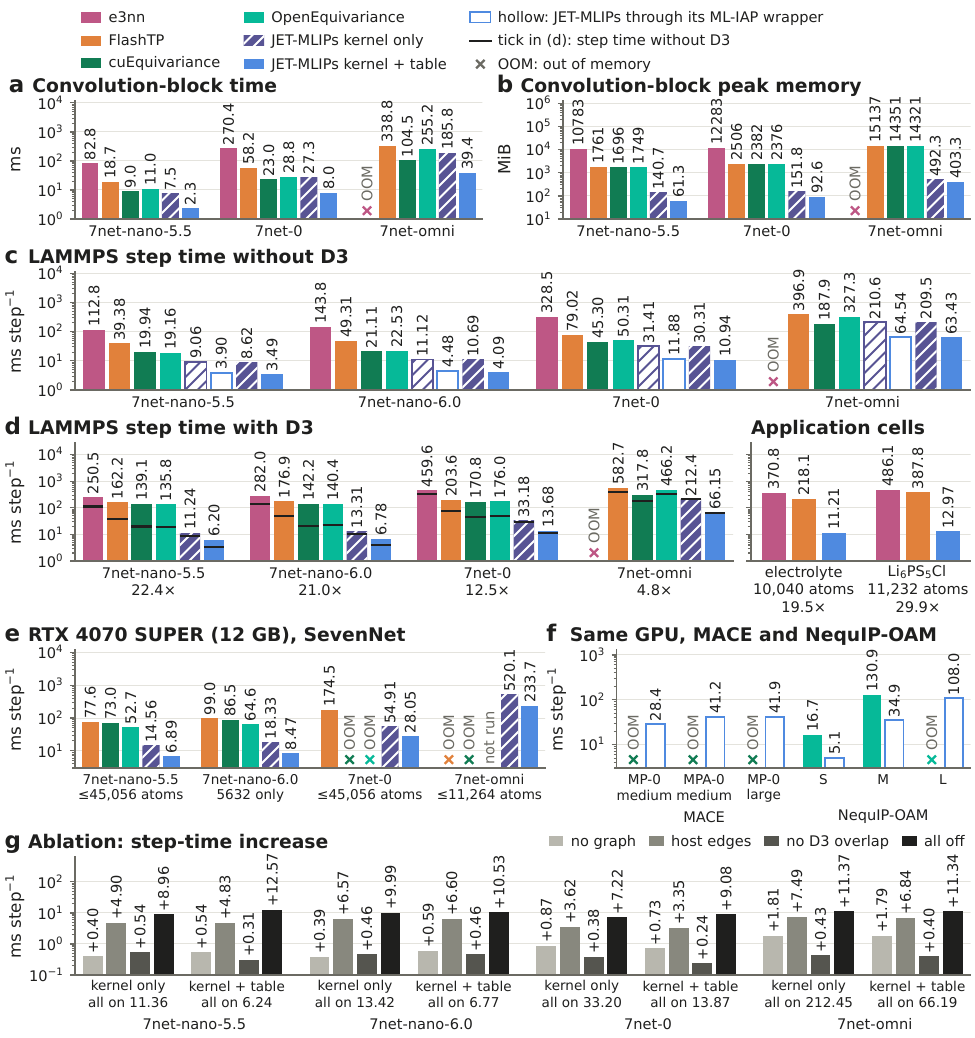}
\caption{\textbf{SevenNet backends, MD integration and consumer-GPU benchmarks.} (a,b) Convolution-block time (forward plus backward, summed over layers) and largest per-layer peak allocation. (c) Rebuild-free LAMMPS step time without D3, where solid \jet{} bars use \texttt{e3gnn}. (d) Rebuild-free step time with D3 (upstream D3 for upstream backends, JET-D3 for \jet{}), annotated with speed-ups of \jet{} kernel + table over cuEquivariance and, for the application cells (7net-nano-6.0), over FlashTP. (e,f) Mean step time including rebuilds on a 12~GB RTX 4070 SUPER without D3, with the largest accommodated system size for \jet{} annotated in (e). (g) Step-time increase relative to the all-on configuration (in ms) when individual \texttt{e3gnn} integration features or all three are disabled. Panels (a--d) and (g) use one H200, and all benchmarks except the application cells use the 5632-atom electrolyte.}
\label{fig:perf_summary}
\end{figure}

Main Table~\ref{tab:mace} compares each model with its upstream model and its existing TP-accelerated backend, and Fig.~\ref{fig:perf_summary}c presents the additional SevenNet backends. The unaccelerated e3nn backend executes on the GPU without specialized TP kernels, and FlashTP uses the pair style of upstream SevenNet. The cuEquivariance and OpenEquivariance backends use ML-IAP with Kokkos, upstream SevenNet 0.13.0 and upstream D3, a combination referred to below as the upstream ML-IAP deployment. Both \jet{} modes execute through \texttt{e3gnn} or through the \jet{} ML-IAP wrapper (Sec.~\ref{app:companion}, hollow bars in Fig.~\ref{fig:perf_summary}c).

Against upstream FlashTP without D3, \jet{} kernel + table is 6.3--12.1$\times$ faster and \jet{} kernel only through \texttt{e3gnn} is 1.9--4.6$\times$ faster across the four SevenNet models (Fig.~\ref{fig:perf_summary}c). Among the upstream backends, cuEquivariance is the fastest except for 7net-nano-5.5, where OpenEquivariance is slightly faster (19.16 against 19.94~ms). Across the repeated runs of main Table~\ref{tab:mace} and Fig.~\ref{fig:perf_summary}c, the max--min spread of the rebuild-free step time remains at most 3.4\% of the median for SevenNet. The exceptions are cuEquivariance for 7net-nano-5.5 (10\%, with runs clustered near 18.1 and 20.0~ms) and for 7net-nano-6.0 (5.2\%). For the other four families, the spread is at most 0.9\% for \jet{} and 0.6\% for the existing TP-accelerated backends, except 7.2\% for the existing TP-accelerated backend of NequIP-OAM S.

\paragraph{Neighbor-list rebuilds.}
\label{sec:rebuilds}
The rebuild penalty of upstream cuEquivariance (Sec.~\ref{sec:nequip}) appears in the mean step time including rebuilds. Without D3, including the 9--10 rebuilds of the electrolyte in the 450 timed steps raises its mean step time from 19.94, 21.11, 45.30 and 187.9~\si{\milli\second\per\step} to 33.33, 36.35, 62.27 and 397.3~\si{\milli\second\per\step} for 7net-nano-5.5, 7net-nano-6.0, 7net-0 and 7net-omni, respectively. The other backends of Fig.~\ref{fig:perf_summary}c remain within 0.7~ms of their rebuild-free values. MACE's existing TP-accelerated backend uses the same cuEquivariance library and shows no such penalty, as its mean step time including rebuilds remains within 0.6~ms of the rebuild-free time for the four MACE models. Given this contrast, the penalty appears specific to the upstream SevenNet integration, and including rebuilds would further increase the speed-ups reported in Table~\ref{tab:mace}.

This penalty grows with system size. In the size scans, which use the first run per size, the cost per rebuild is $(t_\text{all}-t_\text{pure})\times450/N_\text{rebuild}$, where $t_\text{all}$ is the mean step time including rebuilds, $t_\text{pure}$ is the rebuild-free step time and $N_\text{rebuild}$ is the number of rebuilds in the 450 timed steps. At 5632 atoms, each rebuild costs 0.61, 0.75, 0.86 and 9.5~s for 7net-nano-5.5, 7net-nano-6.0, 7net-0 and 7net-omni, respectively. At \num{45056} atoms, it reaches 4.0~s for 7net-nano-5.5, 4.5~s for 7net-0 and 4.7~s for 7net-nano-6.0. For 7net-omni, it reaches 16.1 and 27.5~s at \num{11264} and \num{22528} atoms. The solid scans involve no neighbor-list rebuilds.

\subsection{JET-D3, a GPU dispersion correction}
\label{sec:d3}
JET-D3 (\texttt{pair\_style jet/d3}) implements the D3(BJ) dispersion correction as a LAMMPS pair style in JET-MLIPs and prevents dispersion evaluation from dominating the MD step once the MLIP is accelerated. It is independent of the model family, although all D3 timings below use the SevenNet models, and upstream D3 denotes SevenNet's \texttt{pair\_style d3}.

\paragraph{Method.}
\label{sec:jetd3_method}
Atom pairs are processed in $32\times32$ tiles with one warp per tile, and each unordered pair is visited once. In simulation cells that are at least 1.5 cutoffs long in every periodic direction, the atoms are sorted into spatial bins, and tiles whose atom pairs cannot lie within the cutoff are skipped. The coordination-number-dependent $C_6$ interpolation is factorized per atom and partner type, which reduces a pair to approximately 10 FMA instead of a $5\times5$ reference loop with a division. Buffers are ordinary device allocations, whereas the upstream code uses unified memory per atom. D3 executes on an independent CUDA stream. When combined with \texttt{e3gnn}, it is launched prior to edge construction, enabling concurrent execution so that dispersion evaluation does not serialize with the convolution.

\paragraph{Agreement with upstream D3.}
JET-D3 agrees with upstream D3 at 5632, \num{11264} and \num{22528} atoms within $|\Delta E|/N\le\num{1.4e-9}~\si{\electronvolt\per\atom}$, $|\Delta P|\le\num{8.9e-4}~\si{\bar}$ and $|\Delta F|\le\num{1.2e-7}$~\si{\electronvolt\per\angstrom}. The force difference lies an order of magnitude below the fp32 force spread. Both implementations differ from the fp64 reference s-dftd3 by \SI{4.4e-8}{\electronvolt\per\angstrom} rms (Sec.~\ref{app:accuracy}).

\paragraph{Step time with D3.}
Including D3 increases the speed-ups of \jet{} kernel + table over the upstream SevenNet backends. They reach 22.4, 21.0, 12.5 and 4.8$\times$ against cuEquivariance and 26.2, 26.1, 14.9 and 8.8$\times$ against FlashTP for 7net-nano-5.5, 7net-nano-6.0, 7net-0 and 7net-omni, respectively (Fig.~\ref{fig:perf_summary}d). JET-D3 adds only 2.6--\SI{2.7}{\milli\second\per\step}, whereas upstream D3 adds 119--\SI{130}{\milli\second\per\step} to cuEquivariance and 123--\SI{186}{\milli\second\per\step} to FlashTP. Consequently, upstream FlashTP with D3 spends 76\% of its 7net-nano-5.5 step time in dispersion evaluation.

\paragraph{Standalone scaling.}
In standalone D3 benchmarks on electrolyte replicas (first run per size, D3 settings of Sec.~\ref{sec:system}), JET-D3 takes 3.16, 7.24 and \SI{18.8}{\milli\second\per\step} at 5632, \num{11264} and \num{22528} atoms, respectively. Upstream D3 takes 92.2, 145 and \SI{926}{\milli\second\per\step}, which corresponds to speed-ups of 29, 20 and 49$\times$. Over these three sizes, the device memory of JET-D3 remains nearly constant at 533--545~MiB, whereas upstream D3 requires 783, 1295 and 3471~MiB. At \num{45056} atoms, upstream D3 terminates with \texttt{cudaErrorInvalidAddressSpace}, whereas JET-D3 scales to \num{360448} atoms in \SI{499}{\milli\second\per\step} and 797~MiB.

\subsection{Step timeline and convolution blocks}
\label{sec:time}
\begin{figure}[!htbp]
\centering
\includegraphics{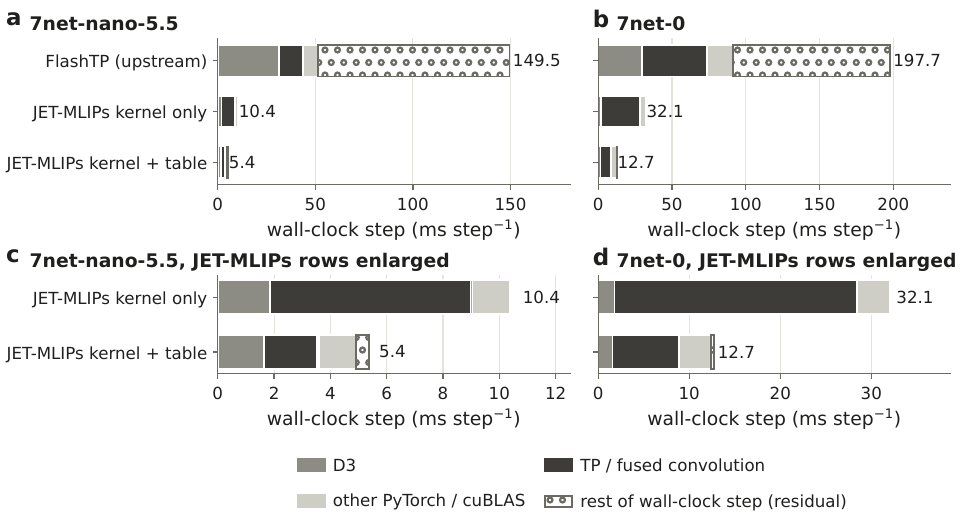}
\caption{\textbf{GPU timeline of one MD step.} (a,b) GPU kernel time by category and residual wall-clock time
for 7net-nano-5.5 and 7net-0 (5632-atom electrolyte, one H200, NVE with D3,
Nsight Systems, last 50 of 60 steps). (c,d) Expanded view of the \jet{} rows.
Residuals subtract summed profiled kernel time from unprofiled mean step time including rebuilds
and include host overhead rather than measured idle time. Bar-end labels give wall-clock time, and the legend defines categories.
No residual is drawn for 7net-0 \jet{} kernel only, whose kernel sum equals its step time,
or below 0.3~ms, including 7net-nano-5.5 kernel only at 0.1~ms.
The upstream residuals are approximate because the upstream wall-clock time varies between runs (Sec.~\ref{sec:time}).
The \SI{0.06}{\milli\second\per\step} GPU edge build is a gray hairline in (c).}
\label{fig:timeline}
\end{figure}

Figure~\ref{fig:timeline}a,b presents Nsight Systems timelines over the last 50 of 60 NVE steps with D3. Upstream FlashTP uses 52 and 91~ms of GPU kernel time per step for 7net-nano-5.5 and 7net-0, of which D3 accounts for 31 and 30~ms and the TP kernels for 12.5 and 44~ms. The host issues 500 and 825 launches per step, accounting for the remainder of the wall-clock steps of 149.5 and 197.7~ms. These wall-clock times are five-run medians of the mean step time including rebuilds with spreads of 13--15\%, while the rebuild-free medians of the same runs are 162.2 and 203.6~ms. With \jet{} (Fig.~\ref{fig:timeline}c,d), the host issues one CUDA-graph launch and four kernel launches per step, rendering the step GPU-bound with 0.8--1.0~ms of idle time. GPU kernel time is 10.4 and 32.0~ms in kernel-only mode and 4.9 and 12.4~ms with the table, of which the convolution takes 1.9 and 7.3~ms with the table. D3 accounts for 1.6--1.9~ms, the GPU edge build for 0.06~ms and the remaining 290 and 530 PyTorch and cuBLAS kernels for 1.3 and 3.5~ms. After tabulation, the remaining modules of 7net-nano-5.5 excluding D3 (1.4~ms) thus approach three quarters of its convolution time (1.9~ms).

\begin{figure}[!htbp]
\centering
\includegraphics{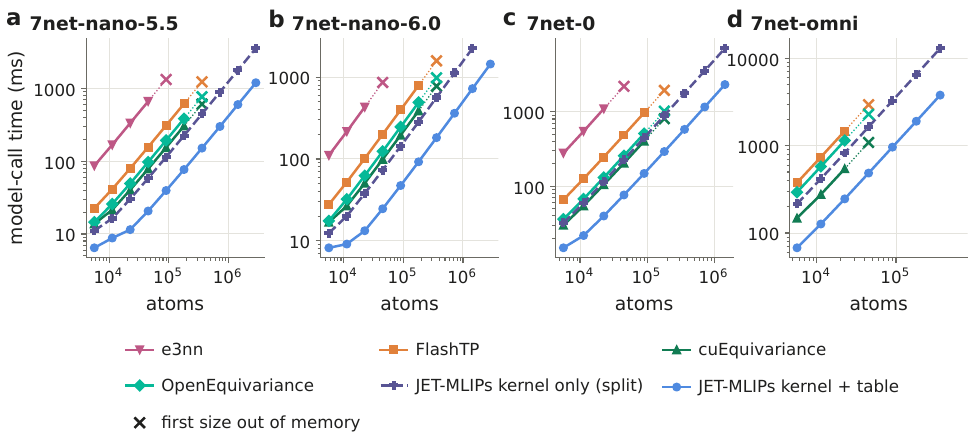}
\caption{\textbf{Model-call time against system size.} Mean forward-plus-force CUDA-event time in ms versus atom
count (5632-atom electrolyte replicas, one H200, first benchmark run per configuration).
(a--d) SevenNet models labeled in each panel, with \jet{} kernel only using default \texttt{split}
and \jet{} kernel + table using default $N$. Crosses mark first out-of-memory sizes at times extrapolated
linearly from the last fit, joined by thin dotted lines without measurements. In (d), e3nn exceeds GPU memory already at
5632 atoms and has no curve. Backend markers are defined in the legend, with kernel only dashed and other
configurations solid. Timings measure standalone model calls excluding LAMMPS integration.}
\label{fig:call_scaling}
\end{figure}

Figure~\ref{fig:perf_summary}a,b reports the execution time and peak allocation of the convolution blocks. The e3nn, FlashTP, cuEquivariance and OpenEquivariance blocks operate on precomputed edge embeddings, whose evaluation requires 1.0--2.6~ms once per model call and is excluded. The \jet{} blocks evaluate the edge geometry inside the kernel (using \texttt{split} in kernel-only mode). With e3nn, FlashTP, cuEquivariance and OpenEquivariance, the convolution blocks account for 73--98\% of a model call for 7net-nano-5.5 and 7net-0 when the edge embedding is counted once (call times in Fig.~\ref{fig:call_scaling}). With \jet{}, this fraction decreases to 67--85\% (kernel only) and 36--53\% (kernel + table). After tabulation, the remaining modules thus account for 47--64\% of the model call, consistent with the shift noted in Sec.~\ref{sec:perf}. Replacing only the e3nn TP with FlashTP reduces the convolution time of 7net-nano-5.5 from 82.8 to 18.7~ms. The unfused TP and aggregation therefore account for at least three quarters of that time despite representing the smaller share of arithmetic (Sec.~\ref{sec:cost}), indicating that their cost is governed by memory traffic rather than arithmetic.

For MACE, Nsight Systems profiles of the 5632-atom electrolyte on the H200 with $N=4096$ show that the fused convolution takes 65\% of GPU kernel time per step for MACE-MP-0 medium (6.4 of 9.7~ms) and 44\% for MACE-MP-0 large (5.7 of 13.0~ms). In these profiles, the species-grouped self-connection kernel of Sec.~\ref{sec:md} was enabled.

\subsection{Model-call scaling}
\label{sec:callscaling}
The model-call scan of Sec.~\ref{sec:memory} replicates the 5632-atom cell until each backend exceeds GPU memory, establishing the capacity limits of the remaining SevenNet backends (Fig.~\ref{fig:memory}a--d). For 7net-nano-5.5, FlashTP and OpenEquivariance accommodate \num{180224} atoms and e3nn \num{45056} atoms. For 7net-0, FlashTP and OpenEquivariance accommodate \num{90112} atoms, matching cuEquivariance, while e3nn reaches \num{22528} atoms. For 7net-omni, the TP accelerators accommodate \num{22528} atoms, whereas e3nn exceeds GPU memory at 5632 atoms. All these capacities remain below those achieved by \jet{} kernel + table in Sec.~\ref{sec:memory}.

In the model-call timings from the same scan, \jet{} kernel + table achieves the shortest model-call time at every size across all four SevenNet models, including 7net-nano-6.0 with its longer cutoff (Fig.~\ref{fig:call_scaling}). At 5632 atoms, it requires 6.4, 14.9 and 67.7~ms for 7net-nano-5.5, 7net-0 and 7net-omni, respectively. \jet{} kernel only outperforms FlashTP throughout but is slower than cuEquivariance for 7net-0 (32.1 against 30.0~ms) and 7net-omni (215 against 148~ms).
Against \jet{} kernel only, the table shortens the call at 5632 atoms by 3.2$\times$ for 7net-omni, 2.2$\times$ for 7net-0 and 1.7$\times$ for 7net-nano-5.5, which is the range quoted in Sec.~\ref{sec:perf}.

\subsection{MD integration: ablation and share of the step}
\label{app:integration}

\paragraph{Ablation.}
In the ablation study, graph replay, GPU edge construction and D3 overlap were disabled individually in \texttt{e3gnn} with D3 enabled (Fig.~\ref{fig:perf_summary}g, medians of three runs). Benchmarks used a timing-only input with a \SI{0.1}{\femto\second} timestep and \texttt{fix nve/limit 0.01} to suppress neighbor-list rebuilds. GPU edge construction provides the largest speed-up, as disabling it increases the 7net-nano-5.5 table step from 6.24 to \SI{11.07}{\milli\second}. Graph replay reduces step time by 2.7--8.7\% with the table and by 2.6--3.5\% with kernel only for the three smaller models, while D3 overlap reduces step time by 1.7--6.8\% with the table for the three smaller models. Outside SevenNet, enabling graph replay in the \jet{} ML-IAP wrapper reduces the 352-atom EquFlash V1 step time by 20\% (from 74.0 to \SI{59.0}{\milli\second\per\step}) and lowers whole-GPU memory from 4155 to 3347~MiB.

\paragraph{Kernel-attributable speed-up.}
\label{sec:intshare}
To separate kernel performance from MD integration, we executed \jet{} kernel only inside the upstream SevenNet ML-IAP deployment (Sec.~\ref{sec:lmp_backends}). It requires 14.04, 15.99, 44.93 and \SI{243.5}{\milli\second\per\step} for 7net-nano-5.5, 7net-nano-6.0, 7net-0 and 7net-omni, respectively, corresponding to 1.42, 1.32, 1.01 and 0.77 times the speed of cuEquivariance and 1.36, 1.41, 1.12 and 1.34 times that of OpenEquivariance. For 7net-omni, this deployment incurs a 34.0~ms overhead relative to \texttt{e3gnn} (\SI{209.5}{\milli\second\per\step}), of which Nsight Systems attributes 26.1~ms (77\%) to Kokkos kernels that pack and unpack ghost-atom data.

\paragraph{Small-cell floor of the upstream step.}
Profiling 7net-nano-5.5 at 360 atoms (LiCoO$_2$) and 5632 atoms (electrolyte) isolates two fixed costs behind the small-cell floor of main Sec.~\ref{sec:scaling}. First, the upstream ML-IAP deployment exchanges ghost-atom features in every layer without graph replay. Relative to \texttt{e3gnn}, this exchange adds 5.4--6.9~ms per step at 360 atoms and 5.4--5.5~ms at 5632 atoms, of which approximately 3.7~ms arises from 60 Kokkos ghost-exchange kernels. Second, the eager cuEquivariance model is host-bound. At 360 atoms, its 13.7--13.8~ms step occupies the GPU for only 6.1~ms while the host issues 587 model kernels, and step time remains near 14~ms up to 1152 atoms for both nano models. Through \texttt{e3gnn}, \jet{} kernel + table eliminates both overheads and requires 1.25 and \SI{3.49}{\milli\second\per\step} at the two sizes. The \jet{} ML-IAP wrapper likewise avoids them by folding ghost atoms into their owners and replaying the call as a single graph, leaving a residual overhead of 0.4--\SI{1.1}{\milli\second\per\step} over \texttt{e3gnn} due to host synchronizations and ML-IAP input preparation.

\paragraph{NequIP-OAM GPU time and wrapper control.}
In Nsight Systems traces of the 5632-atom electrolyte (last 60 of 160 steps), \jet{} kernel + table lowers GPU busy time relative to OpenEquivariance from 5.47 to \SI{1.60}{\milli\second} for NequIP-OAM S (3.4$\times$) and from 43.4 to \SI{11.4}{\milli\second} for M (3.8$\times$). Dense matrix products, which include the radial MLP, account for 25\% (1.36~ms) and 36\% (15.6~ms) of the OpenEquivariance GPU time for S and M, respectively. To isolate kernel performance from MD integration, OpenEquivariance was also evaluated inside the same \jet{} CUDA-graph wrapper, requiring 3.8, 37.7 and \SI{125.7}{\milli\second\per\step} for NequIP-OAM S, M and L compared with 2.0, 11.7 and \SI{37.9}{\milli\second\per\step} for \jet{} kernel + table (rebuild-free medians of three runs, same-harness column of main Table~\ref{tab:mace}).

\subsection{System size and dense solids}
\label{sec:solids}
To test scaling across materials (main Sec.~\ref{sec:scaling} and main Fig.~\ref{fig:scaling}), we compared the electrolyte with LiCoO$_2$, diamond Si and amorphous SiO$_2$. Their atomic densities are 0.097, 0.124, 0.050 and \SI{0.070}{\atom\per\cubic\angstrom}, respectively. The scans use the cells of Sec.~\ref{sec:system} and rebuild-free timings of the first run (Sec.~\ref{sec:runprotocol}). The static accuracy of SevenNet in these solids is given in Table~\ref{tab:solids_acc}.

\begin{figure}[!htbp]
\centering
\includegraphics{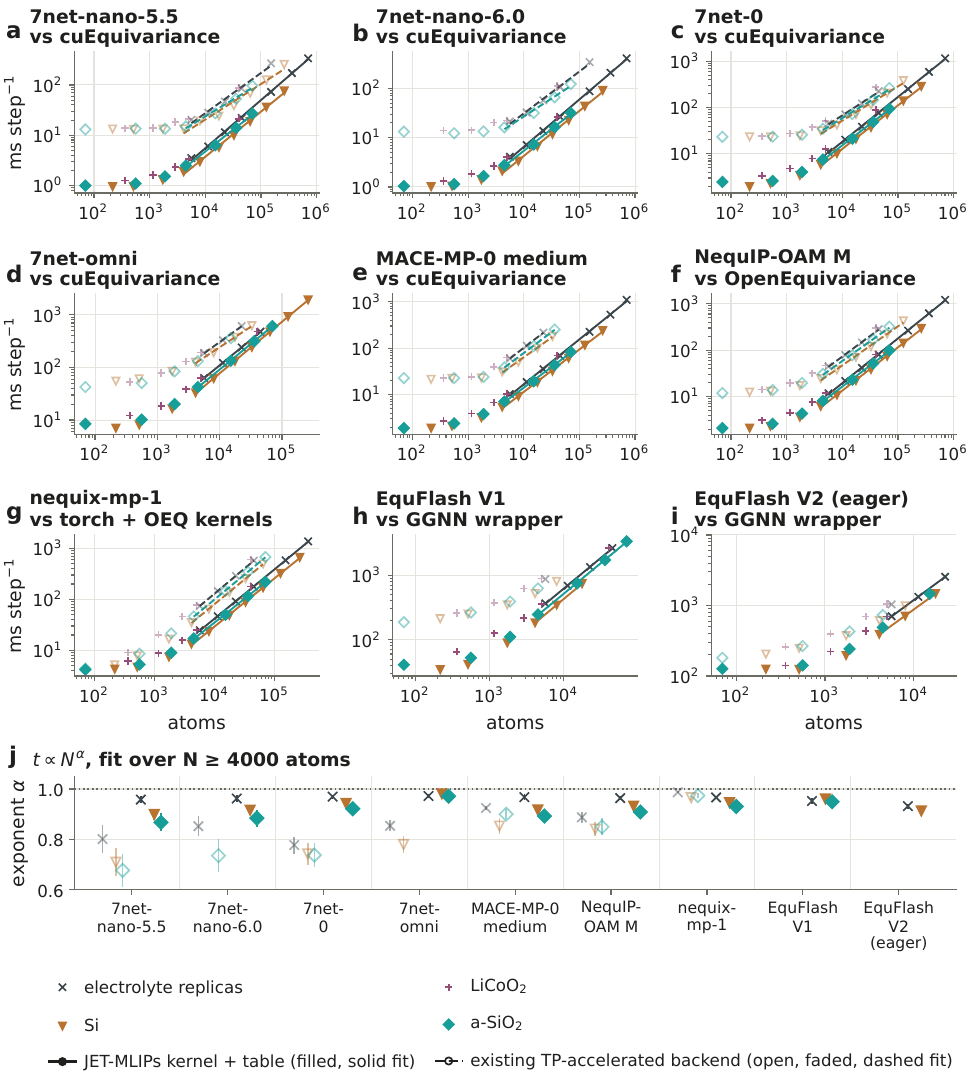}
\caption{\textbf{Growth of the step time with system size.} (a--i) Rebuild-free LAMMPS step time versus atom count
for the labeled models with \jet{} kernel + table and the existing TP-accelerated backends (one H200, first run, replicated electrolyte, LiCoO$_2$, Si and a-SiO$_2$).
System markers and backend fill/line encodings are defined in the legend.
Lines fit $\log t=\alpha\log N+c$ by least squares over $N\ge4000$ atoms, with at least three points
and no LiCoO$_2$ fit. (j) Exponents $\alpha$ of these fits with standard errors, where open markers on the left of each model denote the existing TP-accelerated backend and filled markers on the right \jet{} kernel + table. The dotted line marks $\alpha=1$.}
\label{fig:exponent}
\end{figure}

The power-law fits $\log t = \alpha \log N + c$ (Fig.~\ref{fig:exponent}) use ordinary least squares over the points with $N \ge 4000$ atoms and require at least three such points. LiCoO$_2$, which has only two, and the other scans below this threshold receive no fit. Standard errors are estimated from residuals with $n-2$ degrees of freedom and exclude run-to-run variation. Including the cells below 4000 atoms lowers $\alpha$ to 0.28--1.00 because of the fixed per-step overhead. In contrast, the slope between the two largest sizes approaches linearity (for example, 0.95 for 7net-nano-5.5 with cuEquivariance in Si between \num{125000} and \num{262144} atoms).

\subsection{Memory: backward variants and limits of the existing TP-accelerated backends}
\label{app:memory}
Table~\ref{tab:memory} compares the kernel-only backward variants of SevenNet (\texttt{recompute} and the default \texttt{split}, Sec.~\ref{app:fused}) with \jet{} kernel + table. Relative to \texttt{recompute}, \texttt{split} reduces MD step time by 7.5\% for 7net-omni and 3.8\% for 7net-0 while producing no measurable change for the two nano models. Its \SI{256}{\byte}-per-edge buffer increases model-call memory for 7net-nano-5.5 by 36\% at 5632 atoms and 55\% at \num{360448} atoms (12.2 against 7.9~GiB, main Fig.~\ref{fig:memory}a). For 7net-nano-6.0, this buffer halves the largest accommodated system size from \num{2883584} to \num{1441792} atoms. The table needs neither per-edge MLP storage nor MLP recomputation. It adds only the table itself (\SIrange{17}{227}{\mebi\byte} per model) and a transient harmonics-gradient buffer (Sec.~\ref{sec:groups}). It fits the same largest systems as \texttt{recompute} and runs MD 2.5--3.3$\times$ faster than \texttt{split}.

In the SevenNet LAMMPS scans, cuEquivariance encounters an out-of-memory failure at \num{360448} atoms for the two nano models, \num{152064} for 7net-0 and \num{45056} for 7net-omni. Its largest accommodated LAMMPS sizes (\num{152064}, \num{45056} and \num{22528} atoms) remain at or below those of the model-call scan (\num{180224}, \num{90112} and \num{22528} atoms, Sec.~\ref{sec:callscaling}), consistent with the additional memory required for ghost atoms and ML-IAP buffers. For the other four families, Table~\ref{tab:family_oom} lists the OOM limits and the whole-GPU memory shares behind the 4--55\% range of main Sec.~\ref{sec:memory}. In the 5632-atom cell, the existing TP-accelerated backend of EquFlash V1 already consumes 94\% of the H200 memory (whole-GPU peak, not listed in Table~\ref{tab:family_oom}).

\subsection{Consumer GPU}
\label{app:consumer}
On the 12~GB RTX 4070 SUPER under WSL2, we benchmarked SevenNet (Fig.~\ref{fig:perf_summary}e), MACE and NequIP-OAM (Fig.~\ref{fig:perf_summary}f) on the 5632-atom electrolyte, whereas Nequix and EquFlash were not evaluated on this GPU. Reported step times are means including rebuilds, taken as medians over three runs (seven for NequIP-OAM).

The SevenNet benchmarks underlie the consumer-GPU throughput rates of main Sec.~\ref{sec:memory}. A run is classified as out of memory when whole-GPU usage exceeds \SI{11800}{\mebi\byte} of \SI{12282}{\mebi\byte}, prior to WSL2 spilling to host memory. Both \jet{} modes execute 7net-nano-5.5 and 7net-0 up to the largest system tested (\num{45056} atoms), and accommodate 7net-omni up to \num{11264} atoms before failing at \num{22528} atoms. Among upstream backends, only FlashTP accommodates 7net-0 at 5632 atoms, while neither FlashTP nor cuEquivariance accommodates 7net-omni. OpenEquivariance, which already exceeds memory for 7net-0, was not evaluated for 7net-omni. All upstream backends accommodate 7net-nano-5.5 at \num{11264} atoms and fail at \num{22528} atoms, while FlashTP fails for 7net-0 at \num{11264} atoms. With JET-D3, the 5632-atom kernel + table step requires 13.62, 34.92 and \SI{240.5}{\milli\second\per\step} for 7net-nano-5.5, 7net-0 and 7net-omni, representing an increment of 6.7--\SI{6.9}{\milli\second\per\step} over the D3-free runs. Upstream \texttt{pair\_style d3} was not benchmarked on this GPU because it required several seconds per step under WSL2 compared with 0.12--0.19~s on the H200.

For MACE and NequIP-OAM, both backends execute through ML-IAP, with $N=4096$ for MACE. The existing TP-accelerated backend cannot accommodate any of the three MACE models or NequIP-OAM L, whereas \jet{} kernel + table is 3.3 and 3.8$\times$ faster for NequIP-OAM S and M. On the 352-atom cell, \jet{} kernel + table reproduces the forces of the existing fp32 backend of NequIP-OAM within \SIrange{1.15e-5}{1.30e-5}{\electronvolt\per\angstrom}. Whole-run peak PyTorch allocations of \jet{} kernel + table are 735, 3094 and 1480~MiB for MACE-MP-0 medium, MACE-MPA-0 medium and MACE-MP-0 large and 202, 656 and 1413~MiB for NequIP-OAM S, M and L, compared with 677 and 6015~MiB for OpenEquivariance on NequIP-OAM S and M. The OpenEquivariance step time for NequIP-OAM M was measured under the default allocator and decreases to \SI{121.9}{\milli\second\per\step} when the PyTorch allocator is capped at 80\% of GPU memory.

\clearpage
\section{Application examples}
\label{app:applications}
This section provides simulation and analysis details for the two transport applications of main Sec.~\ref{sec:applications} and main Tables~\ref{tab:viscosity} and~\ref{tab:argyrodite}. Experimental values serve only for comparison and were used neither to fit nor to select inputs.

\paragraph{Shear viscosity of carbonate liquids.} The three simulation cells contain pure EMC (670 molecules, \num{10050} atoms), a solution of 65 LiPF$_6$ in 292 EC and 440 EMC (\num{10040} atoms) and a solution of 148 LiPF$_6$ in the same solvent (\num{10704} atoms). The solvent has an EC:EMC ratio of 3:7 by volume, corresponding to 36~wt\% EC compared with 30~wt\% in the measured 1~M electrolyte used for comparison \cite{si:moganty_patent_2020}. At the mean NPT volumes, the salt concentrations are 0.94 and \SI{2.02}{\mole\per\liter} and the densities are 1.179 and \SI{1.280}{\gram\per\centi\metre\cubed}. Simulations were performed with 7net-nano-6.0 plus D3 (\texttt{e3gnn}, kernel + table, JET-D3) at \SI{298.15}{\kelvin} with a \SI{1}{\femto\second} timestep. Each liquid was equilibrated for \SI{1}{\nano\second} in NPT and subsequently sampled in three NVT production replicas of \SI{10}{\nano\second}.

The shear viscosity was evaluated from the Green--Kubo integral over five shear-stress components sampled every \SI{2}{\femto\second} and averaged over six blocks of \SI{1667}{\pico\second} in each replica. Viscosity values were extracted at integration cutoffs of \SI{50}{\pico\second} for pure EMC, \SI{100}{\pico\second} for the \SI{0.94}{\mole\per\liter} solution and \SI{150}{\pico\second} for the \SI{2.02}{\mole\per\liter} solution, with uncertainties reported as standard errors over the three replicas. All 148 PF$_6^-$ anions in the \SI{2.02}{\mole\per\liter} cell remained intact during production. In a \SI{50}{\pico\second} NVE verification at \SI{1}{\femto\second} with D3 on the \SI{0.94}{\mole\per\liter} cell (mean temperature \SI{300.7}{\kelvin}), the energy drift remained at \SI{2.8e-4}{\milli\electronvolt\per\atom\per\pico\second}.

\paragraph{Li diffusion analysis in Li$_6$PS$_5$Cl.}
For the Li diffusion coefficients of main Sec.~\ref{sec:applications}, the MSD was evaluated from unwrapped Li coordinates saved every \SI{1}{\pico\second}, with every frame used as a time origin. The displacement of the Li center of mass was subtracted prior to MSD evaluation. A trajectory was accepted when the lag time to reach \SI{14}{\angstrom\squared} did not exceed one quarter of the run length. The relative standard deviations follow He et al. (cited in the main text) as $3.43/\sqrt{N_\text{eff}}+0.04$. Here $N_\text{eff}$ is the maximum MSD in the fit window multiplied by the number of Li atoms and divided by $a^2$, where $a$ is the nearest-neighbor distance of the Li 24g sites (3.13--\SI{3.14}{\angstrom}). Across the ten trajectories ($T=300$--\SI{400}{\kelvin}, $x=0.25$--0.615), the lattice parameter of the fixed production cell spans 9.751--\SI{9.779}{\angstrom} and the production runs last 4.5--20.0~ns. The MSD fit windows range from 161--2622~ps at \SI{400}{\kelvin} to 3796--10000~ps at \SI{300}{\kelvin}, and the relative standard deviations are 5.2--6.7\%. A consistency check at \SI{400}{\kelvin} and $x=0.615$ repeated the run with FlashTP using the same model, JET-D3, initial structure and analysis. Its mean NPT volume is \SI{201592}{\angstrom\cubed}, compared with \SI{201604}{\angstrom\cubed} for \jet{}.

\paragraph{Application-cell timing and memory.} Step times on both application cells are shown in Fig.~\ref{fig:perf_summary}d and were measured with the constant-volume production inputs (\SI{1}{\femto\second} for the electrolyte and \SI{2}{\femto\second} for Li$_6$PS$_5$Cl) as medians over three runs. Peak GPU memory is 82.9, 12.8 and 1.3~GiB with e3nn, FlashTP and \jet{} kernel + table for the electrolyte and 37.4, 5.4 and 1.2~GiB for Li$_6$PS$_5$Cl.

\section{Supported layouts and limits}
\label{app:limits}

The kernels support path-and-channel TPs with $l_{\max}\le4$ and channel multiplicities in multiples of 8 up to 128 per layer, including both parities of a degree in 7net-omni. Wider layers require adapter splits (two 128-channel halves for 256-channel MACE-MP-0 large and 32-channel slices for 256-channel EquFlash). The kernel-only mode supports SevenNet's Bessel(8) basis and two hidden layers of 64 units, whereas the table kernel accepts radial functions of edge length and element pair. Execution is restricted to precompiled layer shapes. The released shapes of the five families are compiled ahead of time for the shared-memory and register limits of the build GPU (Sec.~\ref{sec:groups}), and supporting a new shape requires rebuilding the library. Graph replay is restricted to a single MPI rank, and multi-GPU execution was not benchmarked.

For MACE-MPA-0 and MACE-OMAT-0, whose radial transform depends on the element pair, the table kernel stores one table per unordered pair of the elements present. These tables occupy 112~MiB per pair across the two layers at $N=4096$ ($4N_W$ fp32 values per node, $N_W=512$ and 1280). The six elements of the electrolyte yield 21 unordered pairs and 2352~MiB, accounting for most of the difference between the 3.07~GiB whole-run peak of these models and the 0.76~GiB of MACE-MP-0 medium, which requires a single table set (main Table~\ref{tab:mace}). This memory scales as $s(s+1)/2$ with the number $s$ of elements present.
\clearpage
\renewcommand{\floatpagefraction}{0.05}\setcounter{totalnumber}{10}\setcounter{topnumber}{10}\renewcommand{\topfraction}{1}\renewcommand{\textfraction}{0}
\section*{Supplementary tables}

\begin{table}[!htbp]
\centering
\caption{\textbf{Per-edge arithmetic of the deployed models.} Layer-wise input irreducible representations (irreps), live tensor-product (TP) paths retained by the gate, radial-MLP output widths $N_W$ and per-edge fused multiply-add (FMA) counts of the radial MLP and TP for three SevenNet checkpoints. For comparison, all convolution paths are listed for MACE-MP-0 medium ($N_B=10$, three hidden layers of 64 units) and NequIP-OAM M ($N_B=8$, one hidden layer of 128 units). The radial-MLP count is $N_BH+(n_h-1)H^2+HN_W$ (main Sec.~\ref{sec:cost}), and the TP count is $\sum_p f_p\,m_{l_1(p)}$, where $f_p$ is the number of nonzero Clebsch--Gordan coefficients (termed fibers, Fig.~\ref{fig:sevennet_layer}b,c) of path $p$, each requiring one FMA per channel. Consequently, the radial MLP dominates whenever a path averages fewer than $H$ fibers. Irreps are denoted $m\times lp$ for $m$ channels of degree $l$ and parity $p$ (e even, o odd). The SO(3) model 7net-0 assigns even parity (e) to all irreps without inversion parity. Overall model attributes are summarized in Table~\ref{tab:family_models}.}
\label{tab:models}
\footnotesize\setlength{\tabcolsep}{4pt}
\resizebox{\linewidth}{!}{%
\begin{tabular}{lllrrrr}
\toprule
model & layer & input irreps & paths & $N_W$ & MLP FMA & TP FMA \\
\midrule
7net-nano-5.5 \cite{si:oh_lightweight_2026} & L0 & $32\times0e$ & 3 & 96 & \num{10752} & 288 \\
 & L1 & $32\times(0e+1o+2e)$ & 11 & 352 & \num{27136} & \num{2656} \\
 & L2 & $32\times(0e+1o+2e)$ & 3 & 96 & \num{10752} & 288 \\
\addlinespace
7net-0 \cite{si:park_scalable_2024} & L0 & $128\times0e$ & 3 & 384 & \num{29184} & \num{1152} \\
 & L1--L3 & $128\times0e+64\times1e+32\times2e$ & 15 & 960 & \num{66048} & \num{6848} \\
 & L4 & $128\times0e+64\times1e+32\times2e$ & 3 & 224 & \num{18944} & 480 \\
\addlinespace
7net-omni \cite{si:kim_optimizing_2026} & L0 & $128\times0e$ & 4 & 512 & \num{37376} & \num{2048} \\
 & L1 & $128\times0e+64\times1o+32\times2e+32\times3o$ & 34 & \num{1760} & \num{117248} & \num{24864} \\
 & L2 & $128\times0e+64\times(1o+1e)+32\times(2o+2e+3o+3e)$ & 64 & \num{3008} & \num{197120} & \num{47680} \\
 & L3 & $128\times(0o+0e)+64\times(1o+1e)+32\times(2o+2e+3o+3e)$ & 34 of 68 & \num{1760} & \num{117248} & \num{24864} \\
 & L4 & $128\times0e+64\times1o+32\times2e+32\times3o$ & 4 & 256 & \num{20992} & 704 \\
\addlinespace
MACE-MP-0 medium & L0 & $128\times0e$ & 4 & 512 & \num{41600} & \num{2048} \\
 & L1 & $128\times(0e+1o)$ & 10 & \num{1280} & \num{90752} & \num{11008} \\
\addlinespace
NequIP-OAM M & L0 & $32\times0e$ & 3 & 96 & \num{13312} & 288 \\
 & L1--L2 & $128\times0e+64\times1o+32\times2e$ & 11 & 768 & \num{99328} & \num{4416} \\
 & L3 & $128\times0e+64\times1o+32\times2e$ & 3 & 224 & \num{29696} & 480 \\
\bottomrule
\end{tabular}}
\end{table}

\begin{table}[!htbp]
\centering
\caption{\textbf{Architectural specifications and size groups of the 15 released models.} Total parameter counts and radial-MLP parameter fractions (percentage of parameters residing in the radial MLPs) refer to the released, unconverted checkpoints, and $l_{\max}$ denotes the highest spherical-harmonic degree in the convolution. Edge counts per atom are evaluated on the 5632-atom electrolyte cell. Models are grouped by parameter count into small ($<10^6$), medium ($10^6$--$10^7$) and large ($>10^7$) categories. MACE-OMAT-0 was trained on OMat24 \cite{si:barroso_omat24_2024}.}
\label{tab:family_models}
\small\setlength{\tabcolsep}{4pt}
\begin{tabular}{llrrrrrr}
\toprule
group & model & \begin{tabular}[b]{@{}r@{}}parameters\\(M)\end{tabular} & \begin{tabular}[b]{@{}r@{}}radial\\MLP (\%)\end{tabular} & layers & $l_{\max}$ & \begin{tabular}[b]{@{}r@{}}cutoff\\(\AA)\end{tabular} & \begin{tabular}[b]{@{}r@{}}edges\\per atom\end{tabular} \\
\midrule
small & 7net-nano-5.5 & 0.10 & 55.0 & 3 & 2 & 5.5 & 68.1 \\
 & 7net-nano-6.0 & 0.10 & 55.0 & 3 & 2 & 6.0 & 88.2 \\
 & NequIP-OAM S & 0.62 & 5.6 & 2 & 1 & 4.5 & 35.5 \\
 & nequix-mp-1 & 0.71 & 31.9 & 4 & 3 & 6.0 & 88.2 \\
 & nequix-oam-1 & 0.71 & 31.9 & 4 & 3 & 6.0 & 88.2 \\
 & 7net-0 & 0.84 & 29.2 & 5 & 2 & 5.0 & 50.2 \\
\addlinespace
medium & NequIP-OAM M & 3.19 & 7.6 & 4 & 2 & 6.0 & 88.2 \\
 & MACE-MP-0 medium & 4.69 & 2.8 & 2 & 3 & 6.0 & 88.2 \\
 & MACE-MPA-0 medium & 9.06 & 1.5 & 2 & 3 & 6.0 & 88.2 \\
 & MACE-OMAT-0 medium & 9.06 & 1.5 & 2 & 3 & 6.0 & 88.2 \\
 & NequIP-OAM L & 9.61 & 7.7 & 6 & 3 & 6.0 & 88.2 \\
\addlinespace
large & MACE-MP-0 large & 15.8 & 1.6 & 2 & 3 & 4.5 & 35.5 \\
 & 7net-omni & 25.8 & 2.3 & 5 & 3 & 6.0 & 88.2 \\
 & EquFlash V1 & 28.7 & 11.1 & 10 & 3 & 6.0 & 88.2 \\
 & EquFlash V2 & 44.9 & 6.4 & 10 & 4 & 6.0 & 88.2 \\
\bottomrule
\end{tabular}
\end{table}

\begin{table}[!htbp]
\centering
\caption{\textbf{Single-point force and energy errors relative to fp64 and repeated-call force differences.} Force-component errors (\si{\electronvolt\per\angstrom}) and rms per-atom energy errors (\si{\electronvolt\per\atom}) of each fp32 backend relative to the fp64 reference of the same model across ten electrolyte MD frames (5632 atoms, or 352 atoms for 7net-omni, NequIP-OAM L and EquFlash, on one H200). The fp64 reference is fp64 e3nn for SevenNet and each released fp64 model for the other families.
For the non-SevenNet families, all evaluations, including the fp64 reference, receive identical fp32-rounded positions and cell vectors while constructing edge vectors in their native precision (fp32 or fp64). Maximum force errors are therefore comparable only within each family, whereas rms errors are less sensitive to this effect. \jet{} kernel + table uses $N=4096$ for MACE and $N=2048$ for all other families, with an additional MACE-MP-0 medium entry at $N=2048$. SevenNet evaluations return fp32 total energies, and MACE's fp32 atomic-energy summation affects all backends equally. For EquFlash, force accuracy serves as the primary validation metric and the fp64 reference accumulates atomic energies in fp64. The final two columns report the maximum and rms force-component differences (\si{\electronvolt\per\angstrom}) between two consecutive fp32 evaluations on the same frame. For SevenNet, the reported maxima span all ten frames at 5632 atoms across all four models, while rms repeated-call differences were not recorded.}
\label{tab:family_acc}
\label{tab:g8F}
\footnotesize\setlength{\tabcolsep}{4pt}\renewcommand{\arraystretch}{0.84}%
\resizebox{\ifdim\width>\linewidth\linewidth\else\width\fi}{!}{%
\begin{tabular}{@{}llccccc@{}}
\toprule
 & & \multicolumn{3}{c}{error against fp64} & \multicolumn{2}{c}{repeated call, $|\Delta F|$} \\
\cmidrule(lr){3-5}\cmidrule(l){6-7}
model & backend & rms $|\Delta F|$ & max $|\Delta F|$ & rms $|\Delta E|/N$ & max & rms \\
\midrule
7net-nano-5.5 & e3nn & \num{1.42e-6} & \num{1.56e-5} & \num{1.52e-7} & \num{5.48e-6} & -- \\
 & FlashTP & \num{1.42e-6} & \num{1.44e-5} & \num{1.26e-7} & \num{6.44e-6} & -- \\
 & \jet{} kernel only & \num{1.36e-6} & \num{1.30e-5} & \num{1.73e-7} & \num{2.15e-6} & -- \\
 & \jet{} kernel + table & \num{1.42e-6} & \num{1.28e-5} & \num{1.26e-7} & \num{2.38e-6} & -- \\
\addlinespace
7net-nano-6.0 & e3nn & \num{1.51e-6} & \num{1.51e-5} & \num{9.26e-8} & \num{5.90e-6} & -- \\
 & FlashTP & \num{1.51e-6} & \num{1.46e-5} & \num{1.20e-7} & \num{6.20e-6} & -- \\
 & \jet{} kernel only & \num{1.47e-6} & \num{1.12e-5} & \num{1.72e-7} & \num{2.62e-6} & -- \\
 & \jet{} kernel + table & \num{1.48e-6} & \num{1.30e-5} & \num{1.04e-7} & \num{2.38e-6} & -- \\
\addlinespace
7net-0 & e3nn & \num{1.29e-6} & \num{1.21e-5} & \num{1.68e-7} & \num{3.13e-6} & -- \\
 & FlashTP & \num{1.29e-6} & \num{1.21e-5} & \num{1.82e-7} & \num{3.58e-6} & -- \\
 & \jet{} kernel only & \num{1.18e-6} & \num{1.12e-5} & \num{2.08e-7} & \num{1.91e-6} & -- \\
 & \jet{} kernel + table & \num{1.27e-6} & \num{1.25e-5} & \num{1.74e-7} & \num{1.67e-6} & -- \\
\addlinespace
7net-omni (352) & e3nn & \num{1.27e-6} & \num{8.72e-6} & \num{3.29e-7} & -- & -- \\
 & FlashTP & \num{1.27e-6} & \num{8.63e-6} & \num{3.42e-7} & \num{4.17e-6} & -- \\
 & \jet{} kernel only & \num{1.24e-6} & \num{8.70e-6} & \num{2.72e-7} & \num{2.15e-6} & -- \\
 & \jet{} kernel + table & \num{1.27e-6} & \num{7.61e-6} & \num{3.43e-7} & \num{1.79e-6} & -- \\
\midrule
MACE-MP-0 medium & existing TP-accelerated fp32 & \num{4.54e-06} & \num{1.14e-04} & \num{2.73e-04} & \num{4.0e-6} & \num{4.7e-7} \\
 & \jet{} kernel + table & \num{4.54e-06} & \num{1.10e-04} & \num{2.72e-04} & \num{1.5e-6} & \num{2.3e-7} \\
 & \jet{} kernel + table, 2048 nodes & \num{5.50e-06} & \num{1.28e-04} & \num{2.67e-04} & -- & -- \\
MACE-MP-0 large & existing TP-accelerated fp32 & \num{4.82e-06} & \num{1.14e-04} & \num{1.45e-04} & \num{3.6e-6} & \num{5.4e-7} \\
 & \jet{} kernel + table & \num{4.67e-06} & \num{1.13e-04} & \num{1.46e-04} & \num{1.2e-6} & \num{2.0e-7} \\
MACE-MPA-0 medium & existing TP-accelerated fp32 & \num{6.77e-06} & \num{1.12e-04} & \num{2.87e-05} & \num{7.2e-6} & \num{1.3e-6} \\
 & \jet{} kernel + table & \num{4.80e-06} & \num{1.13e-04} & \num{2.66e-05} & \num{2.4e-6} & \num{3.4e-7} \\
MACE-OMAT-0 medium & existing TP-accelerated fp32 & \num{8.96e-06} & \num{1.12e-04} & \num{3.24e-05} & \num{1.8e-5} & \num{2.1e-6} \\
 & \jet{} kernel + table & \num{4.86e-06} & \num{1.06e-04} & \num{2.99e-05} & \num{3.3e-6} & \num{4.7e-7} \\
\addlinespace
NequIP-OAM S & existing TP-accelerated fp32 & \num{7.70e-06} & \num{1.06e-04} & \num{3.21e-07} & \num{2.1e-5} & \num{2.0e-6} \\
 & \jet{} kernel + table & \num{5.33e-06} & \num{1.05e-04} & \num{2.35e-07} & \num{5.5e-6} & \num{6.7e-7} \\
NequIP-OAM M & existing TP-accelerated fp32 & \num{5.54e-06} & \num{1.08e-04} & \num{5.51e-07} & \num{5.5e-6} & \num{7.3e-7} \\
 & \jet{} kernel + table & \num{5.38e-06} & \num{1.15e-04} & \num{2.77e-07} & \num{3.6e-6} & \num{5.2e-7} \\
NequIP-OAM L & existing TP-accelerated fp32 & \num{3.50e-06} & \num{2.68e-05} & \num{1.78e-07} & \num{5.4e-6} & \num{1.1e-6} \\
 & \jet{} kernel + table & \num{3.58e-06} & \num{2.86e-05} & \num{6.05e-07} & \num{2.5e-6} & \num{5.1e-7} \\
\addlinespace
nequix-mp-1 & existing TP-accelerated fp32 & \num{4.64e-06} & \num{1.11e-04} & \num{1.42e-07} & \num{1.9e-6} & \num{3.0e-7} \\
 & \jet{} kernel + table & \num{4.59e-06} & \num{1.13e-04} & \num{1.11e-07} & \num{1.5e-6} & \num{2.2e-7} \\
nequix-oam-1 & existing TP-accelerated fp32 & \num{4.63e-06} & \num{1.10e-04} & \num{2.45e-07} & \num{2.8e-6} & \num{3.8e-7} \\
 & \jet{} kernel + table & \num{4.57e-06} & \num{1.09e-04} & \num{2.45e-07} & \num{2.1e-6} & \num{2.8e-7} \\
\addlinespace
EquFlash V1 & existing TP-accelerated fp32 & \num{1.77e-06} & \num{2.55e-05} & \num{1.65e-06} & \num{1.4e-6} & \num{3.0e-7} \\
 & \jet{} kernel + table & \num{1.81e-06} & \num{2.24e-05} & \num{1.30e-06} & \num{1.2e-6} & \num{2.1e-7} \\
EquFlash V2 & existing TP-accelerated fp32 & \multicolumn{3}{c}{reference (fp64 does not fit)} & \num{2.3e-6} & \num{3.7e-7} \\
 & \jet{} kernel + table & \num{1.49e-06} & \num{1.13e-05} & \num{1.49e-06} & \num{1.1e-6} & \num{2.1e-7} \\
\bottomrule
\end{tabular}}
\end{table}

\begin{table}[!htbp]
\centering
\caption{\textbf{Single-point accuracy relative to fp64 in dense solids.} Rms and maximum force-component errors (\si{\electronvolt\per\angstrom}) and rms per-atom energy errors (\si{\electronvolt\per\atom}) of fp32 e3nn, \jet{} kernel only (k.o.) and \jet{} kernel + table (k.+t.) are evaluated relative to fp64 e3nn in the units indicated in the column headers. The isolated fp64 table rms force error is evaluated relative to the fp64 radial MLP, and the ratio denotes the fp32 e3nn rms force error divided by this isolated table error ($N=2048$, ten frames per system and model, one H200). Configurations were sampled every \SI{0.5}{\pico\second} from \SI{300}{\kelvin} NVT trajectories with a \SI{0.5}{\femto\second} timestep using \jet{} kernel + table and reinitialized velocities after supercell replication. Parentheses indicate atom counts for LiCoO$_2$, diamond Si ($3\times3\times3$ supercell of the 8-atom cubic cell) and amorphous SiO$_2$ ($2\times2\times2$ supercell of the 70-atom cell).}
\label{tab:solids_acc}
\footnotesize\setlength{\tabcolsep}{3.5pt}
\begin{tabular}{@{}ll*{11}{r}@{}}
\toprule
 & & \multicolumn{3}{c}{rms $|\Delta F|$ ($10^{-7}$)} & \multicolumn{3}{c}{max $|\Delta F|$ ($10^{-6}$)} & \multicolumn{3}{c}{rms $|\Delta E|/N$ ($10^{-7}$)} & \multicolumn{2}{c}{table alone, fp64} \\
\cmidrule(lr){3-5}\cmidrule(lr){6-8}\cmidrule(lr){9-11}\cmidrule(l){12-13}
system & model & e3nn & k.o. & k.+t. & e3nn & k.o. & k.+t. & e3nn & k.o. & k.+t. & rms ($10^{-8}$) & ratio \\
\midrule
LiCoO$_2$ (360) & 7net-nano-5.5 & 6.47 & 6.73 & 6.04 & 3.59 & 3.63 & 2.91 & 3.17 & 2.46 & 2.97 & 1.40 & 46 \\
 & 7net-0 & 5.71 & 5.61 & 5.94 & 3.49 & 3.10 & 3.17 & 2.60 & 2.69 & 2.73 & 0.784 & 73 \\
 & 7net-omni & 7.60 & 7.50 & 7.20 & 3.92 & 3.86 & 3.39 & 3.39 & 3.07 & 2.95 & 0.693 & 110 \\
\addlinespace
Si (216) & 7net-nano-5.5 & 8.81 & 8.45 & 8.19 & 3.10 & 3.14 & 2.87 & 2.12 & 1.70 & 1.95 & 0.934 & 94 \\
 & 7net-0 & 7.25 & 7.17 & 7.20 & 3.53 & 2.57 & 3.28 & 2.01 & 2.17 & 2.17 & 0.385 & 188 \\
 & 7net-omni & 8.05 & 7.53 & 8.38 & 3.24 & 3.09 & 3.13 & 2.46 & 2.66 & 2.66 & 0.765 & 105 \\
\addlinespace
a-SiO$_2$ (560) & 7net-nano-5.5 & 17.9 & 17.7 & 19.2 & 9.67 & 14.6 & 10.9 & 3.74 & 3.74 & 4.21 & 4.17 & 43 \\
 & 7net-0 & 15.7 & 15.7 & 16.5 & 9.23 & 10.1 & 8.85 & 3.69 & 3.69 & 3.69 & 2.04 & 77 \\
 & 7net-omni & 18.1 & 18.4 & 17.7 & 10.3 & 13.2 & 9.47 & 5.49 & 6.13 & 4.88 & 2.22 & 82 \\
\bottomrule
\end{tabular}
\end{table}

\begin{table}[!htbp]
\centering
\caption{\textbf{Thermodynamic averages and NVE trajectory divergence.} (a) Averages over the second half (50--100~ps) of 100-ps NVE runs with upstream FlashTP, \jet{} kernel only and \jet{} kernel + table, where PE denotes potential energy. (b) Trajectory divergence across pairs of 10-ps 7net-nano-5.5 NVE runs initialized from an identical state, where \jet{} denotes kernel + table (intact 5632-atom electrolyte, one H200, \SI{0.5}{\femto\second} timestep, no D3). RMSD is the root-mean-square atomic position difference between trajectories, and the rms force magnitude is \SI{1.41}{\electronvolt\per\angstrom}. The $|\Delta T|$ columns report first-passage times above the indicated thresholds. Step-0 force differences in (b) are evaluated within LAMMPS MD runs and therefore differ slightly from the standalone model-call spreads in Sec.~\ref{app:accuracy}.}
\label{tab:md}
{\small\textbf{(a)} Thermodynamic averages and energy drift}\\[2pt]
\small\setlength{\tabcolsep}{4pt}
\begin{tabular}{llcccc}
\toprule
model & backend & drift (\si{\milli\electronvolt\per\atom\per\pico\second}) & $T$ (K) & PE (\si{\electronvolt\per\atom}) & $P$ (bar) \\
\midrule
7net-nano-5.5 & FlashTP (upstream) & \num{-6.8e-6} & 305.8 & $-5.74572$ & \num{12477} \\
 & \jet{} kernel only & \num{-2.2e-5} & 304.7 & $-5.74557$ & \num{12452} \\
 & \jet{} kernel + table & \num{-1.5e-5} & 304.9 & $-5.74560$ & \num{12494} \\
\addlinespace
7net-0 & FlashTP (upstream) & \num{1.4e-5} & 300.9 & $-5.74597$ & 8375 \\
 & \jet{} kernel only & \num{1.4e-5} & 301.5 & $-5.74606$ & 8373 \\
 & \jet{} kernel + table & \num{1.2e-5} & 301.1 & $-5.74600$ & 8348 \\
\bottomrule
\end{tabular}
\par\medskip
{\small\textbf{(b)} Divergence of repeated runs}\\[2pt]
\footnotesize\setlength{\tabcolsep}{2.7pt}
\begin{tabular}{@{}lccccccc@{}}
\toprule
 & \multicolumn{2}{c}{step-0 $|\Delta F|$ (\si{\electronvolt\per\angstrom})} & \multicolumn{2}{c}{first $|\Delta T|$ above (ps)} & RMSD & \multicolumn{2}{c}{RMSD} \\
\cmidrule(lr){2-3}\cmidrule(lr){4-5}\cmidrule(lr){7-8}
pair of runs & max & rms & \SI{e-3}{\kelvin} & \SI{1}{\kelvin} & e-fold (fs) & 1 ps (m\AA) & 10 ps (\AA) \\
\midrule
FlashTP vs FlashTP (3 pairs) & \num{6.7e-6} & \num{1.2e-6} & 0.74--0.79 & 1.77--1.89 & 110 & 0.9--1.1 & 2.07--2.09 \\
\jet{} vs \jet{} (3 pairs) & \num{2.4e-6} & \num{4.9e-7} & 0.68--0.74 & 1.82--1.95 & 109 & 1.0--1.4 & 2.02--2.08 \\
FlashTP vs \jet{} (9 pairs) & \num{1.8e-5} & \num{2.8e-6} & 0.68--0.76 & 1.60--1.85 & 115 & 1.0--2.0 & 1.98--2.09 \\
\bottomrule
\end{tabular}
\end{table}

\begin{table}[!htbp]
\centering
\caption{\textbf{Comparison of kernel-only backward variants and radial tabulation in SevenNet.} Columns report rebuild-free LAMMPS step time, peak PyTorch allocation of a single model call in GiB and the largest accommodated system size in atoms (5632-atom electrolyte for step time and peak allocation, one H200, NVE, no D3). \jet{} kernel only uses \texttt{recompute} or the default \texttt{split} variant, and \jet{} kernel + table uses default $N$.}
\label{tab:memory}
\small
\begin{tabular}{llccc}
\toprule
model & variant & \si{\milli\second\per\step} & peak (GiB) & largest run (atoms) \\
\midrule
7net-nano-5.5 & kernel only, \texttt{recompute} & 8.60 & 0.187 & \num{2883584} \\
 & kernel only, \texttt{split} & 8.62 & 0.254 & \num{2883584} \\
 & kernel + table & \textbf{3.49} & 0.205 & \num{2883584} \\
\addlinespace
7net-nano-6.0 & kernel only, \texttt{recompute} & 10.68 & 0.196 & \num{2883584} \\
 & kernel only, \texttt{split} & 10.69 & 0.292 & \num{1441792} \\
 & kernel + table & \textbf{4.09} & 0.213 & \num{2883584} \\
\addlinespace
7net-0 & kernel only, \texttt{recompute} & 31.51 & 0.312 & \num{1441792} \\
 & kernel only, \texttt{split} & 30.31 & 0.327 & \num{1441792} \\
 & kernel + table & \textbf{10.94} & 0.419 & \num{1441792} \\
\addlinespace
7net-omni & kernel only, \texttt{recompute} & 226.4 & 1.665 & \num{360448} \\
 & kernel only, \texttt{split} & 209.5 & 1.784 & \num{360448} \\
 & kernel + table & \textbf{63.43} & 1.918 & \num{360448} \\
\bottomrule
\end{tabular}
\end{table}

\begin{table}[!htbp]
\centering
\caption{\textbf{GPU memory consumption of the existing TP-accelerated backends at their system-size limits.} Whole-GPU memory shares of the existing TP-accelerated backends and \jet{} kernel + table are compared at identical system sizes on one H200 (nvidia-smi, \SI{143771}{\mebi\byte} total capacity). Listed atom counts correspond to the first CUDA out-of-memory (OOM) failure of the existing backend, except for NequIP-OAM M and nequix-mp-1 in LiCoO$_2$ and amorphous SiO$_2$, which complete the largest scanned size. The existing TP-accelerated backends are cuEquivariance for MACE, OpenEquivariance for NequIP-OAM, the PyTorch backend with OpenEquivariance kernels for nequix-mp-1 and GGNN for EquFlash V1. EquFlash V2 is omitted from this comparison.}
\label{tab:family_oom}
\small\setlength{\tabcolsep}{4pt}
\begin{tabular}{llrrrr}
\toprule
 & & & existing TP-accel. & \multicolumn{2}{c}{\jet{} kernel + table} \\
\cmidrule(lr){5-6}
model & system & atoms & share (\%) & MiB & share (\%) \\
\midrule
MACE-MP-0 medium & electrolyte & \num{45056} & OOM & 6353 & 4.4 \\
 & LiCoO$_2$ & \num{39936} & OOM & 6043 & 4.2 \\
 & Si & \num{64000} & OOM & 8539 & 5.9 \\
 & a-SiO$_2$ & \num{70000} & OOM & 8675 & 6.0 \\
\addlinespace
NequIP-OAM M & electrolyte & \num{152064} & OOM & \num{20957} & 14.6 \\
 & LiCoO$_2$ & \num{39936} & 96 & 6193 & 4.3 \\
 & Si & \num{262144} & OOM & \num{31459} & 21.9 \\
 & a-SiO$_2$ & \num{70000} & 99 & 9397 & 6.5 \\
\addlinespace
nequix-mp-1 & electrolyte & \num{152064} & OOM & \num{34051} & 23.7 \\
 & LiCoO$_2$ & \num{39936} & 80 & 9997 & 7.0 \\
 & Si & \num{125000} & OOM & \num{28309} & 19.7 \\
 & a-SiO$_2$ & \num{70000} & 84 & \num{16779} & 11.7 \\
\addlinespace
EquFlash V1 & electrolyte & \num{11264} & OOM & \num{22817} & 15.9 \\
 & LiCoO$_2$ & \num{39936} & OOM & \num{79593} & 55.4 \\
 & Si & \num{17576} & OOM & \num{33891} & 23.6 \\
 & a-SiO$_2$ & \num{15120} & OOM & \num{29747} & 20.7 \\
\bottomrule
\end{tabular}
\end{table}

\clearpage
\section*{Supplementary figures}


\clearpage
{\renewcommand{\refname}{Supplementary references}
}